\documentclass[a4paper,12pt]{article}
\usepackage[margin=2.5cm]{geometry}
\usepackage{amsmath,amssymb}
\usepackage{enumitem}
\usepackage{xurl}
\usepackage[colorlinks=true,linkcolor=blue,citecolor=blue,urlcolor=blue]{hyperref}
\title{Quantum Chip Paradigm Framework}
\author{
\begin{minipage}{0.96\textwidth}
\centering
\normalsize
Cai\textsuperscript{\textdagger},
Ling Qiao\textsuperscript{1,2,\textdagger},
Bin Yang\textsuperscript{1,2},
Fumin Luo\textsuperscript{1,2},
WeiGui Guo\textsuperscript{1,2},\\
GuoRong Zhang\textsuperscript{1,2},
XueFei Liu\textsuperscript{1,2},
Fan Xu,
Qinglang Guo\textsuperscript{1,2,*}, and
Bin Wu\textsuperscript{*}
\\[0.8em]
\small
\textsuperscript{1}Yangtze Delta Industrial Innovation Center of Quantum Science and Technology, Suzhou, China, 215100\\
\textsuperscript{2}China Academy of Electronics and Information Technology, No. 11 Shuangyuan Road, Shijingshan District, Beijing, China, 100041\\[0.4em]
\small \textsuperscript{\textdagger}These authors contributed equally to this work.\\
\small E-mail: qiaoling@tgqs.net; 1763098000@qq.com\\
\small \textsuperscript{*}Corresponding authors: Qinglang Guo (gql1993@mail.ustc.edu.cn) and Bin Wu
\end{minipage}
}
\date{June 16, 2026}

\begin{document}
\maketitle
\tableofcontents
\newpage

\section{Abstract}

The development of Quantum Electronic Design Automation (Q-EDA) is accelerating alongside the transition of quantum chips from laboratory prototypes to scalable engineering systems. Drawing on the historical experience of classical integrated circuits (ICs) and Electronic Design Automation (EDA), this paper argues that current superconducting quantum chip development is at a stage analogous to the early "SPICE moment" of classical EDA: the scale of physical qubits, control complexity, frequency planning, packaging interconnects, process variation, and cryogenic measurement feedback are jointly driving the industry from experience-driven design toward model-driven design, data closed loops, and standardized engineering systems ([1]; [2]; [5]; [7]; [22]; [23]; [24]; [38]).

This paper systematically proposes the Quantum Chip Paradigm Framework and regards Q-EDA as part of the quantum chip development paradigm itself, rather than merely as a collection of software tools. Taking as references the abstraction hierarchy of classical EDA, the Gajski-Kuhn Y-chart, PDK/standard cell/PCell methodologies, and the evolution of RTL/IP/SoC, this framework points out that quantum chips cannot simply replicate the classical HDL-first route. Instead, they must start from underlying physical objects such as Josephson junctions, resonators, couplers, readout structures, control lines, and packaging environments, and establish bottom-up parameterized modeling, quantum circuit simulation, and manufacturing feedback loops ([35]; [83]; [88]; [89]; [90]; [107]; [108]).

Under this framework, this paper focuses on analyzing PCell, SPICE-Q, Quantum PDK, design-technology-measurement co-optimization (DTCO), and the hierarchical Q-EDA system. The paper further compares the fundamental differences between classical circuits and quantum circuits in terms of signal properties, environmental sensitivity, wiring complexity, and process tolerance, explaining that quantum chip design must simultaneously handle constraints such as coherent evolution of quantum states, open-system noise, measurement backaction, the no-cloning property, frequency crowding, and three-dimensional packaging interconnects ([13]; [18]; [82]; [85]; [91]; [92]; [94]; [110]; [111]).

Finally, this paper proposes a hierarchical Q-EDA framework extending from the physical structure layer, the qubit PCell layer, the logical qubit layer, the quantum arithmetic layer, and the functional quantum IP layer to the Quantum SoC system layer. This hierarchy emphasizes that the key to engineering quantum chips is not to establish in advance an abstract language detached from physical reality, but to transform design knowledge, physical models, layout rules, simulation results, manufacturing data, and measurement feedback into executable, reusable, and auditable engineering objects. The infrastructure formed by PCell, SPICE-Q, Quantum PDK, and the DTCO loop is crucial for supporting the future engineering development of large-scale quantum processors, logical qubits, and fault-tolerant quantum computing systems ([19]; [52]; [56]; [57]; [58]; [112]; [113]; [114]).

\section{The Historical Development of Classical EDA and the Necessity of the Quantum Chip Paradigm Framework}

\subsection{Overview}

Superconducting quantum computing is gradually moving from the era of noisy intermediate-scale quantum (NISQ) devices toward the era of fault-tolerant quantum computing (FTQC). This transition requires not only continued growth in the number of physical qubits, but also the formation of a more systematic engineering methodology for quantum chips across design, fabrication, verification, testing, and operational calibration. Surface-code and logical-qubit roadmaps indicate that fault-tolerant systems typically require large numbers of physical qubits, stable gate fidelities, and repeatable measurement feedback; therefore, quantum chip development cannot remain for long at the stage of laboratory-scale manual layout and case-by-case parameter tuning ([1]; [2]; [19]).

Drawing on the developmental history of classical integrated circuits (ICs), the current quantum chip industry is at a critical stage analogous to the early ``SPICE Moment'' of the classical semiconductor industry. In the 1970s, the emergence of SPICE (Simulation Program with Integrated Circuit Emphasis) shifted circuit design away from a mode dependent on engineers' experience and manual approximate calculations toward an engineering system based on device models, circuit equations, and numerical simulation. Subsequently, the development of Electronic Design Automation (EDA), the Mead-Conway VLSI methodology, Process Design Kits (PDKs), standard cell libraries, and Parameterized Cells (PCells) further advanced chip design from workflows dominated by individual experience to reusable, verifiable, and manufacturable industrial processes ([22]; [23]; [24]; [38]).

By contrast, current superconducting quantum chip design still relies to a large extent on electromagnetic-field simulation software, empirical parameter adjustment, and design rules accumulated within individual research teams. Tools that have emerged in recent years, such as Qiskit Metal / Quantum Metal, KQCircuits, MQT-DASQA, SQuADDS, and EDA-Q, have begun to cover parameterized layout generation, design databases, layout-to-simulation workflows, fabrication mapping, and automated design-space exploration. These efforts show that the basic technical conditions for Q-EDA are taking shape, but the industry as a whole has not yet established stable process interfaces, device-model libraries, verification rules, or cross-institutionally transferable design data structures comparable to those in classical EDA ([54]; [55]; [82]; [84]; [85]; [86]; [87]; [102]).

Therefore, this paper proposes a ``Quantum Chip Paradigm Framework'' centered on Quantum Electronic Design Automation (Q-EDA). Unlike mature-stage classical digital EDA, which often takes HDL and logic synthesis as its entry point, superconducting quantum chips at the current stage require a bottom-up development path driven by underlying physical devices. For superconducting qubits, the Josephson junction (JJ) is the key element that provides nonlinearity and artificial-atom energy-level structure; capacitors, inductors, transmission lines, resonators/cavities, couplers, and readout structures jointly determine qubit frequencies, coupling strengths, loss channels, and crosstalk structures. Thus, quantum chip design must start from these physical devices and electromagnetic boundary conditions, continuously iterate models through experimental data and fabrication feedback, and gradually establish reusable PCells, quantum-circuit simulation models, quantum SPICE models (SPICE-Q), and Quantum PDKs ([5]; [7]; [8]; [83]; [88]; [89]; [90]).

This paper further argues that if the current quantum industry introduces an HDL-first synthesis route analogous to the later stages of classical EDA too early, it may obscure unresolved issues in device models, process variation, packaging parasitics, frequency crowding, and measurement feedback. In the absence of mature device models, standardized logic cells, and stable fabrication processes, directly adopting high-level abstraction can easily disconnect design intent from the actual physical system. By contrast, establishing a closed loop of Design Technology Co-Optimization (DTCO) based on experimental data, physical models, and fabrication feedback better fits the current developmental stage of quantum chips ([82]; [83]; [85]).

In the framework proposed in this paper, Q-EDA is not merely a set of software tools, but also represents an engineering development paradigm for future large-scale quantum processing units (QPUs). Its goal is to establish, in the domain of quantum chip design, a standardized engineering system analogous to that of the classical semiconductor industry, thereby realizing the softwareization of design knowledge, the automation of development processes, and the scaling of fabrication capabilities. The essence of softwareization is to transform the experience accumulated by engineers during design, simulation, fabrication, testing, and operational calibration into executable code, reusable models, and auditable processes, thereby reducing the dependence of R\&D on individual experience, repeated trial and error, computational resources, and capital investment. Related topics will be further discussed in subsequent specialized articles on fabless quantum chip design and the commercial production of quantum chips.

\subsection{Introduction to Quantum Chips}

Over the past half century and more, the evolution of classical integrated circuits (ICs) has been driven not only by Moore's law and device scaling, but also by the establishment of highly standardized engineering paradigms. The emergence of SPICE marked the gradual shift of circuit design from empirical and manual calculation toward an engineering system supported by device models, mathematical equations, and computer simulation. Subsequently, the maturation of the Mead-Conway VLSI methodology, Electronic Design Automation (EDA), and Process Design Kits (PDKs) further promoted the wide application of standard cells and Parameterized Cells (PCells) in large-scale integrated-circuit design ([22]; [23]; [24]; [35]; [38]).

Superconducting quantum computing is now facing a similar, though not identical, paradigm transition. As the number of physical qubits expands from dozens to hundreds and even to the scale of thousands, quantum chip design will find it increasingly difficult to support the development needs of large-scale systems if it continues to rely mainly on isolated electromagnetic simulation, manual parameter tuning, and experiential knowledge accumulated within research teams. Existing studies show that the design process of superconducting quantum chips is essentially a closed-loop process that starts from target Hamiltonians, frequency allocation, and coupling relationships, works backward to identify physical layouts, and then continuously corrects models through electromagnetic simulation, fabrication outcomes, and experimental measurement. In this process, device layout, circuit schematics, effective Hamiltonians, packaging structures, and control signals are mutually coupled, and crosstalk suppression, radiation shielding, frequency collisions, and fabrication-deviation compensation must all be incorporated into a unified design flow ([5]; [7]; [51]; [83]). Therefore, the current industry is closer to its ``SPICE moment'' than to a mature EDA stage: it urgently needs a standardized automation framework capable of unifying physical layout, simulation, fabrication, measurement, and model updating ([52]; [85]).

In recent years, several design tools and workflows for superconducting quantum chips have appeared, including Qiskit Metal / Quantum Metal, KQCircuits, MQT-DASQA, SQuADDS, and EDA-Q, as well as the design databases, physical-layout mapping, simulation flows, and process-mapping mechanisms built around them. These tools constitute important starting points for Q-EDA, but the related work still mainly covers only parts of the quantum chip design chain and has not yet formed an industrial-grade full-stack system covering device mapping, process mapping, layout generation, simulation analysis, fabrication feedback, and measurement calibration ([82]; [84]; [85]; [86]; [87]; [102]). This also shows that quantum design automation at the current stage is more suitably understood as a component of a ``quantum chip development paradigm,'' rather than as a simple transfer of high-level abstractions from classical EDA into quantum hardware design.

Directly transplanting the classical HDL-level EDA paradigm into quantum design is immature, and may even weaken the necessary exploration of fundamental physical devices, fabrication processes, packaging parasitics, and measurement feedback in the early stage. A more reasonable path is to start from the basic components of superconducting quantum circuits and adopt a bottom-up physical modeling approach: beginning with Josephson junctions (JJs), capacitors, inductors, transmission lines, resonators/cavities, couplers, and readout structures, and gradually establishing reusable PCells, quantum-level SPICE models (SPICE-Q), and Quantum PDKs directly coupled to process technology. The corresponding superconducting-circuit Hamiltonian can be expressed in terms of node charge, node flux, and Josephson phase variables as

\[
H = \sum_i \frac{Q_i^2}{2C_i} + \sum_j \frac{\Phi_j^2}{2L_j} - \sum_k E_{J,k}\cos(\varphi_k),
\]

where $Q_i$, $\Phi_j$, and $\varphi_k$ denote charge, flux, and Josephson phase variables, respectively. Establishing a continuous design--simulation--fabrication--measurement iterative closed loop around this class of physical models is more consistent with the actual stage of engineering development for superconducting quantum chips ([7]; [8]; [51]; [52]; [88]; [89]).

This paper therefore proposes the ``Quantum Chip Paradigm Framework,'' treating Q-EDA as part of the quantum chip development paradigm itself rather than merely as a set of software tools. This framework emphasizes bottom-layer modeling centered on physical devices and process realities, uses standardized components, reusable libraries, and automated flows as its handles, gradually promotes the standardization, softwareization, and scalability of quantum chip design, and lays the foundation for the engineering development of future large-scale quantum processors ([52]; [53]; [82]; [85]).

\subsection{Historical Review of Classical EDA}

The development of classical Electronic Design Automation (EDA) was not driven simply by ``stronger tools.'' Rather, it resulted from the synchronous evolution of the scale expansion of the integrated-circuit (IC) industry, increasing design complexity, division of labor in fabrication processes, and standardized interfaces. Early chip design depended heavily on manual drawing, engineers' experience, and local simulation. The absence of a unified interface between design and fabrication led to inconsistencies among institutions in design flows, symbol systems, layout formats, and process descriptions. As circuit scale continued to grow, designers gradually realized that only by abstracting devices, processes, and design rules into standardized engineering languages could chip design move from a ``workshop-style'' process to a replicable, verifiable, and scalable industrial process ([27]; [28]; [38]).

In this historical process, the emergence of SPICE had milestone significance. SPICE advanced analog circuits from empirical approximate calculation into a unified framework based on equation solving and numerical simulation, enabling designers to evaluate circuit behavior before actual fabrication. In other words, SPICE was not only a simulator, but also the key turning point through which classical EDA moved from ``geometric drawing'' toward ``physical modeling.'' Thereafter, the core tasks of EDA gradually expanded from simple circuit drawing to multiple levels such as layout generation, parameter extraction, process-constraint checking, manufacturability analysis, timing verification, and power optimization. For today's IC design, SPICE remains an important foundational tool for analog and mixed-signal circuit verification, and this model-centered engineering methodology has also become the common language of subsequent EDA systems ([22]; [23]; [25]; [35]).

What truly drove the maturation of EDA was the VLSI design paradigm shift represented by Mead-Conway. Its core contribution was not merely the textbook itself, but the transformation of complex semiconductor-physics problems into engineering problems more amenable to collaboration, teaching, and verification through lambda design rules, hierarchical design, modular abstraction, and curricular systems. As this paradigm spread, MOSIS, multi-project wafer (MPW) services, and subsequent PDK systems were gradually established, enabling designers to reuse device models, layout rules, and verification flows under specified process conditions. The significance of PDKs is especially important, because they package material parameters, interlayer-connection rules, minimum-size restrictions, parasitic effects, and verification rules in fabrication processes into callable engineering interfaces, thereby forming a stable data-exchange relationship between design companies and fabrication plants. It was during this period that the abstraction level of EDA began to move upward: designers no longer needed to start from device-physics details each time, but could complete higher-level designs on the basis of standard cells, PCells, and process constraints ([24]; [35]; [38]; [105]; [106]).

From a broader historical perspective, the evolution of EDA can be understood as a process of continuously raising abstraction levels and strengthening industrial collaboration capabilities. The early focus was computer-aided drawing, layout, and local simulation; the middle stage developed toward device-level modeling, standard cell libraries, design-rule checking, and logic synthesis; and later stages entered RTL, system-level modeling, IP reuse, and hardware-software co-optimization. Modern EDA is no longer merely about ``drawing circuits,'' but instead spans the complete flow from specification definition, architecture design, logic synthesis, and physical implementation to test and verification. With CMOS scaling, heterogeneous integration, and the expansion of design spaces, machine learning and data-driven methods have also gradually entered EDA flows, becoming important supplements for layout prediction, timing estimation, parameter optimization, and automated design ([32]; [43]; [44]; [46]).

Therefore, the essence of classical EDA history is not the linear accumulation of several software tools, but a process of paradigm evolution characterized by continuous abstraction, standardization, and engineering formalization. Its key lesson is that only when device models, process rules, design languages, data formats, and verification flows are unified can chip design transform from local optimization dominated by individual experience into a scalable system of industrial collaboration. This historical logic has direct implications for the subsequent discussion of quantum chip development paradigms and the formation of Q-EDA ([32]; [38]).

\subsection{Early EDA and CAD Design}

\subsubsection{Early Chip Development and the Mixed Use of CAD/EDA}

Early chip development had no unified industry standard. For integrated-circuit design in its embryonic stage, the design flow usually depended on hand-drawn circuit diagrams, empirical parameter adjustment, local electromagnetic calculations, and oral communication among process engineers. During this period, the boundary between EDA and CAD was not clear. Many teams in fact alternated among ``computer-aided drawing, local simulation, and manual layout correction,'' rather than using a complete, standardized, and automated EDA flow in today's sense ([27]; [28]; [38]). This also means that early chip design was more a scattered engineering practice than a unified softwareized industrial system.

Historically, 1958 is usually regarded as a key starting point in the development of integrated circuits, because Jack Kilby demonstrated the first integrated circuit that year, proving that multiple electronic devices could be integrated on the same substrate. Kilby's later recollections of the invention of the integrated circuit also became important primary material for understanding this period ([60]; [62]). Subsequently, as transistor counts increased and device interconnect complexity rose, design methods relying solely on manual drawing and experiential judgment quickly exposed problems of efficiency and scalability. In other words, the emergence of the integrated circuit itself objectively pushed subsequent design methods from manualization toward tooling, and from local correction toward system modeling ([60]; [62]; [66]).

In computer-aided design (CAD), Sutherland's Sketchpad is usually regarded as a representative prototype of early interactive CAD. The importance of Sketchpad lies not only in its graphical interface, but also in its demonstration that geometric constraints, object instantiation, and interactive editing could jointly constitute a new engineering representation. However, parameterized CAD, layout-rule checking, and scalable constraint solving truly oriented toward complex chip fabrication constraints matured only later within EDA systems ([27]; [28]; [104]). This shows that the early focus of CAD was mainly to ``help engineers draw faster and more accurately,'' rather than to provide a full-flow modeling platform oriented toward complex constraints, rule checking, and automatic optimization as in the present.

From the perspective of engineering organization, another notable feature of early integrated-circuit development was that teams largely ``worked in their own camps.'' Different companies, research institutions, and laboratories often used their own symbolic conventions, process assumptions, layout habits, and data formats, lacking portable design interfaces. Fabrication information, layout information, and verification information were usually scattered across different documents and software systems, with weak interoperability. This fragmented state not only increased communication costs, but also impeded the reuse of design knowledge. Therefore, the essential problem of early chip development was not only insufficient tools, but also the absence of a unified engineering language, exchangeable model representations, and standardized mechanisms connecting design, fabrication, and verification ([27]; [38]).

Under the framework of this paper, this stage can be understood as the ``pre-standardization period'' of classical EDA: design tools had not yet formed a stable hierarchy, CAD and EDA were often used interchangeably, and design flows depended on individual experience rather than unified software platforms. Precisely for this reason, the subsequent emergence of SPICE, the Mead-Conway VLSI methodology, MOSIS/MPW, PDKs, standard cell libraries, and PCells had genuine paradigmatic significance. They did not simply add several tools, but gradually transformed scattered chip development activities into replicable, collaborative, and verifiable industrial flows ([22]; [23]; [24]; [32]; [38]; [105]; [106]).

\subsubsection{Significance of This Stage for Subsequent EDA Evolution}

The mixed use of early EDA and CAD shows that any complex chip industry, in its initial stage, must first solve the questions of ``what to represent'' and ``how to represent it,'' before addressing ``how to automate.'' That is, the prerequisite for a mature toolchain is that the design objects themselves must first be standardized, modularized, and made computable. For subsequent EDA 1.0, EDA 2.0, and even higher stages, what truly matters is not the enhancement of a single software capability, but whether the entire industry has established unified design syntax, process interfaces, data formats, and verification flows ([28]; [32]; [38]).

Therefore, the birth of the first IC in 1958, the early interactive CAD represented by Sketchpad, the design-automation exchange mechanisms represented by SHARE/DAC, and the workshop-style state of chip design before the 1980s together constituted the starting point of the history of classical EDA. They also show that standardized engineering systems do not appear naturally, but are urgently ``forced out'' by industry after design complexity continues to increase and manual methods can no longer scale ([60]; [62]; [104]).

\subsection{EDA 1.0}

\subsubsection{SPICE and the Starting Point of EDA 1.0 (1969--1972)}

The EDA 1.0 stage can usually be traced to the formation of CANCER, SPICE, and the subsequent SPICE2 series of tools between 1969 and 1972. The emergence of SPICE (Simulation Program with Integrated Circuit Emphasis) marked the shift of integrated-circuit design from an ``experience-driven manual analysis'' paradigm to an engineering paradigm of numerical solution and computer simulation based on circuit equations ([22]; [23]; [25]).

The core idea of SPICE is to uniformly transform device models, topological connections, and external excitations into a solvable system of differential-algebraic equations. A simplified modified nodal analysis form can be written as:

\[
\mathbf{F}(\mathbf{x}, t) + \frac{d}{dt}\mathbf{q}(\mathbf{x}, t) = \mathbf{u}(t),
\]

where $\mathbf{x}$ denotes unknowns such as node voltages and branch currents, $\mathbf{F}$ denotes resistive and nonlinear device contributions, $\mathbf{q}$ denotes dynamic energy-storage terms such as charge and flux, and $\mathbf{u}(t)$ denotes external excitations. Through Newton-Raphson iteration, time stepping, and sparse-matrix solution, complex analog circuits can be approximately predicted and verified before fabrication, thereby realizing the ``computabilization'' of circuit design.

The significance of SPICE lies not only in the simulation tool itself, but also in its provision of a unified circuit-modeling language and exchangeable netlist representation, enabling designers to predict circuit behavior, compare design alternatives, and reuse models before actual tape-out. It was precisely this engineering logic of ``model first, fabricate later'' that laid the foundation for subsequent EDA automation systems.

\subsubsection{The Mead-Conway Revolution and the VLSI Design Paradigm (1980)}

After the late 1970s, as transistor counts grew rapidly, traditional manual layout design methods could no longer scale. In 1980, Mead and Conway systematically proposed the VLSI design methodology in \emph{Introduction to VLSI Systems}, which is regarded as a key turning point in the evolution of EDA 1.0 toward systematic design ([24]).

The core contribution of this methodology was the introduction of lambda-based design rules, which abstracted process dimensions into scalable design rules. At the same time, through hierarchical design, modular layout, rule-driven verification, and teachable design languages, it transformed chip design, which had previously depended heavily on the experience of process experts, into an engineering process more suitable for cross-team collaboration, curricular training, and rapid prototype verification. The subsequent development of MOSIS and multi-project wafer (MPW) services further connected this methodology to practical fabrication access, enabling universities and research institutions to complete closed loops of design, tape-out, and testing at lower cost ([24]; [105]).

Under this paradigm, the goal of EDA began to shift from ``drawing circuits'' to ``constructing, checking, and reusing circuit layouts under explicit rules and process constraints,'' that is, from geometry-centric to rule-centric. This transition directly influenced the subsequent formation of PDKs, standard cell libraries, and PCells: designers no longer merely manipulated geometric figures, but organized manufacturable engineering objects under the constraints of rules, models, and libraries.

\subsubsection{Early Microprocessor Development and the Evolution of EDA Capabilities (1971--1979)}

The development of microprocessors in the 1970s can be viewed as a direct manifestation of the gradual maturation of EDA 1.0 capabilities. Early microprocessors were not completed through fully automated EDA flows in today's sense; their design still relied heavily on engineers' manual logic planning, layout drawing, local circuit simulation, and repeated modification. Nevertheless, from the Intel 4004 to the Motorola 68000, the rapid expansion of transistor counts, instruction-set complexity, bus interfaces, and system-software ecosystems continuously pushed design methods from local experiential optimization toward more systematic hierarchical design, rule checking, simulation verification, and module reuse ([65]; [66]; [77]; [78]).

Therefore, the significance of this period lies not only in the release of several classic products, but also in its clear demonstration of ``how complexity forces toolchain evolution.'' When processors grew from roughly thousands of transistors to tens of thousands, reliance on individual experience and manual layout alone could no longer guarantee correctness, manufacturability, and iteration efficiency. This was the historical background in which SPICE, design rules, standardized layout interfaces, logic verification, and commercial EDA tools gradually moved toward central positions.

\paragraph{1971 Intel 4004 (approximately 2300 transistors)}

The Intel 4004 is usually regarded as the first commercial microprocessor. It originated from the Japanese Busicom calculator project, and its core idea was to replace a set of special-purpose logic chips with a programmable general-purpose processor. This idea was proposed by Ted Hoff, Stanley Mazor, and others; the hardware implementation was led by Federico Faggin and was closely related to Masatoshi Shima's input on system requirements. The 4004 was a 4-bit processor using silicon-gate MOS technology, contained approximately 2300 transistors, and provided a product form of a ``programmable logic core'' for the calculators and embedded control tasks of the time ([66]; [77]; [78]).

From the perspective of EDA capability, the 4004 was still in a highly manual stage: its logic structure, transistor-level implementation, layout arrangement, and timing risks mainly depended on the experience and local checks of the design team, rather than on complete automated synthesis, formal verification, or standard-cell flows. Its scale still belonged to the ``manually manageable range,'' but its real importance was that it transformed processors from special-purpose hardwired logic into reusable and programmable chip products, thereby beginning to amplify the demand for computable modeling, layout rules, device simulation, and design documentation.

\paragraph{1974 Intel 8080 (approximately 6000 transistors)}

The Intel 8080 was an important node in the movement of 1970s microprocessors from specific applications toward general-purpose computing. Compared with the 4004 and 8008, the 8080 used an 8-bit datapath, a 16-bit address space, and a 40-pin package, could address 64 KB of memory, and supported more complex system integration through stronger bus and I/O capabilities. Intel's retrospective materials also describe it as one of the first truly general-purpose microprocessors. The 8080 was later used in early microcomputers such as the Altair 8800, promoting the expansion of microprocessors from calculators, terminals, and embedded control into personal computing and general-purpose software ecosystems ([66]; [78]; [115]).

From the perspective of design methodology, the approximately 6000 transistors of the 8080 already required clearer functional partitioning, control-logic organization, bus-interface planning, and critical-path verification. At this time, circuit simulation tools such as SPICE could be used for local verification of key analog and transistor-level modules, but the processor as a whole was still far from automatic synthesis. Engineering teams still needed to iterate repeatedly among manual logic design, manual layout, test vectors, and system board-level verification. The 8080 therefore embodies the early characteristics of EDA 1.0: tools began to become necessary support, but design correctness still depended heavily on human architectural decomposition and experiential checking.

\paragraph{1975 MOS 6502 (approximately 3510 transistors, \$25)}

The MOS Technology 6502 was a representative low-cost microprocessor of the 1970s. It was promoted by Chuck Peddle, Bill Mensch, and the MOS Technology team, targeting price-sensitive embedded systems, personal computers, and consumer electronics markets. Its launch price of around 25 dollars was significantly lower than that of many competing processors at the time, enabling microprocessors to enter the broader product ecosystems of the Apple I/II, Commodore PET, Atari, and later game consoles. The commonly cited transistor count of the 6502 is approximately 3510, and this relatively small scale, together with concise instruction implementation, high yield, and cost control, formed the basis of its commercial success ([66]; [67]).

From the perspective of EDA and design methodology, the significance of the 6502 was not merely that it was ``cheap,'' but that it demonstrated the strong coupling among design complexity, chip area, fabrication yield, and market price. It cannot be simply equated with mature standard-cell design, but its cost-oriented circuit organization, reuse of repeated structures, and compression of unnecessary logic already embodied early engineering ideas that later appeared in modular design, layout reuse, and manufacturing-oriented optimization. In other words, the 6502 shows that the goal of EDA is not only to make more complex circuits, but also to make usable designs sufficiently small, stable, and manufacturable under constraints of process, area, yield, and testability.

\paragraph{1976 Zilog Z80 (approximately 8500 transistors)}

The Zilog Z80 was developed by Federico Faggin, Masatoshi Shima, and others after leaving Intel, and was an important case in which the early microprocessor industry moved from single-product competition toward architectural compatibility and ecosystem competition. The Z80 maintained strong compatibility with the Intel 8080 software ecosystem while adding more registers, index registers, a more complex interrupt mechanism, and a more convenient single-power-supply system interface. Common sources usually report a scale of approximately 8500 transistors, while Zilog oral histories also record the design process, tape-out rhythm, and differences in transistor-count accounting ([66]; [116]).

From the perspective of EDA evolution, the complexity of the Z80 came not only from the increase in transistor count, but also from the ``compatibility constraint'' itself. To allow existing 8080 software and development tools to migrate, the design team had to maintain predictable behavior in instruction semantics, register behavior, bus timing, and exception/interrupt handling, while also adding new functional extensions. Such constraints meant that verification was no longer merely about confirming whether a single circuit module worked; it also had to confirm whether the interfaces among hardware, instruction set, development tools, and user software were consistent. Thus, the Z80 can be regarded as an important early signal of EDA's movement from circuit simulation toward logic-behavior verification, compatibility verification, and design intellectual-property protection.

\paragraph{1978 Intel 8086 (approximately 29000 transistors)}

The Intel 8086 was the origin of the x86 architecture and a key product in the migration from the 8-bit microprocessor ecosystem to 16-bit systems. It adopted a 16-bit datapath and a 20-bit address space, expanded accessible memory through segmented addressing, and later entered the IBM PC ecosystem through the 8088 version with an external 8-bit data bus. The Computer History Museum regards the 8086/8088 as one of the starting points of the x86 family and the ``microprocessor wars,'' because this architecture, originally transitional in character, ultimately became a long-term industrial standard through software compatibility and system ecosystems ([78]; [117]).

The 8086 is often listed as having approximately 29000 transistors, but this number must be interpreted with caution: subsequent die-level analysis has pointed out that if only actual active transistors are counted, the number is about 20000; if potential sites in ROM and PLA that could be placed but were not necessarily actually used are also counted, the figure approaches the traditional 29000 count ([118]). This difference in accounting itself also shows that by the 8086 stage, microprocessors already included microcode, PLAs, bus interfaces, control state machines, and multiple datapath structures, and design complexity could no longer be fully characterized by a single transistor count. For EDA, the 8086 represents a stage in which correctness expanded from the circuit level to the architectural, microcode, and compatibility levels: module partitioning, control-logic checking, timing constraints, and software-visible behavior all began to become core issues in the design flow.

\paragraph{1979 Motorola 68000 (approximately 68000 transistors)}

The Motorola 68000 was a representative high-complexity microprocessor of the late 1970s. It contained approximately 68000 transistors and adopted an architecture usually described as having mixed 16/32-bit characteristics: the external data bus was 16 bits, while the internal registers and many aspects of the programming model had a 32-bit orientation, and it supported a 24-bit address space. Compared with 8-bit processors such as the 8080, 6502, and Z80, the 68000 targeted more complex operating systems, high-level languages, and graphical computing environments, and was later used in the Apple Lisa, Macintosh, Amiga, Atari ST, and various workstations and embedded systems ([65]; [66]).

From the history of EDA, the significance of the 68000 is that it already approached the limits of manual design methods. Historical accounts indicate that its design still relied heavily on engineers' manual control logic, flowcharts, microcode, and layout reasoning, but control logic, exception handling, addressing modes, bus protocols, and internal datapaths at this scale had already forced teams to adopt stricter hierarchical organization, documented interfaces, local simulation, and rule-based checking. The 68000 was therefore a key pressure point before the industrialization of commercial EDA in the 1980s: when single-chip processors entered the tens-of-thousands-of-transistors scale, automated layout assistance, DRC, netlist verification, timing analysis, and more systematic logic verification were no longer merely efficiency tools, but necessary conditions for reliably completing complex chips ([65]; [66]).

\subsubsection{Summary of the EDA 1.0 Stage}

The essence of EDA 1.0 can be summarized as:

\begin{center}
\textbf{EDA 1.0 = SPICE numerical simulation + VLSI design-rule abstraction + modular design methodology}
\end{center}

This stage gradually shifted circuit design from experience-driven to model-driven: SPICE provided a unified simulation language and an entry point for device models, the Mead-Conway methodology promoted the standardization of design rules and hierarchical abstraction, and the growth of microprocessor scale from thousands to tens of thousands of transistors continuously amplified the demand for automation tools. As a result, design abstraction began to migrate from the device level toward the logic level, and the design flow gradually evolved from local optimization dominated by individual experience into a teachable, reusable, and verifiable engineering system.

Therefore, EDA 1.0 was not the emergence of a single tool, but the ``first systematic engineering revolution in circuit design.'' It directly corresponded to the technical demand created by the scale transition from thousands to tens of thousands of transistors, and also laid the foundation for subsequent HDL, PDKs, standard cell libraries, and the Fabless/Foundry division of labor ([23]; [24]; [65]; [66]).

\subsection{EDA 2.0}

\subsubsection{The Industrialization and Standardization Stage of EDA 2.0}

EDA 2.0 marks the stage in which Electronic Design Automation evolved from a ``collection of tools'' into ``industrial standard infrastructure.'' During this period, EDA was no longer merely a simulator or drawing tool used to assist design, but gradually became the core engineering interface connecting design companies, IP suppliers, EDA platforms, and wafer foundries ([35]; [38]).

The core feature of this stage was that design abstraction rose further from the circuit level to a unified framework of ``Hardware Description Language (HDL) + Process Design Kit (PDK) + verification rules.'' HDL enabled functional designs to be described, simulated, and synthesized at the RTL level, while PDKs packaged device models, layout layers, design rules, parasitic parameters, and verification scripts from fabrication processes into callable interfaces. Together, the two decoupled design and fabrication at the engineering level and laid the foundation for the division of labor between Fabless design and Foundry manufacturing ([35]; [36]; [37]; [38]).

The basic abstraction relationship can be expressed as:

\[
Chip Design = f(HDL, PDK, SPICE Models, PCell Libraries)
\]

where HDL is responsible for functional expression and the entry point for logic synthesis, PDK is responsible for physical constraints and process interfaces, SPICE models provide device-level verification, and PCell supports structured reuse in analog/custom layout.

\subsubsection{Cadence's Acquisition of Gateway and the Standardization of Modern EDA (1989)}

Cadence's acquisition of Gateway Design Automation in 1989 was one of the important events in the integration of modern commercial EDA platforms. Gateway's Verilog ecosystem, commercial simulator, and subsequent standardization process brought RTL design, logic simulation, and logic synthesis into a more unified industrial flow. The expansion of Cadence, Synopsys, Mentor, and other companies during the same period also moved EDA from dispersed simulation, layout, and verification tools toward platformized toolchains ([37]; [38]).

More broadly, the core change of this period was not a particular acquisition itself, but the fact that commercial EDA platforms, HDL standards, logic synthesis, physical implementation, and signoff verification began to be organized around common data models and process interfaces. The basic hierarchy of modern EDA thereby gradually stabilized: the RTL/HDL layer is responsible for functional modeling, the Gate/Netlist layer for logic synthesis and optimization, and the Physical Design layer for placement and routing, parasitic extraction, timing closure, and fabrication-rule verification. This hierarchy allowed design teams to transmit constraints and results among different abstraction layers, and also provided a basis for cross-company collaboration in large-scale chip development.

\subsubsection{HDL Standardization: VHDL and Verilog}

\paragraph{VHDL Standardization (1983--1987)}

VHDL (VHSIC Hardware Description Language) was promoted in the 1980s by the U.S. Department of Defense VHSIC program and became the IEEE 1076 standard in 1987, with several subsequent revisions. Its goal was to provide formal, simulatable, and reviewable descriptive capabilities for complex digital systems, enabling designs to be modeled and verified at the system and behavioral levels ([36]).

The importance of VHDL lies in its incorporation of parallel hardware behavior, signals, processes, hierarchy, and testbenches into a unified language framework, so that complex digital systems no longer had to be expressed only through schematics and gate-level netlists. Its system-level abstraction can be represented as:

\[
System Behavior \rightarrow \mathcal{L}_{VHDL}(Processes, Signals, Concurrency)
\]

\paragraph{Development and Standardization of Verilog (1990s--1995)}

Verilog was initially developed by Gateway Design Automation, subsequently entered a broader commercial EDA ecosystem, and became the IEEE 1364 standard in 1995. Compared with VHDL, Verilog's syntax was closer to procedural description styles familiar to engineers and was also easier to combine with logic simulation, RTL design, and synthesis tools, so it rapidly became widespread in industry ([37]; [38]).

The key role of Verilog was to promote RTL (Register Transfer Level) as the mainstream design level and to enable Logic Synthesis to automatically generate gate-level circuits from behavioral/register-transfer descriptions. Thus, chip development further shifted from ``drawing circuit diagrams'' to a flow of ``describing functions, applying constraints, automatic synthesis, and physical implementation.'' This period marked:

\[
EDA \rightarrow HDL\text{-}centric\ Design\ Flow
\]

\subsubsection{Formation of the PDK and Fabless Model (1990s)}

In parallel with HDL standardization, the emergence of PDKs and the maturation of the Fabless/Foundry model constituted another key element of EDA 2.0. A PDK is not merely a collection of files provided by a fabrication plant, but a structured engineering interface that packages process-layer definitions, device models, parameterized cells, design-rule checking (DRC), layout-versus-schematic (LVS), parasitic extraction (PEX), SPICE model libraries, and reliability constraints together. It enabled design teams to complete chip development under known process rules without directly controlling the fabrication process, thereby providing the engineering basis for industrial division of labor ([35]; [38]; [106]).

The introduction of PDKs meant that designers no longer needed to directly control fabrication processes, but could complete design, verification, and tape-out preparation under models and constraints provided by the process foundry. On the one hand, this change lowered the barrier for design teams to enter advanced processes; on the other hand, it also required foundries to release process capabilities in a verifiable, callable, and version-controllable form. Thus, a stable interface boundary was formed between design companies and wafer fabs, allowing the Fabless model to scale.

The Fabless model can be represented as:

\[
Value\ Chain = Design\ (Fabless) + Manufacturing\ (Foundry)
\]

This structure changed the division of labor in the semiconductor industry and made the separation between design companies and wafer fabs one of the mainstream business models. For the quantum chips discussed in this paper, the historical experience of PDKs and Fabless shows that design and fabrication can be truly decoupled only when fabrication capability can be stably encapsulated by models, rules, PCells, and verification flows.

\subsubsection{Summary of the Overall Characteristics of EDA 2.0}

The essence of EDA 2.0 can be summarized as three structural transformations:

\paragraph{(1) From circuit simulation to language systems}

SPICE remained important, but HDL became the dominant design representation.

\paragraph{(2) From device modeling to platformized design}

PDKs abstracted processes into reusable interfaces, decoupling design from fabrication.

\paragraph{(3) From vertical integration to industrial division of labor}

The Fabless + Foundry model became mainstream, meaning that the key interfaces of the chip industry shifted from internal enterprise experience to auditable, deliverable, and versioned engineering data packages. Overall, EDA 2.0 can be formalized as:

\[
EDA 2.0 = HDL + PDK + Logic\ Synthesis + SPICE\ Verification
\]

The key significance of this stage is that EDA was no longer merely a ``design tool,'' but became infrastructure for the semiconductor industry. Through HDL, PDKs, SPICE models, PCells, signoff verification, and data-exchange formats, it directly defined the organizational form and business model of the modern chip industry ([35]; [36]; [37]; [38]).

\subsection{EDA 3.0}

\subsubsection{EDA 3.0: From Toolchains to the IP Ecosystem}

The core feature of EDA 3.0 is that chip design further developed from ``automation of a single circuit or individual RTL module'' into ``system-level integration centered on reusable IP (Intellectual Property).'' At this stage, the focus of chip design was no longer only how to complete logic synthesis, placement and routing, and timing closure, but how to reuse pre-designed functional modules on a given platform and combine processor cores, memory controllers, accelerators, interconnect structures, and peripheral interfaces into scalable SoC (System-on-Chip) systems ([39]; [41]).

The reuse of pre-designed IP blocks rapidly became widespread in industry and gradually became the mainstream mode of modern chip development. The reason is that as chip scale and heterogeneity continued to increase, designing every functional block completely from scratch not only became extremely costly, but also carried enormous risk. By contrast, encapsulating mature compute units, interconnect units, accelerator units, and communication interfaces into IP blocks can significantly shorten design cycles, improve verification efficiency, and enhance portability and maintainability across different projects ([39]; [40]; [41]).

From the perspective of engineering methodology, EDA 3.0 emphasizes not the ``local optimum'' of a single module, but the ``combinatorial optimum'' within a platform-level architecture. A typical system-integration relationship can be written as:

\[
\mathrm{SoC} = \mathcal{A}\big(\{IP_i\}_{i=1}^{N}, \mathcal{N}, \mathcal{M}, \mathcal{P}\big),
\]

where $\{IP_i\}_{i=1}^{N}$ denotes multiple reusable IP blocks, $\mathcal{N}$ denotes the on-chip interconnect network (NoC or other interconnect structures), $\mathcal{M}$ denotes the memory hierarchy, $\mathcal{P}$ denotes peripherals and I/O interfaces, and $\mathcal{A}$ denotes the system-integration operator. This abstraction shows that the object of EDA 3.0 has risen from ``circuit function'' to ``system function,'' and that the key design problems have become communication, bandwidth, latency, power consumption, and scalability among modules ([39]; [79]).

\subsubsection{Platformized SoC and Reusable Design Methodology}

EDA 3.0 is important because it transformed chip design from a one-off project into platformized engineering. Research represented by ESP (Embedded Scalable Platforms) shows that modern SoCs can adopt modular and tile-based architectural organizations, combining pre-designed IP blocks through standard interfaces into system configurations of different scales, thereby achieving rapid customization while maintaining compatibility ([39]). The key to this approach is not whether a single module is sufficiently complex, but whether the module can be repeatedly integrated, verified, and deployed.

This platformized method differs from the traditional idea of ``designing from scratch.'' Traditional methods emphasize the uniqueness of each design, while platformized design places greater emphasis on reuse, compatibility, and scalability. For high-performance computing, deep-learning acceleration, embedded computing, and heterogeneous SoCs, reusable IP blocks have become one of the most important engineering prerequisites, because most products do not need to reinvent processors, caches, NoCs, DMAs, or standard communication interfaces, but need to reorganize these mature modules in a more appropriate way ([41]; [39]).

From the perspective of design methodology, IP reuse also promoted the evolution ``from component-level reuse to model-level reuse.'' Rincon et al. (2007) pointed out that merely reusing specific components themselves is insufficient; more important is the reuse of abstract models that can describe system behavior, communication structures, and design patterns. In other words, EDA 3.0 does not merely assemble ready-made modules, but reuses architectural patterns, interconnect patterns, and verification patterns at higher levels. This brings design methodology closer to modular and pattern-based development in software engineering.

\subsubsection{Industrial Significance of EDA 3.0}

The industrial significance of EDA 3.0 lies in its movement of chip design from ``single-project delivery'' toward ``continuous platform evolution.'' Under this model, a mature IP block not only serves a particular chip, but can also be reused across multiple product lines; an SoC platform is no longer merely a single product, but a system base that continues to evolve and expand. Recent SoC research further shows that pre-designed IP blocks have become basic constituent elements of heterogeneous SoCs and can migrate flexibly among different system configurations through tile-based architectures ([79]).

Therefore, EDA 3.0 can be summarized as:

\[
EDA 3.0 = IP Reuse + Platform-Based Design + System Integration
\]

Here, IP reuse addresses the problem of functional reuse, platform-based design addresses the problem of architectural organization, and system integration addresses the problem of scale expansion. Unlike EDA 2.0, which mainly solved the problem of ``how to describe circuits,'' EDA 3.0 mainly solves the problem of ``how to organize complex systems,'' namely how to integrate multiple mature modules into a complete chip platform that is manufacturable, verifiable, and maintainable ([39]; [40]).

From the framework of this paper, EDA 3.0 represents a further evolution of the chip design paradigm toward systems engineering, platform engineering, and ecosystem engineering. The essence of this stage is not the enhancement of a single tool capability, but the transformation of chip development from ``designing a circuit'' into ``building a continuously reusable hardware platform.''

\subsection{EDA 4.0: The Next-Generation Design Paradigm Driven by AI and Agentic AI}

The core feature of EDA 4.0 is the deep entry of artificial intelligence (AI) into key stages of chip design, and its gradual evolution from ``assistive intelligent tools'' into ``design agents capable of executing tasks (Agentic AI).'' Unlike EDA 2.0, which mainly solved ``how to describe circuits with HDL,'' and EDA 3.0, which mainly solved ``how to reuse IP and organize SoC platforms,'' the goal of EDA 4.0 is to endow the design flow itself with stronger capabilities for understanding, planning, generation, verification, and repair, so that chip design can move from static toolchains toward dynamic closed-loop systems ([32]; [43]; [44]).

\paragraph{AI in EDA: From Predictive Tools to Generative Tools}

Early AI for EDA mainly focused on prediction and optimization tasks, such as placement-and-routing estimation, timing prediction, congestion analysis, power modeling, and parameter search. The survey by Huang et al. (2021) systematically summarized the applications of machine learning in different design stages according to the EDA hierarchy, showing that AI was initially embedded in traditional EDA flows more as an ``enhanced analyzer'' than as a replacement for the flow itself. With the growth of data scale and the improvement of computing power, AI is no longer used only for classification and regression, but has gradually expanded to higher-level tasks such as automatic generation, automatic repair, and automatic planning ([32]).

The typical goal at this stage can be written as a PPA (Power, Performance, Area)-oriented optimization problem:

\[
x^{*}=\arg\min_{x\in\mathcal{X}} \Big(\lambda_P P(x)+\lambda_F F(x)+\lambda_A A(x)\Big),
\]

where $x$ denotes a design scheme, $P(x)$, $F(x)$, and $A(x)$ denote power, performance, and area metrics, respectively, and $\lambda_P$, $\lambda_F$, and $\lambda_A$ are weights. The change introduced by EDA 4.0 is that optimization no longer relies only on manual search; instead, AI models automatically propose candidate schemes in larger search spaces and use simulation or back-end tools for rapid screening and iteration ([32]; [46]).

\paragraph{LLM4EDA: Natural Language, Scripts, and HDL Generation}

In recent years, large language models (LLMs) have begun to become key technical components of EDA 4.0. Zhong et al. (2024) summarize the applications of LLMs in EDA into three categories: conversational assistants, HDL and script generation, and HDL verification and analysis. They further point out that important future directions include logic synthesis, physical design, multimodal feature extraction, and representation alignment related to circuit structures. In other words, LLMs are not only ``able to write code,'' but may become unified interfaces connecting design intent, engineering scripts, and verification flows ([43]).

This capability changes the entry point of chip design. In the past, engineers first had to master complex tool commands, scripting languages, and design flows. In EDA 4.0, designers may increasingly describe requirements directly in natural language, and the model then generates Verilog, Tcl, constraint files, test scripts, or verification templates. The abstraction relationship can be represented as:

\[
Specification \xrightarrow{\;\mathcal{M}_{\theta}\;} HDL / Script / Constraint,
\]

where $\mathcal{M}_{\theta}$ denotes a parameterized language model. The significance of this transition is that design knowledge is no longer only encapsulated in engineers' minds, but is gradually transformed into callable, auditable, and correctable model capabilities ([43]; [44]).

\paragraph{Agentic AI: From Generating Answers to Executing Tasks}

The higher stage of EDA 4.0 is not ``AI that can generate content,'' but ``AI that can execute tasks.'' The key to Agentic AI is that the model not only outputs results, but can also perform goal decomposition, task planning, tool invocation, feedback correction, and multi-round iteration. Existing surveys of LLM4EDA and circuit foundation models show that LLMs have already been used for HDL generation, EDA script generation, verification assistance, debugging, and design-knowledge retrieval. In the future, if connected with open-source EDA toolchains, simulators, and layout flows, agentic workflows may further take on cross-stage flow orchestration tasks ([43]; [44]; [45]).

A simplified agentic design loop can be written as:

\[
a_t = \pi_{\theta}(o_t, m_t, g), \qquad s_{t+1}=T(s_t,a_t),
\]

where $o_t$ is the current observation, $m_t$ is external memory or context, $g$ is the design goal, $a_t$ is the action selected by the model, and $T$ denotes the state transition brought by tools and the environment. Unlike static generation, the essence of agentic AI is ``continuous action under constraints,'' and it is therefore more suitable for handling the many repetitive, context-dependent, and iteratively corrected tasks in EDA, such as script generation, parameter search, constraint repair, coverage convergence, and back-end flow coordination ([44]; [45]).

\paragraph{Foundation Models and Data-Driven EDA}

Another support for EDA 4.0 is circuit foundation models. The survey by Wang et al. (2025) points out that this direction has already covered a large body of related work, and that a considerable portion of it was published after 2022, indicating rapid growth in recent years. They further divide circuit foundation models into two categories: encoder models for prediction tasks and decoder models for generation tasks. This distinction is important because it shows that AI in EDA is no longer merely ``assistive feature engineering,'' but is gradually forming a general model layer for circuit representation learning and generative design ([44]).

Methodologically, the role of foundation models is to condense previously scattered design experience into transferable representations:

\[
f_{\theta}: \mathcal{D}_{circuit} \rightarrow \mathcal{Z},
\]

where $\mathcal{D}_{circuit}$ denotes cross-stage and cross-design datasets, and $\mathcal{Z}$ denotes a shared circuit representation space. For EDA 4.0, this means that models must not only ``understand'' circuits, but must also be able to reuse representational capabilities across tasks, thereby supporting multiple stages such as prediction, generation, repair, and optimization ([44]).

\paragraph{Engineering Significance of EDA 4.0}

The essence of EDA 4.0 is the upgrading of traditional design automation into ``AI-native design automation.'' At this stage, AI is no longer merely accelerating a local task; rather, it gradually takes over intermediate steps in design flows, freeing human engineers from large amounts of repetitive labor and allowing them to focus more on architectural decisions, goal setting, and critical review. At the same time, agentic AI also brings new problems, such as hallucination, tool misuse, context loss, unclear boundaries of verification responsibility, and security issues. Therefore, EDA 4.0 cannot be simply understood as ``connecting an LLM to tools,'' but should be regarded as a new engineering system requiring constraints, evaluation, and governance ([43]; [45]).

Therefore, EDA 4.0 can be summarized as:

\[
EDA 4.0 = AI-assisted EDA + LLM4EDA + Agentic AI Flows + Foundation Models,
\]

Its goal is not to replace engineers, but to establish a design collaboration system that can continuously learn, plan, and execute. For the future chip industry, this means that EDA will further evolve from a ``toolbox'' into ``intelligent design infrastructure,'' and gradually enter a new stage centered on automated reasoning and autonomous collaboration ([32]; [43]; [46]; [44]).

\section{Quantum Chip Paradigm Framework}

\subsection{Overview}

\subsubsection{Background for Proposing the Quantum Chip Paradigm Framework}

A comparison with the development history of classical integrated circuits (Integrated Circuit, IC) and electronic design automation (EDA) shows that every scale transition in chip technology has essentially been accompanied by a simultaneous reconstruction of the ``design paradigm + toolchain + abstraction hierarchy + manufacturing interface.'' From the 1970s to the 1990s, classical EDA completed a standardized evolution from SPICE simulation, VLSI design rules, and MPW/MOSIS prototype manufacturing to HDL, PDKs, and commercial EDA platform systems, gradually transforming chip design from an experience-driven activity into a system engineering discipline characterized by engineering formalization, softwareization, and platformization ([22]; [23]; [24]; [35]; [38]; [105]).

Against this historical backdrop, current quantum chips, especially superconducting quantum circuits, remain in a transitional stage analogous to the period between the early and middle phases of classical EDA. Although the scale of quantum bits has grown from experimental systems with tens of qubits to systems at the thousand-qubit level ([3]; [61]; [63]), design methodologies still rely heavily on fragmented electromagnetic simulation, empirical parameter tuning, and experimental iteration, while lacking unified device model libraries, process interfaces, parameterized components, verification rules, and data structures that are transferable across institutions. This gap is not the absence of a single tool, but rather evidence that the quantum chip development paradigm has not yet completed its engineering encapsulation.

\subsubsection{The Essence of Q-EDA: Not Only a Tool, but a Paradigm}

In the development of quantum chips, the significance of Quantum Electronic Design Automation (Q-EDA) lies not only in ``providing design tools,'' but also in defining a new quantum chip development paradigm (Quantum Chip Paradigm Framework). This paradigm must simultaneously answer three fundamental questions: what the basic design objects of quantum chips are, how computable mappings can be established among the physical, simulation, manufacturing, and measurement layers, and how scalable quantum systems can be constructed through standardized abstractions.

For superconducting quantum chips, the basic design objects are not Boolean logic gates in classical digital circuits, but Josephson junctions, resonators, couplers, readout structures, control lines, packaging interconnects, and the physical qubits and logical qubits formed by their combinations. Q-EDA therefore cannot be merely an extension of classical EDA; instead, it should redefine ``how quantum hardware is systematically constructed.'' Similar to the role of SPICE in classical EDA, future Q-EDA systems need to establish a unified foundation for quantum circuit modeling and simulation, enabling design to shift from being ``experiment-driven'' to being ``model-driven + data-closed-loop-driven.'' This judgment is jointly supported by reviews of superconducting quantum device design, the SQuADDS database workflow, the scqubits circuit quantization tool, and continuous-time dynamical simulation tools such as Qiskit Dynamics ([5]; [7]; [82]; [83]; [88]; [89]; [90]).

\subsubsection{Paradigm Gap: The Absence of a Standardized Engineering System}

The core bottleneck of the current quantum industry does not lie primarily in algorithmic innovation, but in the absence of a standardized engineering system comparable to the one formed by classical EDA during the 1970s--1990s. This systemic absence is first reflected at the level of design abstraction: different research institutions typically use their own independent parameter models, electromagnetic simulation tools, layout conventions, and naming systems, making design results difficult to migrate, reuse, and cross-validate. Even when multiple teams are all designing transmons, resonators, or couplers, there is often no unified mapping among their geometric parameters, effective Hamiltonians, fabrication deviations, and measurement data.

Second, quantum chips lack a unified process interface analogous to a PDK. Superconducting Josephson junctions, thin films, dielectrics, interconnects, and packaging processes all exhibit significant batch-to-batch variation, and these variations directly affect qubit frequencies, decoherence, coupling strengths, and readout fidelities. Without versionable Quantum PDKs, PCell libraries, statistical process models, and verification rules, it is difficult to form a stable mapping relationship between design and fabrication ([5]; [7]; [8]; [83]).

Third, current quantum chips still lack cross-layer closed-loop design mechanisms. Classical EDA forms a closed loop from model to fabrication and then back to model through ``SPICE / HDL / PDK / Tapeout / Silicon Feedback,'' whereas current quantum chip design often remains at a non-systematic closed-loop stage of ``electromagnetic simulation $\rightarrow$ fabrication $\rightarrow$ experimental measurement $\rightarrow$ manual parameter tuning.'' For large-scale quantum processors, Q-EDA must incorporate electromagnetic simulation, quantum dynamical simulation, fabrication data, cryogenic testing, and operational calibration results into the same data closed loop in order to support predictable, reusable, and scalable engineering development ([51]; [82]; [85]; [90]).

Therefore, the development of Q-EDA is essentially not a matter of optimizing individual tools, but of reconstructing an engineering paradigm. It must unify physical devices, fabrication processes, layout rules, simulation models, measurement data, and system architectures into an executable, verifiable, and iterative engineering system.

\subsubsection{Core Ideas of the Quantum Chip Paradigm Framework}

The Quantum Chip Paradigm Framework proposed in this paper does not mechanically copy the hierarchical structure of classical EDA into quantum hardware, but instead establishes a set of engineering abstractions suited to the physical constraints of quantum chips. Its core ideas include three interrelated threads: bottom-up modeling starting from physical devices such as Josephson junctions, resonators, couplers, readout structures, and control lines; continuous model correction through a design-technology-characterization closed loop; and the encapsulation of reusable experience into PCell, SPICE-Q, Quantum PDK, and standardized data flows ([5]; [7]; [82]; [83]; [85]).

\paragraph{(1) Bottom-up physics-driven modeling}

The basic construction units of quantum chips are not Boolean logic gates, but physical devices such as Josephson Junctions, capacitors, inductors, coplanar waveguides, microwave resonators, couplers, and readout structures. For superconducting platforms, Josephson junctions provide nonlinearity, resonators and transmission lines determine microwave modes, and couplers and readout chains affect gate operations, crosstalk, and measurement fidelity. Therefore, the first layer of abstraction in Q-EDA must begin with physical Hamiltonians, electromagnetic boundary conditions, and material parameters, rather than only with ideal quantum-gate symbols ([5]; [7]; [8]; [83]).

A simplified superconducting-circuit Hamiltonian can be written as:

\[
H = \sum_i \frac{Q_i^2}{2C_i} + \sum_j \frac{\Phi_j^2}{2L_j} - \sum_k E_{J,k}\cos(\varphi_k),
\]

where $Q_i$, $\Phi_j$, and $\varphi_k$ denote charge, flux, and Josephson phase variables, respectively. Establishing automated quantization, energy-spectrum analysis, parameter sweeping, and electromagnetic simulation interfaces around such models is a prerequisite for transforming laboratory experience into reusable design objects; work such as scqubits, computer-aided quantization, SQuADDS, and EDA-Q has already provided early support from the perspectives of circuit quantization, design databases, and layout-simulation workflows, respectively ([82]; [85]; [88]; [89]).

\paragraph{(2) Design-technology-characterization closed loop (DTCO)}

Quantum chips must achieve stable scalability through closed-loop optimization among design (Design), technology (Technology), and experimental characterization (Characterization). This process is similar to the DTCO concept in classical EDA, but in quantum systems it places greater emphasis on noise modeling, frequency drift, statistical process variation, packaging parasitics, and cryogenic control chains. Fabrication deviations are not errors to be handled only during the later testing stage; rather, they should enter the optimization flow during the design stage in the form of parameter distributions, process corners, yield models, and measurement feedback ([5]; [8]; [51]; [83]).

Within this framework, layout geometry, electromagnetic simulation, equivalent Hamiltonians, process data, cryogenic measurements, and operational calibration should not be regarded as mutually isolated files, but should constitute a continuously iterative data closed loop. SQuADDS, EDA-Q, Qiskit Dynamics, and reviews of superconducting quantum chip design concerns all indicate that the key to engineering quantum chips is shifting from single-device optimization to cross-layer model updating and data-driven verification ([82]; [83]; [85]; [90]).

\paragraph{(3) Standardization and softwareization}

The goal of standardization and softwareization is to transform quantum design knowledge into executable, reusable, and auditable software objects through PCell, SPICE-Q, Quantum PDK, GDSII/manufacturing data flows, and verification rules. PCell enables transmons, CPW resonators, couplers, Purcell filters, readout lines, and test structures to be generated parametrically; Quantum PDK encapsulates material layers, layout layers, process rules, device models, DRC/LVS constraints, and manufacturing interfaces into an engineering environment callable by designers ([56]; [57]; [58]; [85]; [106]).

The essence of this process is to transform quantum chip development from laboratory engineering into a programmable engineering system. In this system, design experience no longer exists only in researchers' personal memories and local scripts, but is deposited into versioned libraries, models, workflows, and data structures. Only when these objects can be reused across projects, teams, and process lines will quantum chip development possess a collaborative foundation comparable to that of the classical semiconductor industry ([82]; [85]; [106]).

\subsubsection{Scope of Applicability and Research Boundaries}

This framework is primarily directed at superconducting quantum chip systems, but its methodology can be extended to other quantum computing implementation paths, including photonic quantum systems, trapped-ion quantum systems, and semiconductor spin qubits. Differences among physical platforms are mainly reflected in the implementation methods at the device layer, but they share a high degree of commonality across the three dimensions of ``abstraction hierarchy + engineering closed loop + standardized interface''; this is consistent with research directions in reviews of superconducting qubit engineering, quantum design automation frameworks, and hardware-agnostic control software such as Qibolab ([5]; [81]; [83]; [84]; [85]).

\subsubsection{Summary}

The core viewpoint of the Quantum Chip Paradigm Framework is that the scaling path for quantum computing should not rely solely on algorithmic breakthroughs, but must rely on the systematic reconstruction of the engineering paradigm. Just as classical EDA defined the structural boundaries of the modern semiconductor industry, Q-EDA and its paradigm framework will determine whether the future quantum chip industry can move from experimental science toward engineering and industrialization.

\subsection{Early Q-EDA}

\subsubsection{Origins of Early Q-EDA and the CAD Stage}

Early quantum electronic design automation (Early Quantum EDA, Early Q-EDA) can be understood as the stage in which quantum chip design transitioned from laboratory-level manual electromagnetic design toward preliminary computer-aided design (CAD), parameterized layout generation, and simulation workflows. This stage had not yet formed a complete standardized Q-EDA system, but features similar to those of early classical EDA had already begun to emerge: preliminary parameterization of device modeling, the nascent softwareization of design flows, and a trend from single-point experiments toward system-level integration ([5]; [6]; [54]; [55]; [82]).

At this stage, the core problem in quantum chip design remained how to start from physical devices, construct predictable quantum Hamiltonian models, and relate them to actual fabricated structures. Taking Josephson junctions, transmon qubits, and coplanar waveguide resonators as examples, designers must not only provide layout geometries, but also understand how geometric dimensions affect capacitance, inductance, resonant frequency, coupling strength, and loss channels. Tools such as Qiskit Metal, KQCircuits, Quantum Metal, and SQuADDS show that the focus of Early Q-EDA is shifting from isolated drawing to parameterized components, simulation interfaces, and design databases ([82]; [87]; [102]).

\subsubsection{2009: Early Two-qubit Superconducting Quantum Chips}

Around 2009, superconducting quantum circuits realized early two-qubit (2-qubit) coupled systems, one of the important milestones in the transition of quantum chips from single-qubit manipulation to basic quantum logic operations ([12]; [74]). Such experiments demonstrated the ability to generate two-qubit entangled states and perform gate operations between controllably coupled qubits, proving that superconducting circuits can not only realize coherent control of individual qubits, but also serve as a candidate route for scalable quantum computing platforms.

From an engineering perspective, this stage still relied heavily on a mode of ``manual design + electromagnetic simulation + experimental parameter tuning,'' and had not yet formed a unified design automation flow. Its design process can be summarized as:

\[
\begin{gathered}
\text{Device Geometry} \rightarrow \text{Electromagnetic Simulation}
\rightarrow \text{Hamiltonian Extraction}\\
\rightarrow \text{Experimental Calibration}
\end{gathered}
\]

This flow is similar to the experience-driven design of the pre-SPICE era in early classical EDA: designers had already begun to use computational tools, but models, process rules, and experimental feedback had not yet been encapsulated into stable reusable workflows.

\subsubsection{2023: IBM Condor and Thousand-qubit Systems}

In 2023, IBM released the ``Condor'' superconducting quantum processor, with 1121 physical qubits, marking an important milestone in the scaling development of quantum chips ([61]; [63]). The significance of this system lies not only in the increase in qubit count, but also in the significant rise in system engineering complexity: multilayer routing and packaging integration, large-scale crosstalk suppression, frequency planning, statistical control of process deviations, and co-design of control electronics and quantum chips all began to become unavoidable system-level problems ([15]; [17]; [61]; [63]).

At this scale, the traditional method of tuning qubits one by one has become unsustainable, and system design has begun to exhibit complexity explosion problems similar to those of the classical VLSI era. Recent research on modular superconducting quantum computing, cryogenic control electronics, and million-qubit resource estimation also shows that future bottlenecks will arise simultaneously from qubit quality, interconnect density, cryogenic thermal budget, control-stack scalability, and calibration data management ([95]; [96]; [97]). Therefore, this stage can be regarded as the transition point at which Q-EDA moves from experiment-driven design toward system-level engineering design.

\subsubsection{Summary of Engineering Features of Early Q-EDA}

The Early Q-EDA stage can be summarized as an engineering germination period in the transition from laboratory manual design to parameterized CAD/EDA prototypes. This stage had already produced capabilities such as parameterized layouts, local simulation interfaces, and device databases, but these capabilities had not yet formed cross-institutionally reusable PCell, Quantum PDK, SPICE-Q, and complete DTCO closed loops.

\paragraph{(1) Initial emergence of CAD-ization}

Design began to rely on parameterized tools and layout generation frameworks, such as Qiskit Metal / Quantum Metal, KQCircuits, and some KLayout/Python workflows. These tools can reduce the cost of manual drawing and support geometric parameter sweeps, GDSII output, and partial simulation interfaces, but overall they still lack unified PDKs, cross-process rules, and standard signoff flows ([54]; [55]; [87]; [102]).

\paragraph{(2) Physical modeling remains dominant}

Quantum systems still need to be verified through electromagnetic simulation, equivalent circuit models, and effective Hamiltonian extraction. Their basic relationship can be summarized as:

\[
H_{\mathrm{eff}} = f(geometry, \epsilon_r, L, C, E_J),
\]

where geometric dimensions, dielectric constant, inductance, capacitance, and Josephson energy jointly determine qubit frequency, anharmonicity, coupling strength, and readout response. Tools such as scqubits, computer-aided quantization, and Qiskit Dynamics show that quantum chip design still strongly depends on mappings from physical models to system behavior ([88]; [89]; [90]).

\paragraph{(3) A complete EDA closed loop has not yet formed}

Unlike classical EDA, Early Q-EDA had not yet formed a complete closed loop jointly composed of standard PCell libraries, Quantum PDK, automated placement and routing, SPICE-Q simulation, and data-driven DTCO. Existing tools more often address local links, such as parameterized layouts, local electromagnetic simulation, or design databases; the full chain from design intent to manufacturing signoff, cryogenic measurement, and model updating is still in formation ([82]; [85]).

\subsubsection{Paradigm Significance}

The essence of Early Q-EDA can be compared to the ``eve of SPICE'' stage in classical EDA:

\[
Early\ Q\text{-}EDA \approx Classical\ CAD\ Era\ (pre\text{-}SPICE).
\]

The core significance of this stage is that it demonstrates the early possibility of engineering scalability for quantum systems, preliminarily forms the idea of parameterized design, and lays the foundation for the subsequent SPICE-Q, Quantum PDK, and PCell systems in Q-EDA 1.0. Early Q-EDA is not a mature automation system, but a proof-of-engineering phase of quantum chip engineering, whose main role is to verify whether quantum systems can be systematically designed, modeled, fabricated, and repeatedly calibrated like classical chips.

\subsection{Q-EDA 1.0}

\subsubsection{Q-EDA 1.0: Establishing the Engineering Paradigm for Quantum Chips}

Q-EDA 1.0 marks the entry of quantum chip development from the experimental verification and CAD prototype stage of Early Q-EDA into an initial industrial stage oriented toward system scaling and engineering standardization. The core feature of this stage is that quantum chips gradually shift from controllable experimental apparatuses to scalable engineering systems, and their design complexity begins to exhibit a growth trend similar to that of the early industrialization stage of classical VLSI. Reviews of superconducting qubit engineering and design concerns, together with toolchain work such as SQuADDS / EDA-Q, all indicate that device parameters, layout geometry, Hamiltonians, simulation, and fabrication feedback are being incorporated into a more unified engineering flow ([5]; [82]; [83]; [85]).

Taking superconducting quantum computing as an example, the scale of physical qubits has expanded from tens of qubits to systems at the thousand-qubit level; routes such as IBM Condor indicate that quantum chips are entering a stage that requires support from systematic toolchains ([3]; [61]; [63]). This scale transition can be compared with the early VLSI stage in classical semiconductor history, such as the evolution from the Intel 4004 to the 8086, which marked a structural shift in design methodology from manual circuit design to EDA-driven design ([24]; [66]; [77]).

Therefore, Q-EDA 1.0 can be formally described as:

\[
\mathcal{S}_{Q} : \mathcal{O}(10^1) \rightarrow \mathcal{O}(10^3),
\]

where $\mathcal{S}_{Q}$ denotes the quantum chip system-scale mapping function. The key issue at this stage is no longer whether single-qubit control is feasible, but whether coupling, crosstalk, error accumulation, process uncertainty, and calibration closed loops in multiqubit systems can be modeled in a unified manner.

\subsubsection{The Dual-path Structure of the Classical EDA Modeling Paradigm and Its Quantum Counterpart}

The development of classical EDA formed two complementary modeling paths:

\paragraph{(1) Bottom-up Device Physics Driven modeling}

This method starts from physical devices, such as transistors or MOS structures, and establishes the ability to predict circuit behavior through SPICE-level models ([23]). Its core advantages are physical consistency and verifiability.

In quantum systems, the corresponding basic units are Josephson Junctions and superconducting resonant structures, whose dynamics are described by the RCSJ model:

\[
C \frac{d^2 \varphi}{dt^2} + \frac{1}{R} \frac{d \varphi}{dt} + I_c \sin \varphi = I_{\mathrm{ext}}
\]

This equation forms the basis of superconducting quantum circuit modeling ([12]).

\paragraph{(2) Top-down Architectural / HDL-driven modeling}

This method starts from system behavior and decomposes it layer by layer into logic units and circuit implementations ([36]; [37]). Its advantage lies in scalability and design abstraction capability, but it depends on the accuracy of underlying models.

In quantum systems, however, because of measurement collapse and decoherence effects, high-level abstractions must be strictly constrained by physical models; otherwise, they lead to unverifiable designs ([4]).

Therefore, at the Q-EDA 1.0 stage, the two paths must coexist:

\[
Valid Q-EDA = Bottom-up Physical Model \cap Top-down System Abstraction
\]

\subsubsection{The Key Role of SPICE in Classical EDA and Its Quantum Analogy}

The role of SPICE (Simulation Program with Integrated Circuit Emphasis) in classical EDA is not merely that of a circuit simulation tool, but that of a unified model layer connecting device physics and system behavior ([23]). Its core contributions include:

\begin{itemize}[leftmargin=2em]
\item Providing a numerical solution framework for nonlinear circuits
\item Establishing mappings between device-level parameters and system-level behavior
\item Supporting process-corner (PVT variation) analysis
\end{itemize}

In quantum systems, the corresponding requirement can be formalized as the quantum dynamical equation:

\[
i \hbar \frac{d}{dt} |\psi(t)\rangle = \hat{H} |\psi(t)\rangle
\]

where the effective Hamiltonian usually includes driving and noise terms:

\[
\hat{H}_{\mathrm{eff}} = \hat{H}_0 + \hat{H}_{\mathrm{drive}} + \hat{H}_{\mathrm{noise}}
\]

Current quantum EDA still lacks a unified solution framework similar to SPICE, and this gap is precisely the core motivation for proposing ``SPICE-Q'' ([5]; [7]; [83]). Existing tools already cover several subproblems: scqubits and computer-aided quantization target superconducting circuit quantization and energy-spectrum analysis, Qiskit Dynamics targets time-dependent Hamiltonians and Lindblad dynamics, and SQuADDS connects layouts, simulation, and searchable design databases; superconducting SPICE-like tools such as JoSIM also show that Josephson junctions and superconducting circuits can obtain specialized transient simulation support within a SPICE syntax framework ([52]; [59]; [82]; [88]; [89]; [90]).

It should be noted that, in quantum circuits, information is usually transmitted and manipulated by high-frequency microwave signals as carriers, which is fundamentally different from the modeling framework of traditional SPICE based on the lumped-element circuit assumption. Therefore, in its modeling method, SPICE-Q must extend from the quasi-static approximation of classical circuits to distributed-parameter systems, and further incorporate quantum dynamical descriptions to characterize system evolution. The specific technical parameters and implementation details of SPICE-Q will be given in detail in subsequent independent work; this chapter serves only as a conceptual explanation to present its basic processing framework.

\subsubsection{Toolchain Divergence: HDL-first vs Device-centric Q-EDA}

There is a clear divergence in the current quantum EDA toolchain:

\paragraph{(A) HDL-first high-abstraction path}

The HDL-first path attempts to describe quantum circuits directly from logical or system-level descriptions, such as quantum HDL, control description languages, or high-level quantum compilation models. This path is necessary for long-term system-level design, because future quantum processors will ultimately require programmable system descriptions, resource scheduling, and hardware-software co-design interfaces ([80]; [81]). However, at the current Q-EDA 1.0 stage, it still faces problems such as insufficiently stable device models, the absence of unified process interfaces, difficulty in standardizing noise models, and difficulty in expressing packaging/measurement feedback in advance.

Therefore, treating HDL-first as the only main route at the current stage can easily lead to premature abstraction: although high-level descriptions appear to improve design efficiency, without verifiable JJ, qubit, resonator, coupler, readout, and control-line models, it is difficult to ensure that logical designs can be mapped into real manufacturable and calibratable quantum chips ([5]; [7]; [83]).

\paragraph{(B) Device-centric Q-EDA path}

The Device-centric Q-EDA path takes physical devices as its core and prioritizes the construction of reusable components such as qubit PCells, resonator PCells, coupler PCells, readout PCells, control-line templates, and test structure libraries. Its goal is not to remain at manual layout, but to encapsulate common quantum devices and their electromagnetic parameters, port definitions, manufacturing-layer information, and simulation entry points into callable objects, and gradually build reusable design libraries, design databases, and process-compatible models ([56]; [57]; [58]; [82]; [86]; [87]; [102]).

This path is more consistent with the actual current stage of superconducting quantum chips: device models and process fluctuations still need to be continuously corrected through experimental feedback, and therefore high-level system abstractions must be built upon underlying PCell, SPICE-Q, Quantum PDK, and measurement-data closed loops. In other words, device-centric design does not exclude subsequent HDL or system-level orchestration; rather, it provides a verifiable physical foundation for them.

\subsubsection{Necessity of SPICE-Q and Quantum PDK}

By analogy with classical EDA, SPICE and PDK together constitute the infrastructure of the modern semiconductor industry: SPICE provides the core for device-level simulation, while PDK provides the interface for process and design rules. In quantum systems, the corresponding structure can be expressed as:

\[
SPICE\text{-}Q + Quantum\ PDK + PCell\ Library.
\]

SPICE-Q needs to cover capabilities such as numerical solution of many-body quantum systems, noise and decoherence modeling, parameter drift, statistical process modeling, and measurement-feedback reinjection; Quantum PDK needs to define material layers, layout layers, parameterized components for JJ/CPW/resonators/couplers, design rules, manufacturing constraints, and verification flows ([56]; [57]; [58]; [82]; [85]).

Open quantum systems are commonly described by the Lindblad master equation:

\[
\frac{d\rho}{dt} = -\frac{i}{\hbar}[\hat{H}, \rho] + \sum_k \mathcal{D}[L_k]\rho,
\]

where $\rho$ is the density matrix, $\hat{H}$ is the system Hamiltonian, and $L_k$ denotes dissipation-channel operators. This framework is widely used for modeling open quantum systems, and it also shows that SPICE-Q cannot be merely a syntactic extension of classical SPICE, but must be capable of handling the coupled relationships among quantum-state evolution, decoherence, and experimental measurement ([13]; [90]).

\subsubsection{Systemic Significance of Q-EDA 1.0}

The essence of Q-EDA 1.0 is not a tool upgrade, but a turning point in the engineering paradigm: the design flow shifts from experiment-driven to model-driven, from single-device design to system-level design, and from fragmented tools toward an initially standardized ecosystem. Formally, it can be expressed as:

\[
Q\text{-}EDA\ 1.0 = \{PCell, SPICE\text{-}Q, Emerging\ PDK, Device\ Calibration\ Loop\}.
\]

The mark of completion for this stage is not the appearance of any single piece of software, but the formation of stable and reusable logical-qubit PCells, corresponding SPICE-Q models, and a process closed-loop verification system. Only when these objects can be reused across projects, verified across processes, and continuously corrected with measurement data will quantum chip development begin to enter a structurally stable period analogous to the early industrialization stage of classical EDA, thereby laying the foundation for subsequent system-level co-design and DTCO maturity in Q-EDA 2.0.

\subsection{Q-EDA 2.0}

\subsubsection{Q-EDA 2.0: From Device-level Automation to System-level Orchestration}

At this stage, quantum chip development begins to move beyond the device-level parameterized design and basic simulation of Q-EDA 1.0, and further toward engineering orchestration at the system, process, and platform levels. Unlike classical EDA 2.0, in which HDL, PDK, and logic synthesis jointly constituted industrial standards, the focus of Q-EDA 2.0 is how to introduce higher-level description languages, standardized tool invocation interfaces, and scalable design flows oriented toward process and topological constraints into quantum chip development. QHDL, Qibolab, MQT-DASQA, and EDA-Q provide early references from the directions of hardware description, control orchestration, application-driven architectural mapping, and full-flow design automation, respectively ([80]; [81]; [84]; [85]).

It should be emphasized that Q-EDA 2.0 is still a stage in formation, rather than an already fully mature industrial standard. Although existing tools already cover several key links such as layout generation, simulation, database management, and control interfaces, most of them remain solutions with only ``partial-chain usability,'' rather than integrated full-flow systems ([81]; [82]; [84]; [85]; [86]; [87]).

\subsubsection{Q-HDL: Early Attempts at High-level Quantum Hardware Description}

At this stage, the quantum hardware description language Q-HDL is expected to become one of the mainstream high-level interfaces for development. In 2023, Netzer and Markidis proposed QHDL, a low-level circuit description language for gate-based quantum computing systems, whose design goal is to provide quantum computing systems with VHDL-like description capability and to establish a common framework for low-level integration between FPGAs and quantum systems ([80]).

However, it should be noted that QHDL is currently still early prototype work, rather than a standard language that has been widely accepted by industry. In other words, the significance of Q-HDL in Q-EDA 2.0 does not lie in its having already completely replaced device-level design, but in its indication that quantum chip development is beginning to exhibit a new hierarchical interface: designers no longer describe systems only through geometric parameters and electromagnetic simulation, but are beginning to attempt to write system behavior, component relationships, and control logic as programmable hardware descriptions ([80]).

Its abstraction relationship can be written as:

\[
\mathcal{D}_{Q2}=
\left\{
\begin{array}{l}
\text{Q-HDL},\ \text{SDK orchestration},\\
\text{PDK-aware layout},\ \text{process-constrained routing}
\end{array}
\right\}.
\]

This indicates that the key to Q-EDA 2.0 is not merely ``one more tool,'' but the formation of a unified representation layer capable of transmitting constraints among design, simulation, control, and manufacturing.

\subsubsection{Tool Invocation Languages and the SDK Orchestration Layer}

In Q-EDA 2.0, the importance of tool invocation languages rises significantly. Whether Tcl-style flow scripts or Python-based SDK interfaces, their role is no longer merely to ``assist drawing,'' but to become an intermediate orchestration layer running through layout generation, simulation submission, parameter sweeps, control configuration, and data retrieval. The work of Qibolab shows this trend especially clearly: it provides software-layer support for self-built quantum hardware platforms and realizes programmatic access to quantum control through pulses-oriented drivers, transpilers, and optimization algorithms ([81]).

From an engineering perspective, the significance of this tool invocation layer is that it unifies workflows originally scattered across different software, different experimental scripts, and different control consoles, thereby reducing repetitive manual operations and improving portability between experiment and design. That is, Q-EDA 2.0 does not rely solely on a stronger description language, but on an SDK ecosystem capable of connecting Q-HDL, layout tools, simulation engines, and control systems ([80]; [81]; [84]; [85]).

\subsubsection{From Planar Routing to Topology-aware Routing and Three-dimensional Interconnects}

It should be noted that the underlying capabilities invoked by relevant SDKs differ significantly across different stages of Q-EDA. In the Q-EDA 1.0 stage, the main goal of routing is usually to reduce interference with qubits and preserve microwave-signal fidelity as much as possible; whereas in the later stage of Q-EDA 2.0, as system density and functional complexity increase, routing rules may gradually expand from simple planar layouts to stronger topology-aware and even vertical interconnect design ([53]; [21]; [38]).

This change does not arise in isolation, but is closely related to the 3D integration route of superconducting quantum systems. Existing work has shown that flip-chip structures, indium bump bonds, superconducting through-silicon vias (TSVs), and multilayer interconnect structures can support higher-density readout and control while maintaining good quantum coherence properties ([91]; [92]; [94]). Subsequent studies further indicate that three-dimensional stacking and TSV interconnects are expected to reduce planar routing congestion and provide space for more complex control and packaging architectures ([91]; [94]).

Therefore, the routing optimization objective in Q-EDA 2.0 can be written as:

\[
\min_{\pi}\ \alpha\,\mathrm{crosstalk}(\pi)+\beta\,\mathrm{loss}(\pi)+\gamma\,\mathrm{violations}(\pi)+\delta\,\mathrm{routing\ cost}(\pi),
\]

where $\pi$ denotes a candidate routing scheme, and the objective function simultaneously constrains crosstalk, loss, rule violations, and overall routing cost. Compared with Q-EDA 1.0, the key change here is that the optimization object is no longer merely ``avoiding interference,'' but begins to consider the co-design of device interconnects, packaging hierarchy, vertical channels, and system topology ([21]; [38]).

\subsubsection{From PDK to the Possibility of a Fabless-style Ecosystem}

However, only after the formation of the first logical qubit, SPICE-Q, and PDK will the large-scale emergence of Fabless waferless quantum chip design companies become a genuinely discussable possibility. The judgment here is not that this ecosystem is already mature, but that once process interfaces, device parameters, layout rules, and simulation flows are encapsulated as reusable standards, a division-of-labor structure similar to that in the classical semiconductor industry may emerge between design and manufacturing. EDA-Q's discussion of fabrication process mapping, SQuADDS's construction of design databases, KQCircuits / Quantum Metal's support for parameterized layouts and simulation interfaces, QPDK's open-source implementation of superconducting quantum RF PCell/PDK, and GDSII-to-wafer's discussion of DRC, LVS, DFM, and MDP manufacturing data flows all indicate that an early technical foundation for this direction is being formed ([56]; [57]; [58]; [82]; [85]; [87]; [102]).

This definition is highly consistent with the role of PDK in the classical chip industry. As recent reviews of quantum supercomputing roadmaps have pointed out, in traditional chip design, a PDK is a key process interface provided by the foundry, determining which device models, rules, and constraints design engineers can use; if quantum chips are to move toward scalable engineering systems, they also need a similar process encapsulation layer that transforms manufacturing realities into standard information callable by design ([5]; [64]; [83]; [85]). In this sense, the endpoint of Q-EDA 2.0 is not merely a more complex design software package, but an industrial environment capable of supporting ``design-process separation, flow standardization, and cross-project module reuse.''

This will also change the currently fragmented landscape of quantum chip research and development. A more standardized Q-EDA 2.0 ecosystem may reduce fragmentation among different research institutions and companies in toolchains, layout conventions, and process assumptions, thereby promoting a more unified quantum chip market and a clearer division-of-labor structure. This judgment is an inference based on the existing PDK and open tool ecosystem, but it is consistent with the current trend of toolchains developing toward databaseization, platformization, and workflowization ([81]; [82]; [84]; [85]).

\subsubsection{Stage Definition of Q-EDA 2.0}

Taken together, the core features of Q-EDA 2.0 can be summarized as:

\[
\begin{aligned}
\text{Q-EDA 2.0}={}&\text{Q-HDL}+\text{SDK orchestration}\\
&+\text{topology-aware routing}+\text{emerging PDK stack}.
\end{aligned}
\]

It represents a transitional stage in which quantum chip development moves from ``device-level automation'' toward ``system-level orchestration and standardized interfaces.'' Compared with Q-EDA 1.0, the focus of Q-EDA 2.0 is no longer only the local design of a particular qubit or resonator, but enabling Q-HDL, control scripts, layout tools, simulation systems, and process rules to work together under the same engineering syntax ([80]; [81]; [84]; [85]).

\subsection{Q-EDA 3.0}

\subsubsection{Q-EDA 3.0: An IP-based, Systematized, and AI-native Quantum Chip Design Paradigm}

Q-EDA 3.0 marks the further evolution of the quantum chip design paradigm from toolchain-driven development (Q-EDA 1.0--2.0) into a stage of ``IP-based system engineering + AI-native design automation.'' At this stage, quantum chips are no longer designed primarily around individual devices (qubit, resonator, JJ junction) as core design objects, but gradually shift toward a system-level development mode in which functionalized quantum IP (Quantum IP blocks) serves as the basic design unit.

This trend has structural similarity to the evolution path of classical EDA from RTL/IP reuse to SoC design: design objects rise from individual devices and local wiring to composites of functional modules, interconnect structures, error models, control logic, and system scheduling ([39]; [40]; [41]; [68]; [69]; [70]). In the quantum domain, this means that design objects will also gradually shift from ``geometry and electromagnetic structures'' to composites of ``functional modules + error models + control logic'' ([5]).

Recent quantum system roadmaps and research on modular superconducting quantum computing have also clearly pointed out that future scalable quantum computing systems will depend on modular design, standardized control stacks, and cross-layer collaborative optimization ([61]; [95]; [96]; [97]).

\subsubsection{Functionalized Quantum IP (Quantum IP Blocks)}

In Q-EDA 3.0, functionalized quantum IP becomes the core abstraction unit. Such IP is no longer a single physical structure, but includes:

\begin{itemize}[leftmargin=2em]
\item Logical qubits (logical qubits with error correction encoding)
\item Encoding modules (surface code / bosonic encoding blocks)
\item Control and measurement-control systems (microwave control + readout stack)
\item Quantum algorithm modules (variational / QAOA / VQE subcircuits)
\end{itemize}

Its abstract form can be expressed as:

\[
Quantum System = \sum_i IP_i(\theta_i, \epsilon_i, \mathcal{H}_i)
\]

where $\theta_i$ denotes control parameters, $\epsilon_i$ denotes error models, and $\mathcal{H}_i$ denotes effective Hamiltonians.

The important change at this level is that the error budget begins to become a ``first-class constraint'' in IP design, rather than a later correction term ([2]; [5]).

\subsubsection{Integration of AI and Agentic EDA}

Another core feature of Q-EDA 3.0 is the AI-native design flow. In recent years, Agentic AI has been used in EDA automation workflows, including design generation, verification, tuning, and feedback closed loops ([45]).

Recent studies show that AI has begun to enter chip design automation, quantum optimal control, experimental feedback optimization, and other links; in superconducting quantum hardware, machine-learning modeling, deep reinforcement learning, and neural-network control pulses have all been used to improve gate design and control robustness ([43]; [98]; [99]; [101]). This mechanism can naturally be extended in the quantum domain as:

\[
Agentic Q-EDA = f_{AI}\big(Q-HDL, SPICE-Q, PDK, Experimental Feedback\big)
\]

where the AI agent not only executes design tasks, but can also perform dynamic optimization based on experimental data (such as $T_1$, $T_2$, gate fidelity, crosstalk matrix).

In addition, quantum control itself is also a key domain for deep AI involvement. Quantum Optimal Control has been widely used to improve quantum gate fidelity and reduce decoherence errors; machine-learning characterization, deep reinforcement learning, and neural-network pulse design in recent years further show that control models can gradually move from ``manual calibration'' toward data-driven closed-loop optimization ([73]; [98]; [99]; [100]; [101]). Together, these results indicate that quantum systems are entering a stage of ``control intelligence.''

\subsubsection{Modular Quantum Architectures and the IP Economic Structure}

As quantum systems scale up, single-chip architectures gradually evolve toward modular quantum systems (modular QPUs / quantum data centers). Multinode superconducting quantum computing architectures, million-qubit-level modular resource estimation, and distributed error-correction module schemes all indicate that future system design must simultaneously address local gate performance, intermodule link fidelity, cryogenic interconnects, and error-correction overhead ([95]; [96]; [97]).

This trend further abstracts the design object of Q-EDA 3.0 as a ``quantum system IP economy'':

\[
Quantum SoC = \mathcal{F}(\{Quantum IP Library\}, Control Stack, Interconnect Fabric)
\]

where:
\begin{itemize}[leftmargin=2em]
\item IP Library: standardized quantum functional modules
\item Control Stack: AI-enhanced control and compilation system
\item Interconnect Fabric: cross-QPU quantum interconnect network
\end{itemize}

Relevant studies show that modular architectures can alleviate single-chip scaling bottlenecks and support cross-chip entanglement and distributed quantum computing ([95]; [96]; [97]).

\subsubsection{From Control Plane to System-level Closed Loop}

Recent studies indicate that the control layer (control plane) of quantum systems is undergoing differentiation between openness and closure, with different vendors showing markedly different degrees of openness regarding pulse-level interfaces, hardware control APIs, and experimental feedback data. Hardware-agnostic control layers such as Qibolab and dynamical simulation tools such as Qiskit Dynamics show that the control stack is becoming part of the quantum chip design stack ([81]; [90]). This further strengthens the trend in Q-EDA 3.0 that ``the control stack is the design stack.''

At the same time, quantum system roadmaps emphasize three core indicators:

\[
Scale + Quality + Speed
\]

as the unified objective function for future system design optimization. Research on modular superconducting architectures and multinode quantum computing also shows that this objective function must further incorporate cross-layer metrics such as interconnect links, cryogenic power consumption, error-correction overhead, and system scheduling ([61]; [95]; [96]; [97]).

\subsubsection{Summary: The Essence of Q-EDA 3.0}

The essence of Q-EDA 3.0 can be summarized as:

\[
\text{Q-EDA 3.0} =
\left\{
\begin{array}{l}
\text{Quantum IP Economy},\ \text{Agentic AI Design Loop},\\
\text{Modular Architecture},\ \text{Closed-loop DTCO}
\end{array}
\right\}
\]

Its core transformations include:

\begin{itemize}[leftmargin=2em]
\item From device design $\rightarrow$ functional IP design
\item From toolchains $\rightarrow$ AI-autonomous systems
\item From single chips $\rightarrow$ modular quantum systems
\item From static design $\rightarrow$ closed-loop adaptive optimization
\end{itemize}

This stage means that quantum chip design is beginning to enter a structured evolutionary path integrating ``system engineering + artificial intelligence + automatic control,'' and may become a key intermediate form for the future scaling of fault-tolerant quantum computing (FTQC).

\section{Logical Layering of Classical-Circuit EDA}

The logical layering of classical electronic design automation (Electronic Design Automation, EDA) is not a single linear stack of tools "from low to high"; rather, it is a multidimensional abstraction system jointly constituted by the behavioral view, the structural view, and the physical view. The Gajski-Kuhn Y-chart describes VLSI design as a process of continuous transformation and refinement among the behavioral, structural, and physical domains; this methodology shows that the core of EDA is not simply drawing layouts, but maintaining consistency among function, structure, and manufacturing constraints across different abstraction layers ([35]; [107]).

From the perspective of historical evolution, classical EDA includes both bottom-up and top-down paths. The bottom-up path starts from transistors, interconnects, SPICE models, and layout rules, and gradually forms PCells, standard-cell libraries, and process design kits (PDK); the top-down path starts from system specifications, algorithms, RTL, logic synthesis, and physical implementation, mapping high-level functions into manufacturable circuits. The two are not mutually exclusive, but are mutually closed in modern EDA flows through netlists, timing models, library mapping, constraint files, and signoff verification ([23]; [35]; [36]; [37]; [38]; [108]).

In classical computer architecture, common abstraction levels usually include the physical layout layer (layout/circuit level), the device parameterized-cell layer (device/PCell level), the logic gate level, the arithmetic level, the register-transfer and functional IP layer (RTL / IP level), and the system and algorithm level. Together, these levels support the design flow from transistor networks to SoC platforms, and also explain why standard-cell libraries, logic synthesis, IP reuse, and high-level synthesis (High-Level Synthesis, HLS) have become core EDA capabilities at different stages ([39]; [40]; [41]; [68]; [70]; [109]).

However, this classical layering scheme cannot be directly transplanted into quantum-circuit design. Classical digital circuits hide low-level device nonidealities through voltage thresholds, noise margins, and Boolean logic, whereas coherence, measurement collapse, crosstalk, decoherence, and frequency crowding in quantum systems create stronger coupling between low and high levels. Therefore, the objective of the "Quantum Chip Paradigm Framework" is not simply to copy the classical EDA layering, but to redefine the hierarchical structure of functional abstraction so that classical semiconductor circuits and quantum circuits can be related through mappings that are comparable but not conflated ([2]; [5]; [7]; [83]).

\subsection{Part Level - Layout / Early Stage (Physical Layout Level)}

This level is usually referred to as the physical layout level, and its core concern is the geometric structure and physical implementation of the circuit. At this level, the design object is no longer an abstract logical function, but concrete conductor paths, device shapes, local interconnects, hierarchical relationships, and process manufacturability constraints. Circuit function must be realized at this stage through spatial structures, material layers, and manufacturing rules; therefore, this is the engineering form of representation closest to wafer processing.

In the history of classical integrated-circuit development, this level corresponds to early manual drawing and layout design automation tools, as well as to the physical design, mask data, and manufacturing signoff stages in modern EDA flows. Its design results are ultimately mapped to photolithography masks (mask layout) and wafer manufacturing processes, and are usually jointly constrained by layout data formats such as GDSII/OASIS, process-layer definitions, design rules, and verification scripts ([24]; [35]; [38]). From the perspective of the Gajski-Kuhn Y-chart, it belongs to the bottom-level representation of the physical domain and is the manufacturing object to which the behavioral and structural domains must ultimately converge ([107]).

From a physical perspective, the design constraints at this level include linewidth, spacing, interlayer connections, parasitic capacitance and parasitic inductance, interconnect resistance, interconnect delay, and manufacturability windows. In modern EDA flows, these constraints are usually verified through design rule check (Design Rule Check, DRC), layout-versus-schematic consistency check (Layout Versus Schematic, LVS), parasitic extraction (PEX), and design for manufacturability (DFM) analysis ([35]; [106]). Therefore, the physical layout layer is not a "drawing layer," but a signoff interface between design intent and manufacturing reality.

At this level, the circuit structure can be formally represented as a geometric mapping function:

\[
\mathcal{L}: \mathcal{C} \rightarrow \mathbb{R}^2,
\]

where $\mathcal{C}$ denotes the set of circuit functional units, and $\mathbb{R}^2$ denotes the two-dimensional layout space. This mapping emphasizes that, at this stage, function must be implemented through spatial structure rather than merely expressed through symbolic logic. For multilayer processes, $\mathbb{R}^2$ can also be extended into a geometric database with layer information and process constraints.

In superconducting quantum circuits, the importance of the physical layout layer is further amplified. Because qubits are extremely sensitive to their electromagnetic environment, layout not only determines wiring structures, but also directly affects key physical parameters such as decoherence times ($T_1$, $T_2$), crosstalk strength, resonant-frequency drift, packaging modes, and readout fidelity ([5]; [7]; [83]). For example, the length, bends, spacing, and grounding structures of microwave traces all change the effective impedance environment, thereby affecting quantum-gate fidelity and readout signal quality.

Therefore, in the context of quantum chips, this level is not only a manufacturing layer but also a physical modeling layer. Its design objective can be further extended as:

\[
\min_{\mathcal{L}} \; \alpha \cdot \mathrm{loss} + \beta \cdot \mathrm{crosstalk} + \gamma \cdot \mathrm{nonideal\ coupling},
\]

where the optimization variables include not only geometric layout, but also electromagnetic-field distribution, device coupling strength, package boundaries, and process tolerances. Recent work on parameterized layout generation, PCell, QPDK, KQCircuits, Quantum Metal, and EDA-Q for superconducting quantum chips shows that quantum layouts are shifting from manually constructed geometric objects into programmable, verifiable, and reusable engineering modules ([56]; [57]; [58]; [82]; [85]; [87]; [102]).

In summary, the physical layout layer is the shared bottom-level foundation of classical EDA and quantum EDA. Its core characteristics are strong physical constraints, geometry-dominated representation, and manufacturing-driven optimization. In the quantum chip paradigm framework, this level remains the physical anchor layer for all higher-level abstractions.

\subsection{Component Level - Transistor PCell (Device Parameterized-Cell Layer)}

At this abstraction level, the design focus rises from pure geometric layout to device-level modeling. Its core objects mainly include transistors, capacitors, resistors, and closely coupled local interconnect structures. Compared with the layout layer, the key change at this stage is that circuit function is no longer defined only implicitly by geometric structures, but is explicitly expressed through parameterizable device models, layout-generation rules, and SPICE models, thereby enabling a computable mapping between device physics and circuit function ([22]; [23]; [24]; [35]; [106]).

In classical EDA, the Transistor PCell layer is a key interface connecting the physical domain and the structural domain. It is neither a purely geometric shape nor a high-level logical function, but a generative engineering object with parameters, ports, layout shapes, electrical models, and process constraints. The formation of this level enables designers to manage large numbers of transistor instances through a unified device abstraction, and provides a foundation for subsequent reuse of standard cells, analog-circuit modules, and process libraries.

\paragraph{Formation of Device-Level Abstraction}

During the EDA 1.0 stage, transistors gradually evolved from structural geometric objects into standardized functional devices. The core basis of this transformation was the establishment of semiconductor device models: the current-voltage relationship of MOSFETs, parasitic capacitance, threshold voltage, channel-length modulation, and process parameters were formalized as models usable for SPICE simulation, thereby allowing device behavior to be embedded in circuit-level simulation flows ([22]; [23]; [35]).

For example, a simplified MOSFET strong-inversion saturation-region model can be expressed as:

\[
I_D \approx \frac{1}{2}\mu C_{ox}\frac{W}{L}(V_{GS}-V_{TH})^2,
\]

where $W$ and $L$ denote the channel width and length, respectively, $C_{ox}$ denotes the gate-oxide capacitance per unit area, and $V_{TH}$ denotes the threshold voltage. This expression is only an approximate form used in teaching and early analysis, while modern processes usually rely on more complex compact models such as BSIM; nevertheless, it illustrates how device parameters enter predictions of circuit behavior and also explains why consistency must be maintained between PCells and SPICE models ([35]).

\paragraph{Engineering Significance of Parameterized Cells (PCell)}

At this level, the introduction of parameterized cells (Parameterized Cell, PCell) marks the shift of device design from fixed layout instances to generative models. The same class of device can be defined through a parameter set, for example:

\[
\theta = \{W, L, t_{ox}, N_d, layout\ constraints\},
\]

where the parameters determine not only the electrical behavior of the device, but also the layout geometry, port positions, layer structure, and design-rule constraints. Design tools can automatically generate the corresponding layout structures according to these parameters and maintain consistency with SPICE models, symbolic views, and verification rules.

The PCell mechanism transforms foundational objects such as transistors, capacitors, resistors, and local interconnects from manually drawn graphics into library-driven reusable cells, thereby significantly improving the scalability of design. It is especially important in analog/custom circuits, memories, I/O, radio-frequency circuits, and process test structures, because these modules often require extensive parameter variation while still having to comply with unified process rules and layout templates ([35]; [106]). This mechanism has key significance in the industrialization of EDA, because it structures the relationship between device design and layout implementation as a repeatedly callable generation function rather than a one-off manual design process.

\paragraph{Rising Importance of Interconnect Structures}

At this abstraction level, interconnect (routing) begins to shift from an auxiliary structure into an important factor affecting circuit performance. As device dimensions shrink, interconnect delay gradually approaches or even exceeds the switching delay of transistors themselves, making system performance increasingly dependent on routing topology and the distribution of parasitic parameters ([35]).

System delay can be approximately expressed as:

\[
T_{total} = T_{device} + T_{wire}(R, C)
\]

where $T_{wire}$ is dominated by interconnect resistance and capacitance, and further affects signal integrity, power consumption, and timing margin. Therefore, the design problem at this stage has expanded from simple device selection into the joint optimization of "device + interconnect."

\paragraph{Structural Positioning in the EDA 1.0 System}

Component Level - Transistor PCell is an important component of the classical EDA 1.0 system. Its core role is to use SPICE-level physical models of devices as the behavioral basis, parameterized device cells (PCell) as the core design abstraction, and explicit incorporation of interconnect parasitic effects into performance analysis. As PCells, standard-cell libraries, and process rules gradually matured, circuit design shifted from experience-driven manual engineering to a model-driven engineering system.

This level provided foundational structural support for the subsequent scaling of VLSI. It enables designers to invoke verified device templates and process models without repeatedly drawing the geometric details of every transistor, and to further combine these templates into logic gates, standard cells, and higher-level functional modules ([22]; [23]; [24]; [35]; [106]).

\subsection{Logical Space Level - Gate / Logical Gate (Logic-Gate Abstraction Layer)}

At this abstraction level, the core mode of circuit-design representation is further elevated from the transistor level to the logic level. The design object no longer directly concerns the physical behavior of individual transistors or capacitors, but instead takes logic gates and standard cells as the basic computational units, such as AND, OR, NOT, NAND, NOR, XOR, and their composite structures. The key feature of this level is that circuit behavior begins to establish a strict correspondence with Boolean algebra, thereby realizing the first systematic abstraction from physical implementation to mathematical logical expression ([35]; [68]; [71]; [108]).

The logic-gate layer can be established because classical digital circuits use voltage thresholds, noise margins, and regenerative properties to approximately encapsulate low-level continuous circuit behavior into discrete 0/1 states. This abstraction is extremely important in classical EDA because it allows designers to temporarily ignore transistor-level details and perform design and optimization among logical function, timing constraints, and standard-cell libraries.

\paragraph{Formal Foundation of Logical Abstraction}

The core significance of the logic-gate layer lies in its introduction of a representation for a "discrete-state system." Each logic gate can be regarded as a mapping function from input Boolean variables to an output Boolean variable. For example:

\[
y = f(x_1, x_2, ..., x_n), \quad f \in \{AND, OR, NOT, NAND, NOR, XOR\}
\]

This form of representation transforms circuit design from an analog physics system into a discrete mathematical system, allowing Boolean algebra to be used for formal derivation and optimization ([68]).

The composability of logic gates is the key property of this level, namely that complex circuits can be constructed through combinations of a finite set of basic logic units:

\[
F_{circuit} = \bigcirc_{i=1}^{N} g_i
\]

where $g_i$ denotes a basic logic-gate function.

\paragraph{Standardized Logic Cells and Industrialized Design}

In the evolution from EDA 1.0 to EDA 2.0, logic gates gradually changed from design concepts into standardized industrial cells. The establishment of standard cell libraries enables each logic gate to possess not only a Boolean function, but also engineering properties such as layout height, pin positions, input capacitance, output drive strength, propagation delay, dynamic power, leakage power, and process-corner models. These properties are usually jointly provided by process libraries, Liberty timing libraries, layout libraries, and the PDK, allowing logic design to proceed without directly handling transistor-level physical details ([35]; [108]).

Therefore, a standard cell is not equivalent to an abstract logic gate. A logic gate answers "what Boolean function does this circuit compute," whereas a standard cell answers "how is this Boolean function implemented under a particular process in a manufacturable, timing-characterized, routable, and signoff-ready manner." The maturity of this level directly supports logic synthesis, static timing analysis, automatic placement and routing, and large-scale digital chip design flows.

\paragraph{Relationship Between the Logic Level and the Device Level}

A logic gate is essentially a combinational structure implemented by an underlying transistor network. For example, a CMOS NAND gate can be composed of four transistors, and its function is jointly determined by the on and off states of those transistors. However, in the logic-level abstraction, this device complexity is completely hidden, thereby realizing a "function-first" design paradigm.

This abstraction relationship can be expressed as:

\[
\mathcal{G}: Transistor Network \rightarrow Boolean Function
\]

This mapping is one of the most important abstraction transitions in the EDA system. It shifts design complexity from exponential growth at the physical level to combinational growth at the logic level, thereby supporting the development of medium- and very-large-scale integration (MSI/VLSI) ([24]; [35]).

\paragraph{Structural Significance in the EDA 1.0 System}

In the EDA 1.0 stage, the Logical Space Level constitutes the core of design-language standardization. It formalizes circuit behavior as a Boolean logic system and connects logical function, layout constraints, timing models, and device implementation through standard-cell libraries. As a result, device implementation and logical function are partially decoupled in engineering practice: designers can use gate-level networks to describe functions, and synthesis, mapping, and physical-implementation tools then lower the logical structures into a specific process.

The maturity of this level directly promoted the formation of modern digital-circuit design methodology, and became the theoretical foundation for subsequent HDL languages (such as VHDL and Verilog), logic synthesis, and library-mapping algorithms ([36]; [37]; [68]; [71]; [108]). For the quantum chip paradigm framework of this paper, this point also provides an important comparison: classical logic gates can become stable abstractions because their underlying device and process models are already sufficiently standardized; if quantum logic units are to acquire a similar status, verifiable physical models and process interfaces must first be established.

\paragraph{Summary}

The essence of Logical Space Level can be summarized as:

\[
Physical Devices \rightarrow Boolean Abstraction \rightarrow Composable Logic System
\]

This level marks the transition of EDA from a "device engineering system" into a "logical mathematical system," realizing the key leap in circuit design from physical implementation to mathematical expression.

\subsection{Arithmetic Level - Gate / Operation Gate (Arithmetic Operation Level)}

At this abstraction level, the focus of circuit design rises further from logic gates to arithmetic operation units. Unlike the logic layer, which mainly processes Boolean variables, the core objective of this level is to implement composable numerical computing functions, such as addition, subtraction, multiplication, and comparison. Therefore, this level can be regarded as a key transition layer "from logical representation to numerical representation" ([68]; [70]).

\paragraph{Structured Formation of Arithmetic Units}

At this level, basic arithmetic operations are constructed from combinations of logic gates into standard structural modules. For example, adders, subtractors, and arithmetic logic units (ALU, Arithmetic Logic Unit) become core functional units. These structures are usually composed of repeatable logic units, thereby realizing scalable computing capability.

For example, the basic structure of an $n$-bit adder can be recursively constructed from full adders, with the basic relationship:

\[
S_i = A_i \oplus B_i \oplus C_i
\]

\[
C_{i+1} = (A_i B_i) + (C_i (A_i \oplus B_i))
\]

where $S_i$ denotes the sum bit and $C_i$ denotes the carry signal. This structure reflects the basic mechanism by which arithmetic operations are mapped from Boolean logic combinations to numerical computation ([71]).

\paragraph{Arithmetic-Level Abstraction and Combinational Complexity}

Compared with the logic-gate layer, the key change at the arithmetic layer is a significant increase in combinational complexity. Logic gates usually have local input-output relationships, whereas arithmetic units introduce cross-bit dependencies (carry propagation), causing the system to exhibit chain-like or tree-like dependency structures.

Taking addition as an example, its time complexity is related to the carry-propagation path:

\[
T_{adder} \propto O(n)
\]

In a simple serial structure, carry propagation determines the overall delay; in an optimized structure (such as a carry-lookahead adder), this complexity can be reduced to:

\[
T_{CLA} \propto O(\log n)
\]

This optimization reflects an important feature of the EDA 2.0 stage: the shift from "functional-correctness design" to "structural-optimization design" ([70]).

\paragraph{Systematic Emergence of the Arithmetic Logic Unit (ALU)}

In the EDA 2.0 system, a typical representative of arithmetic-level abstraction is the ALU (Arithmetic Logic Unit). The ALU unifies logical and arithmetic operations within the same hardware structure, enabling a processor to execute basic operations such as addition, subtraction, logical operations, shifts, comparisons, and conditional selection according to control signals. Internally, it is usually composed of adders, logic-operation units, shifters, multiplexers, and status-flag generation logic; these submodules are not isolated circuits, but cooperate under a unified control path.

Its overall function can be expressed as:

\[
F_{ALU} = \sum_{k} \alpha_k f_k(A, B, control),
\]

where $f_k$ denotes different operation modes and the $control$ signal determines the specific execution path. This structure embodies the design idea of control-driven functional selection and is one of the foundations of modern processor microarchitecture ([68]; [70]; [71]).

\paragraph{System-Level Significance in EDA 2.0}

In the EDA 2.0 stage, arithmetic-level abstraction marks the entry of circuit design from logical-function implementation into computational-structure implementation. This level extends Boolean logic to numerical computation and introduces cross-bit dependencies, global delay optimization, area-power-performance tradeoffs, and synthesizable HDL descriptions. Modules such as adders, multipliers, comparators, shifters, and ALUs gradually become reusable structured computing units, and can be described and optimized through VHDL, Verilog, or later high-level synthesis tools ([36]; [37]; [68]; [70]; [108]).

The importance of this level lies in its expansion of "whether the circuit is correct" into "whether the computational structure is efficient." For example, an adder must not only generate the correct sum bits, but also make structural choices among carry propagation, area, power consumption, and timing closure; this is also an important sign of the transition of EDA from purely logical-correctness tools to architecture-level optimization tools.

\paragraph{Summary}

The essence of Arithmetic Level can be summarized as:

\[
Logic Gates \rightarrow Numerical Operations \rightarrow Structured Computational Blocks
\]

This level marks the transition of EDA from "logic-system design" to "computing-system design," and lays the foundation for subsequent RTL-level and system-level abstractions.

\subsection{Functional IP Level - IP Block / RTL Functional Circuits (Functional IP and RTL-Level Functional Circuits)}

At this abstraction level, the focus of circuit design rises further from arithmetic operation units to reusable functional intellectual-property modules (Intellectual Property, IP blocks). These IPs usually no longer correspond to a single basic operation, such as addition or a logic gate, but instead provide complete functional implementations for complex mathematical or engineering tasks, such as matrix multiplication, fast Fourier transform (FFT), and signal-processing pipelines.

\paragraph{From Arithmetic Structures to Algorithm-Level Hardware Mapping}

The core feature of this level is the systematic mapping from "algorithm to hardware structure." In RTL (Register Transfer Level) design, function is no longer expressed by a single circuit structure, but is jointly defined by register-transfer behavior and combinational logic. For example, matrix multiplication is usually expressed in hardware as:

\[
C_{ij} = \sum_{k=1}^{n} A_{ik} \cdot B_{kj}
\]

In hardware implementation, this expression usually does not directly correspond to a single circuit unit, but is decomposed into a combination of multipliers, adder trees, and pipeline architectures ([72]).

Therefore, matrix multiplication is essentially an "algorithm-level description" rather than a gate-level or arithmetic-level structure in the traditional sense.

\paragraph{IP Reuse and Third-Party Design Ecosystems}

In the evolution from EDA 2.0 to EDA 3.0, one important feature of the Functional IP Level is the formation of a third-party IP supply ecosystem. Complex functional modules are no longer always constructed by chip-design teams level by level from the bottom, but are integrated through reusable IP cores. Typical forms of IP include DSP processing units, matrix-multiplication accelerators, image/video coding modules, high-speed communication interface controllers, memory controllers, and on-chip interconnect modules.

These IPs are usually provided in the form of RTL, synthesizable HDL, gate-level netlists, or encrypted deliverables, together with testbenches, timing constraints, software drivers, and verification documentation. Synthesis tools then map them to target standard-cell libraries or FPGA resources, ultimately lowering them to the physical-implementation layer. The engineering significance of IP reuse is that it advances hardware design from "implementing functions from scratch each time" to a platform-based flow of "combining, verifying, and integrating mature modules under system constraints" ([39]; [40]; [41]).

\paragraph{High-Level Synthesis and Pipeline Structures}

With the development of high-level synthesis (High-Level Synthesis, HLS), algorithm descriptions can be directly transformed into RTL hardware structures. Taking matrix multiplication as an example, HLS tools usually map it, according to target constraints, into MAC arrays, adder trees, systolic arrays, hierarchical pipelines, and on-chip memory-access structures. The optimization objective is no longer only logical correctness, but a comprehensive tradeoff among throughput, latency, resource utilization, memory bandwidth, and power consumption ([70]; [72]; [109]).

Therefore, this level is essentially a unification of the algorithm-optimization problem and the hardware-structure design problem. HLS does not eliminate RTL, logic synthesis, or physical implementation; instead, it moves the design entry point further upward, enabling engineers to explore hardware architectures in a form closer to algorithms, while tools complete scheduling, resource binding, pipeline insertion, and RTL generation.

\paragraph{Clarification of the Layered Attribute of Matrix Multiplication}

It should be emphasized that matrix multiplication is not a circuit structure at a single abstraction level, but a computing paradigm that spans multiple abstraction levels. In different EDA stages, it takes different forms:

\begin{itemize}[leftmargin=2em]
\item Algorithm layer: expression of linear algebra operations
\item RTL layer: register transfer + multiply-accumulate structure
\item Gate level: logic networks of adders and multipliers
\item Physical layer: transistor implementation
\end{itemize}

Therefore, matrix multiplication essentially belongs to an "algorithm $\rightarrow$ hardware mapping problem," rather than a fixed hierarchical structure ([72]).

\paragraph{Structural Significance in the EDA 2.0 System}

Functional IP Level is one of the core features in the transition from EDA 2.0 to EDA 3.0. It marks the movement of hardware design from arithmetic modules to complete algorithmic IP, from single operation structures to reusable functional cores, from local logic optimization to system-level performance optimization, and makes RTL/system-level design one of the main entry points for modern chip development.

The maturity of this level shows that hardware design is no longer only a problem of structural construction, but a systems engineering problem jointly constrained by algorithms, architecture, verification, software interfaces, and physical implementation. It lays the foundation for modern SoC (System-on-Chip) design, platform-based IP reuse, and hardware-software co-design methodology ([39]; [40]; [41]; [70]; [109]).

\paragraph{Summary}

Functional IP Level can be summarized by the following mapping relationship:

\[
Algorithm \rightarrow RTL IP Block \rightarrow Synthesizable Hardware Structure
\]

This level marks the further evolution of EDA from "functional-unit design" to "algorithm-driven system-level hardware design," realizing a unified representation of computation and architecture.

\subsection{Algorithm / System / SoC Level - Algorithm to System to RTL / SoC / Subsystem (System-Level and System-on-Chip Abstraction Layer)}

At this abstraction level, the core object of circuit design expands further from a single functional IP module to a complete system-level architecture. The design focus is no longer limited to local computing units or reusable IP cores, but instead centers on the organization of a complete System-on-Chip (System-on-Chip, SoC), including compute subsystems, memory hierarchy, interconnect fabric, peripheral interfaces, software stacks, and power-management strategies.

From the perspectives of the Gajski-Kuhn Y-chart and modern SoC methodology, system-level design requires decisions to be made simultaneously across the behavioral, structural, and physical domains: the behavioral domain describes algorithms, protocols, and software-visible functions; the structural domain determines the organization of processor cores, accelerators, on-chip networks, caches, and controllers; and the physical domain constrains area, timing, power, packaging, and manufacturing feasibility ([39]; [41]; [107]). Therefore, system-level abstraction is not a purely software description detached from low-level implementation, but rather incorporates algorithmic requirements, hardware structures, and physical constraints into design-space exploration.

\paragraph{Mapping Relationship from Algorithm to System}

The core feature of this level is the three-layer mapping structure of "algorithm-system-hardware implementation." In the EDA 3.0 paradigm, a complex algorithm (for example, image processing, machine-learning inference, or a communication protocol stack) no longer directly corresponds to a single hardware module, but is decomposed into multiple subsystems that execute cooperatively.

This mapping relationship can be formally expressed as:

\[
\mathcal{A} \rightarrow \{S_1, S_2, ..., S_n\} \rightarrow \{RTL_1, RTL_2, ..., RTL_n\}
\]

where $\mathcal{A}$ denotes a high-level algorithm description, $S_i$ denotes a system-level functional partition, and $RTL_i$ denotes the corresponding register-transfer-level implementation.

This decomposition process usually relies on design space exploration (Design Space Exploration, DSE), whose objective is global optimization among performance, power, and area (PPA, Power-Performance-Area) ([69]).

\paragraph{Systemic Features of SoC Architecture}

At this abstraction level, an SoC is no longer an extension of a single processor core, but a computing platform jointly constituted by multiple heterogeneous subsystems. A typical SoC usually includes a CPU subsystem, a GPU or dedicated accelerator, a memory hierarchy, DMA/cache-coherence structures, peripheral interfaces, and an on-chip interconnect network (Network-on-Chip, NoC). These modules are not simply stacked together, but cooperate through on-chip interconnects, clock/reset structures, power management, software drivers, and system-verification flows ([39]; [42]; [69]).

Therefore, the core problem at the SoC level is not "whether to implement a particular functional module," but how to coordinate computation, communication, storage, power, and software programmability within the system scope. For EDA, this means that design tools need to support cross-IP integration verification, interconnect modeling, power analysis, software-hardware co-debugging, and system-level design-space exploration.

\paragraph{Layered Relationship Between RTL and System-Level Design}

At this stage, RTL design is no longer an independent design layer, but the implementation layer of the system architecture. System-level design first defines functional partitioning and communication protocols, while RTL is responsible for implementing specific register-transfer behavior and timing-control logic.

This relationship can be expressed as:

\[
System Specification \rightarrow Microarchitecture \rightarrow RTL Implementation
\]

where microarchitecture serves as the bridge between the system level and the RTL level, defining pipeline structures, cache policies, and the organization of execution units ([70]).

\paragraph{Key Features of the EDA 3.0 Stage}

Algorithm / System / SoC Level is the core abstraction layer of the EDA 3.0 stage. Its key change is that the design object expands from single-IP design to whole-system design, from functional-module optimization to system-level collaborative optimization, and from local performance metrics to joint optimization of power, performance, area, bandwidth, latency, and software usability. At this point, EDA no longer serves only hardware implementation, but enters the stage of hardware-software co-design ([39]; [41]; [42]; [69]).

At this stage, compilers, operating systems, runtime systems, drivers, and software-development ecosystems begin to directly affect hardware-architecture design. For example, whether an accelerator is effective depends not only on whether its RTL or physical implementation is efficient, but also on whether data movement, software APIs, compiler mapping, and system scheduling can fully exploit the hardware. Therefore, the SoC level can be regarded as a key turning point in classical EDA from "circuit engineering" to "computing systems engineering."

\paragraph{Typical Industrial Example}

The ARM architecture system can serve as a typical industrial example of SoC-level abstraction. Its success does not depend only on the optimization of a particular CPU microarchitecture, but on the long-term coordination among a standardized instruction set architecture (ISA), a scalable IP ecosystem, SoC integration methodology, software toolchains, and operating-system ecosystems. For chip-design companies, the value of the ARM ecosystem lies in its integration of processor cores, bus interfaces, debugging mechanisms, software compatibility, and third-party IP into a reusable platform, thereby enabling different vendors to develop differentiated SoCs around common interfaces ([40]; [41]; [69]; [70]).

This type of system design reflects the core idea of "ecosystem-driven design" in the EDA 3.0 stage: chip design no longer delivers only a hardware module, but a computing platform capable of co-evolving with software, toolchains, IP libraries, and manufacturing flows.

\paragraph{Summary}

Algorithm / System / SoC Level can be summarized as:

\[
Algorithm \rightarrow System Architecture \rightarrow Hardware-Software Co-Designed SoC
\]

This level marks the expansion of EDA from "hardware design methodology" to "computing systems engineering," realizing unified modeling among algorithms, architectures, and implementations.

\subsection{Current Status (Engineering Practice Status of Current EDA Abstraction Levels)}

In current engineering practice for classical electronic design automation (EDA), artificial intelligence (AI) methods are gradually being introduced to assist or optimize the design and search processes of parameterized cells (Parameterized Cell, PCell). Their core objective is to use machine-learning or optimization algorithms to efficiently explore the PCell parameter space under given process constraints and design rules, thereby realizing automatic optimization of circuit structures at higher abstraction levels.

\paragraph{Hierarchical Relationship Between PCell and PDK}

In industrial EDA flows, transistor-level PCells and logic-gate-level PCells are usually provided in a unified form by foundries as process design kits (Process Design Kit, PDK). A PDK includes not only the geometric and electrical models of devices, but also SPICE-level behavioral models, design rule check (DRC) constraints, and parasitic-extraction models, thereby forming the complete infrastructure from devices to circuit simulation ([22]; [66]).

In industrial EDA flows, transistor-level and logic-gate-level PCells are usually provided by foundries through PDKs. A PDK includes not only geometric definitions and electrical models, but also SPICE-level behavioral models, DRC constraints, and parasitic-extraction information, thereby connecting device modeling, layout verification, and circuit simulation into a complete infrastructure ([22]; [35]; [66]).

\paragraph{Engineering Structure of Design Abstraction Levels}

In actual chip-design processes, designers usually do not directly manipulate transistor-level structures, but instead design based on higher-level abstraction units, including:

\begin{itemize}[leftmargin=2em]
\item Operation Gate PCells (operation-level logic units)
\item Functional Circuit PCells (functional circuit cells)
\item IC-level PCells (integrated-circuit-level modules)
\end{itemize}

Together, these abstract units constitute a hierarchical design structure, enabling complex systems to be progressively constructed through modular composition without directly handling the details of low-level physical implementation ([35]).

\paragraph{Automated Mapping Flow from Design to Implementation}

In a standard EDA flow, the design process usually follows the following mapping chain:

\[
\begin{gathered}
\text{High-level design} \rightarrow \text{PCell-based schematic}
\rightarrow \text{Layout generation}\\
\rightarrow \text{GDSII fabrication data}
\end{gathered}
\]

Here, GDSII (Graphic Database System II), as the final physical manufacturing data format, is the standard output form for chip tape-out. For quantum chips, GDSII is also becoming a key carrier connecting layout design and wafer manufacturing data conversion, but it still needs to be further combined with quantum-specific DRC, LVS, DFM, and mask data preparation (MDP) flows ([56]).

The key constraint of this flow is that all abstract designs must satisfy the design rules defined in the PDK and undergo electrical-behavior verification through SPICE simulation, in order to ensure manufacturability and reliability in physical implementation ([35]; [56]; [66]).

\paragraph{Core Role of SPICE Verification in the Current System}

SPICE simulation remains the core tool connecting abstract design and physical implementation in the current EDA system. Whether for PCell-level devices or functional circuit modules, their final electrical behavior must be verified through SPICE models to ensure that metrics such as timing, current, and power consumption satisfy design requirements.

Its basic simulation form can be represented as a system of nonlinear differential equations:

\[
\frac{d\mathbf{x}}{dt} = f(\mathbf{x}, \mathbf{u}, t)
\]

where $\mathbf{x}$ denotes circuit state variables (such as voltage and current), and $\mathbf{u}$ denotes input excitation signals.

\paragraph{Current Positioning of AI-Assisted Design}

At present, the role of AI in EDA is still mainly concentrated on local optimization tasks, such as:

\begin{itemize}[leftmargin=2em]
\item PCell parameter optimization
\item Automatic layout generation (placement \& routing assistance)
\item Timing optimization and power estimation
\item Acceleration of design-space search
\end{itemize}

These methods usually serve as enhancement modules for traditional EDA flows rather than replacements for complete design flows ([47]). Therefore, the current system still takes "PDK + SPICE + PCell" as its core infrastructure.

\paragraph{Summary}

The current EDA engineering system can be summarized by the following structure:

\[
\begin{gathered}
\text{PCell (Device/Logic/IP)}
\xrightarrow{\text{PDK constraints}}
\text{Circuit Design}\\
\xrightarrow{\text{SPICE verification}}
\text{GDSII fabrication}
\end{gathered}
\]

This system shows that classical EDA is still centered on "process-constraint-driven hierarchical design," while AI methods are mainly embedded as optimization and acceleration tools rather than structural replacement technologies.

\section{Comparison between Classical Circuits and Quantum Circuits}

\subsection{Positioning of the Quantum-Chip Paradigm Framework}

Drawing on the developmental experience of classical integrated circuits (ICs) and the EDA ecosystem, the methodology for quantum-chip development can be analyzed through a structural comparison. The evolution of classical EDA has essentially unfolded around the gradual progression from "device-physics constraints $\rightarrow$ elevation of abstraction levels $\rightarrow$ standardized design flows." Its core driving forces have come from the maturation of SPICE modeling systems, standard-cell libraries, logic synthesis, the PDK ecosystem, and physical signoff flows ([22]; [23]; [24]; [35]; [38]; [107]; [108]).

Against this background, quantum electronic design automation (Quantum EDA, Q-EDA) should not be positioned merely as an extension at the tool level, but rather understood as a reconstruction of a complete engineering paradigm for quantum chips. The success of classical EDA stems from stable Boolean abstractions and manufacturable process interfaces; quantum chips, by contrast, must simultaneously handle coherent evolution of quantum states, measurement feedback, open-system noise, microwave control, layout electromagnetic effects, and manufacturing variability. Therefore, the purpose of the comparison in this chapter is not to prove that the two are identical, but to clarify which experiences from classical EDA can be transferred and which must be reconstructed under quantum-physical constraints ([5]; [7]; [13]; [83]; [110]).

\paragraph{Paradigm Comparison and Reconstruction of Abstraction Levels}

In the classical EDA system, abstraction levels typically rise progressively from transistors and interconnects to logic gates, arithmetic units, RTL/IP modules, and SoC-level system design. This process relies on a key premise: the underlying continuous circuit behavior can be stably discretized through thresholds, noise margins, and standard-cell libraries, thereby making Boolean logic a reliable intermediate abstraction ([35]; [68]; [108]).

In the quantum-chip system, the physical carrier of information changes from classical charge/voltage to quantum states, phase, microwave coherent fields, and measurement records. Superposition, entanglement, no-cloning, and measurement disturbance of quantum states prevent quantum circuits from directly inheriting the abstraction path of classical logic gates--RTL--SoC; instead, the mapping relationship among "physical devices--quantum circuits--control systems--logical qubits--system architecture" must be redefined ([2]; [5]; [7]; [18]; [110]; [111]).

Therefore, the core task of Q-EDA is not to simply replicate the classical EDA flow, but to construct a new cross-scale abstraction system, so that quantum devices such as Josephson junctions, resonators, couplers, readout structures, and control lines can progressively construct system-level quantum-circuit structures through PCell-like abstraction, SPICE-Q, Quantum PDKs, and experimental-feedback closed loops ([82]; [83]; [85]; [88]; [89]; [90]).

\paragraph{Core Differences between Paradigms}

The key differences between classical EDA and Q-EDA are first reflected in signal carriers and state descriptions. Classical EDA takes voltage, current, charge, and logic levels as its principal variables, and establishes verifiable flows through KCL/KVL, SPICE equations, Boolean logic, and timing models; Q-EDA, by contrast, must take quantum-state amplitudes, phases, density matrices, Hamiltonians, decoherence channels, and measurement records as its core variables. Second, classical digital circuits typically pursue deterministic logic computation, whereas quantum chips must handle coherent evolution, measurement backaction, noise spectra, frequency drift, and calibration feedback in open quantum systems ([13]; [35]; [110]).

This difference does not mean that the experience of classical EDA becomes entirely invalid. On the contrary, PDKs, PCells, layout signoff, design databases, and automated flows still have important reference value; however, these mechanisms must be redefined around the physical models of quantum devices and experimental feedback. It follows that Q-EDA cannot directly take the classical HDL/RTL-driven design methodology as its sole main line; instead, it must establish a bidirectional iterative framework based on physical models, manufacturing data, and measurement feedback ([51]; [82]; [83]; [85]).

\paragraph{Paradigm Goal: From Tools to an Engineering System}

Under this framework, the ultimate goal of Q-EDA is to establish a standardized engineering system analogous to that formed by classical EDA in the 1980--1990s. Its foundational objects, however, are no longer stable CMOS standard cells, but quantum-device models, parameterized quantum components, Quantum PDKs, SPICE-Q / quantum-dynamics simulators, cryogenic measurement data, and calibration closed loops. Standardized quantum-device models are used to describe Josephson junctions, transmons, resonators, and couplers; Quantum PCells transform these models into design objects that can generate layouts and invoke simulations; and Quantum PDKs encapsulate material layers, process rules, device-parameter distributions, layout verification, and manufacturing interfaces ([56]; [57]; [58]; [82]; [85]).

Its core significance lies in transforming the currently highly fragmented practice of quantum-chip design into a reusable, verifiable, and scalable engineering-system architecture. This goal is not to make quantum chips become classical digital chips, but to endow quantum-chip development with capabilities similar to those of the classical semiconductor industry in model reuse, flow verification, process feedback, and cross-team collaboration.

\paragraph{Summary}

The quantum-chip paradigm framework can be formalized as:

\[
\begin{gathered}
\text{Quantum Device Physics} \rightarrow
\text{Parameterized Quantum Components}\\
\rightarrow \text{System-Level Quantum Architecture}
\end{gathered}
\]

This framework emphasizes that Q-EDA is not merely a collection of design tools, but a structural intermediate layer connecting the realization of quantum physics with the realization of system engineering. Its development path essentially corresponds to the evolutionary process by which classical EDA moved from the SPICE era toward a standardized VLSI system.

\subsection{Signal Properties}

Within the framework of quantum electronic design automation (Q-EDA), the nature of signals undergoes a fundamental change: their information carriers are no longer only the voltages, currents, or logic levels of classical electronics, but quantum states and their coherent evolution in open systems. For implementation platforms such as superconducting quantum chips, information usually exists in the form of microwave photons, resonator modes, transmon energy-level states, and measurement-output records; its dynamics must be jointly described by quantum mechanics, open-system theory, and microwave electromagnetic-field models, rather than relying solely on the KCL/KVL framework of classical circuits ([4]; [7]; [13]; [110]).

This does not mean that classical electromagnetics becomes invalid in quantum chips. On the contrary, layouts, capacitances, inductances, transmission lines, and packages still require support from classical electromagnetic simulation; however, these results must be further mapped into quantum Hamiltonians, decoherence channels, and measurement models. The change in signal properties is one of the most fundamental modeling-level differences between Q-EDA and classical EDA.

\paragraph{Basic Equations of Quantum Dynamics}

The core of quantum-signal evolution is described by the Schr\"odinger equation and the Lindblad master equation. In a closed system, the evolution of the quantum state $|\psi(t)\rangle$ satisfies:

\[
i\hbar \frac{d}{dt}|\psi(t)\rangle = \hat{H}|\psi(t)\rangle,
\]

where $\hat{H}$ is the system Hamiltonian, which determines the interaction structure among qubits, couplers, resonators, and microwave drives ([7]; [110]). In practical quantum chips, however, the system inevitably couples to material defects, residual thermal photons, control lines, readout links, and the packaging environment; hence an open quantum-system description must be adopted. The evolution of its density matrix $\rho$ can be written as:

\[
\frac{d\rho}{dt} = -\frac{i}{\hbar}[\hat{H}, \rho] + \sum_k \gamma_k \left( \hat{L}_k \rho \hat{L}_k^\dagger - \frac{1}{2}\{\hat{L}_k^\dagger \hat{L}_k, \rho\} \right),
\]

where $\hat{L}_k$ is a dissipative-channel operator and $\gamma_k$ is a decoherence-rate parameter ([13]). For Q-EDA, this means that "signal simulation" cannot output only voltage waveforms or logic levels; it must also predict quantum-state evolution, decoherence, readout probability distributions, and experimental feedback.

\paragraph{Physical Nature of Microwave Quantum Signals}

In superconducting quantum circuits, signals usually exist as microwave modes in the GHz frequency range, and their quantized form corresponds to electromagnetic-field modes in resonators or transmission lines. Unlike classical microwave engineering, these signals must simultaneously satisfy quantization conditions and open-system noise constraints.

Quantum-field operators can be expressed as:

\[
\hat{a}, \hat{a}^\dagger
\]

They satisfy the commutation relation:

\[
[\hat{a}, \hat{a}^\dagger] = 1
\]

This structure determines that a signal no longer has a definite amplitude, but instead exists in the form of a probability amplitude, thereby making the measurement process itself part of the system dynamics ([7]).

\paragraph{Essential Differences from Classical Signal Models}

Compared with voltage-current signals in classical EDA, the key differences of quantum signals are that unknown quantum states cannot be copied arbitrarily, the measurement process changes the system state, coherence and phase relationships directly carry information, and noise is not merely externally added white noise but a structural perturbation jointly determined by material defects, control lines, readout links, and environmental spectral density. The no-cloning theorem states that an arbitrary unknown quantum state cannot be perfectly copied; this constitutes a fundamental distinction from classical digital signals, which can be buffered, copied, and fanned out ([110]; [111]).

Therefore, the traditional SPICE framework based on linear systems, deterministic waveforms, and lumped-parameter noise models cannot be directly applied to quantum-signal analysis. Quantum chips need to be extended to SPICE-Q or open-system numerical simulation frameworks that incorporate Hamiltonians, dissipative channels, measurement backaction, microwave drives, and experimental feedback into a unified model ([13]; [52]; [88]; [89]; [90]).

\paragraph{Significance of Signal Modeling in Q-EDA}

In the Q-EDA system, signal modeling is no longer an auxiliary analysis step, but a core component of the entire design flow. Device design, routing structures, control pulses, readout links, and system architectures must all be jointly modeled on the basis of quantum-dynamical equations, so as to ensure that key metrics such as coherence time, decoherence rate, gate fidelity, and readout fidelity are controllable at the system level ([5]; [7]; [83]).

This also explains why quantum-chip design cannot rely only on traditional layout tools or classical microwave simulation. Classical simulation can provide electromagnetic parameters, but Q-EDA must further transform these parameters into Hamiltonians, noise channels, and measurement models, enabling designers to evaluate whether a layout truly corresponds to an operable, calibratable, and scalable quantum circuit.

\paragraph{Summary}

The properties of quantum signals can be summarized as:

\[
Classical Voltage/Current Signals \rightarrow Quantum States \rightarrow Open-System Coherent Dynamics
\]

This transformation marks the entry of Q-EDA from the paradigm of "circuit-level signal processing" into that of "quantum-dynamical system design"; its core computational foundation expands from classical electromagnetic theory to the framework of the Schr\"odinger equation and the Lindblad master equation.

\subsection{Environmental Sensitivity}

Superconducting qubits typically operate in the millikelvin temperature regime provided by dilution refrigerators, with typical temperatures on the order of approximately $10\,\mathrm{mK}$. At this temperature scale, the thermal excitation energy $k_B T$ is far below the microwave qubit energy-level spacing $\hbar \omega_q$, thereby suppressing thermal transitions; at the same time, however, quantum states exhibit extremely high sensitivity to material defects, residual thermal photons, flux noise, charge noise, package modes, and control-line crosstalk ([4]; [5]; [7]; [10]; [12]; [16]).

This differs fundamentally from classical digital circuits. Noise in classical CMOS is usually engineered into the system through noise margins, timing margins, and error-correction mechanisms, whereas small environmental perturbations in quantum chips directly manifest as energy relaxation, phase decoherence, frequency drift, and reduced gate fidelity. Therefore, environmental sensitivity is not a post-processing issue in quantum-chip design, but a core constraint that must be jointly handled by Q-EDA front-end modeling, layout selection, materials processing, and package design.

\paragraph{Thermal-Noise Constraints and Decoherence Mechanisms}

Although thermal noise is significantly suppressed under ideal conditions, practical devices are still affected by residual thermal photons and high-frequency environmental noise. Their contributions to qubit decoherence processes are usually described by the energy-relaxation time $T_1$ and the phase-decoherence time $T_2$:

\[
\frac{1}{T_2} = \frac{1}{2T_1} + \frac{1}{T_\phi}
\]

where $T_\phi$ denotes the pure phase-decoherence time ([10]). In Q-EDA modeling, these time constants are no longer posterior evaluation parameters, but design constraints.

\paragraph{Defect Mechanisms of Two-Level Systems (TLS)}

Two-Level Systems (TLS) are widely present in the dielectric materials and interface layers of superconducting circuits. Their origins include microscopic inhomogeneities such as disordered atomic configurations, oxide-layer defects, interface roughness, and surface-adsorbed molecules. These defects couple to the electric-field participation regions of qubits, leading to energy relaxation, frequency drift, and reduced device-to-device consistency ([10]; [16]).

The coupling between a TLS and a qubit can be approximately expressed as:

\[
\hat{H}_{TLS-q} = g (\hat{\sigma}_+ \hat{\tau}_- + \hat{\sigma}_- \hat{\tau}_+),
\]

where $\hat{\sigma}$ denotes the qubit operator, $\hat{\tau}$ denotes the TLS-system operator, and $g$ is the coupling strength. This coupling gives rise to avoided crossings in spectra, and also causes coherence time and frequency stability to exhibit device-to-device differences. For Q-EDA, TLS should not be treated merely as explanatory factors after experimental measurement, but should enter design constraints as early as possible through material selection, layout participation ratios, surface treatment, and statistical models.

\paragraph{Influence of Materials and Process Variability}

Compared with classical EDA, environmental sensitivity in quantum chips arises not only from external noise, but also from the statistical variability of materials and processes themselves. Variations in oxide-layer thickness, nonuniformity of Josephson-junction area, thin-film stress, interface roughness, residual contamination, and packaging boundary conditions all change the critical current, capacitance, inductance, resonant frequency, and loss channels. These factors lead to significant within-wafer dispersion and run-to-run drift in device parameters, ultimately appearing as qubit frequency dispersion, coupling-strength shifts, and differences in coherence time ([5]; [8]; [10]; [16]; [83]).

Therefore, process variability in quantum chips cannot be described only by a small number of worst-case corners as in classical CMOS. A more reasonable approach is to establish statistical process models and measurement-feedback closed loops that connect manufacturing outcomes, cryogenic test data, and design-parameter updates.

\paragraph{Modeling Status in Q-EDA}

In the Q-EDA framework, environmental sensitivity is no longer regarded as a noise-correction term, but as a core component of system design. Thermal noise, TLS distributions, material statistics, package parasitics, and control-line crosstalk must be explicitly incorporated into simulation models at the design stage and jointly optimized with device geometry design, routing structures, materials processing, and packaging schemes, thereby forming a closed loop for design-process-measurement co-optimization ([51]; [82]; [83]; [85]).

This is also one of the most easily underestimated differences in the comparison between classical circuits and quantum circuits. Classical EDA can usually transform environmental perturbations into noise margins, timing margins, or reliability metrics; Q-EDA, however, must incorporate environmental perturbations directly into Hamiltonians, dissipative channels, and experimental calibration models. Otherwise, even if a design result is manufacturable at the layout level, it may not be usable in the sense of quantum dynamics.

\paragraph{Summary}

The environmental sensitivity of superconducting quantum systems can be summarized as:

\[
Thermal Fluctuations + TLS Defects + Material Disorder \rightarrow Decoherence-limited Quantum Performance
\]

Therefore, Q-EDA must elevate environmental modeling from "post-processing analysis" to "front-end constraint-driven design" to ensure that quantum chips possess predictable coherence performance under actual manufacturing and operating conditions.

\subsection{Routing Complexity}

In superconducting quantum chips, the routing system not only undertakes signal-transmission functions, but must also maintain impedance matching, mode purity, low loss, and low crosstalk in the GHz microwave band. Classical CMOS interconnects primarily focus on resistance-capacitance delay, signal integrity, power consumption, and timing convergence; quantum routing, by contrast, also directly changes the electromagnetic environment, coupling topology, readout links, and decoherence channels of qubits. Therefore, the quantum-routing problem is essentially a "many-body coupling optimization problem under electromagnetic-field constraints," with complexity significantly higher than that of ordinary geometric connectivity optimization ([5]; [7]; [83]; [91]; [92]; [94]).

This difference indicates that routing in Q-EDA cannot be merely a quantum version of classic place-and-route. Routing results must simultaneously satisfy the constraints of microwave engineering, quantum dynamics, packaging, cryogenic control, and process rules.

\paragraph{Microwave Transmission Constraints and Parasitic Coupling Problems}

Signals in quantum routing propagate as distributed electromagnetic modes, rather than as ideal wire models. Therefore, any change in geometric structure may introduce parasitic capacitance, parasitic inductance, impedance discontinuities, radiation loss, and nonlocal coupling effects. These effects can be approximately represented by a frequency-dependent impedance:

\[
Z(\omega) = R + i\omega L + \frac{1}{i\omega C} + Z_{\mathrm{parasitic}}(\omega),
\]

where $Z_{\mathrm{parasitic}}(\omega)$ characterizes the high-frequency response caused by geometric discontinuities, neighboring structures, package boundaries, and parasitic modes. In quantum systems, this term often changes resonant frequencies, readout responses, and residual coupling, and further leads to crosstalk and frequency drift ([5]; [7]; [83]; [91]). Therefore, verification of quantum routing must simultaneously include electromagnetic-field simulation, port models, mode analysis, and assessment of quantum-dynamical effects.

\paragraph{Crosstalk and Multi-Qubit Coupling Constraints}

In multi-qubit chips, routing affects not only individual signal paths, but also introduces undesired inter-qubit coupling, control-line crosstalk, readout crosstalk, and package-mode coupling. A simplified residual coupling term can be written as:

\[
\hat{H}_{\mathrm{cross}} = \sum_{i \neq j} J_{ij} \hat{\sigma}_i^z \hat{\sigma}_j^z,
\]

where $J_{ij}$ denotes the residual coupling strength caused by routing structures, couplers, package boundaries, or parasitic modes. As chip scale increases, the number of potential coupling paths grows rapidly, elevating crosstalk from a local layout issue to a system-level constraint ([5]; [7]; [83]). Therefore, the routing module of Q-EDA cannot output only wires that satisfy geometric rules; it must also output model information usable for Hamiltonian updates, crosstalk-matrix estimation, and calibration-strategy design.

\paragraph{Complexity Introduced by Three-Dimensional Integration and TSVs}

Three-dimensional integration technologies, such as flip-chip, indium bump bonds, through-silicon vias, TSVs, and multilayer interconnects, are often used in classical EDA to increase interconnect density and system integration; in superconducting quantum chips, they are likewise important approaches for alleviating planar-routing congestion and the growth in the number of control lines. However, three-dimensional interconnects also introduce new electromagnetic modes, dielectric loss, interface reflections, package cavity modes, and interlayer coupling problems. These effects may disrupt microwave-mode purity and reduce quantum-state fidelity or readout fidelity ([91]; [92]; [93]; [94]).

Therefore, 3D/TSV structures in quantum chips cannot be evaluated only according to classical package-interconnect rules; they must be co-designed with qubit frequency planning, resonator modes, readout links, package simulation, and compatibility with cryogenic processes.

\paragraph{Nature of the Routing Problem in Q-EDA}

Under the Q-EDA framework, the routing problem is no longer a geometric optimization problem in the traditional sense, but a global electromagnetic-field optimization problem under quantum-dynamical constraints. Its objective function can be abstracted as:

\[
\mathcal{L}_{route} = \alpha \cdot \mathrm{crosstalk} + \beta \cdot \mathrm{loss} + \gamma \cdot \mathrm{decoherence} + \delta \cdot \mathrm{frequency\ shift} + \eta \cdot \mathrm{rule\ violations},
\]

where the terms constrain crosstalk, loss, decoherence, frequency shift, and process-rule violations, respectively. This optimization function reflects the direct coupling relationship between routing design and quantum coherence. Unlike the classical EDA idea that "connectivity is sufficient before timing optimization," quantum routing must coordinate with electromagnetic simulation, Hamiltonian extraction, and measurement-calibration models already at the layout-generation stage ([82]; [83]; [85]).

\paragraph{Summary}

The core sources of quantum-routing complexity can be summarized as:

\[
Microwave Field Propagation + Parasitic Coupling + 3D Integration Effects \rightarrow Nontrivial System-Level Decoherence Constraints
\]

Therefore, the routing module in Q-EDA must be elevated from "geometric-connection optimization" to an "electromagnetic-quantum-state co-simulation optimization" problem, so as to support scalable design of large-scale quantum chips.

\subsection{Process Variability / Tolerance}

Classical CMOS circuits likewise exhibit process variability, but its effects are usually handled through PVT corners, statistical timing analysis, yield models, and design margins. For superconducting quantum chips, the effects of process variability enter the quantum-system Hamiltonian more directly: small variations in Josephson-junction area, oxide-layer thickness, thin-film stress, interface roughness, dielectric loss, and package boundary conditions all change qubit frequency, anharmonicity, coupling strength, readout response, and coherence time ([5]; [8]; [10]; [16]; [83]).

Current fabrication processes for superconducting quantum chips exhibit significant statistical variability, with sources including within-wafer and wafer-to-wafer / run-to-run parameter deviations. These deviations affect not only whether a single device meets its target frequency, but also frequency crowding, crosstalk matrices, and calibration complexity in multi-qubit systems. Therefore, process variability in Q-EDA cannot be treated merely as posterior manufacturing statistics, but should become a constraint and optimization variable at the design stage.

\paragraph{Physical Sources of Process Variability}

In superconducting quantum circuits, one of the most critical sensitive structures is the Josephson junction, whose critical current $I_c$ is extremely sensitive to junction area, oxide-barrier thickness, and material-interface quality. A simplified relationship can be written as:

\[
I_c \propto A \cdot e^{-d/\lambda},
\]

where $A$ is the junction area, $d$ is the oxide-barrier thickness, and $\lambda$ is the tunneling decay length. Because $A$ and $d$ exhibit unavoidable random fluctuations in nanoscale processes, $I_c$ follows a statistical distribution and further affects the Josephson energy $E_J$. For a transmon, its frequency is approximately:

\[
\omega_q \approx \sqrt{8 E_J E_C} - E_C,
\]

where $E_C$ is the charging energy. It can therefore be seen that fabrication deviations are directly converted into qubit frequency drift and changes in anharmonicity. This is analogous to the effect of threshold-voltage or channel-length fluctuations on timing in classical CMOS, but in quantum chips it further affects frequency crowding, gate fidelity, and calibratability ([5]; [8]; [83]).

\paragraph{Statistical Process Corners and Design Robustness}

To address the above uncertainties, Q-EDA must introduce statistical process corners and distribution-based modeling methods. Worst-case corners in classical CMOS still have reference value, but quantum chips require a fuller description of parameter distributions, because system performance often depends on the relative relationships among multiple qubit frequencies, coupling strengths, and readout parameters, rather than on whether a single device falls into the worst corner.

Under this framework, process parameters can be modeled as random variables:

\[
p(x) = \mathcal{N}(\mu, \sigma^2),
\]

The design objective is also no longer single-point optimization, but a joint constraint problem involving expectation, variance, yield, and calibratability:

\[
\min \; \mathbb{E}[\mathcal{F}(x)] + \lambda \cdot \mathrm{Var}(\mathcal{F}(x)),
\]

where $\mathcal{F}(x)$ denotes quantum-circuit performance metrics, such as coherence time, gate fidelity, frequency-matching error, crosstalk strength, or readout error rate. For Q-EDA, the objective of statistical process models is not to eliminate all variability, but to predict at the design stage which variations can be compensated by calibration and which will lead to structural unusability of the system ([5]; [8]; [83]).

\paragraph{Process-Design Closed Loop in DTCO}

Within the design-technology co-optimization (DTCO) framework, process tolerance is no longer regarded as a purely manufacturing error, but as part of the design space. Process models, device-geometry design, system-level layout, cryogenic measurement, and calibration results must form a closed feedback loop:

\[
\begin{gathered}
\text{Process Variation} \rightarrow \text{Device Parameter Distribution}
\rightarrow \text{Circuit Performance Spread}\\
\rightarrow \text{Measurement Feedback} \rightarrow \text{Design Update}.
\end{gathered}
\]

This closed loop allows the Q-EDA system to predict yield, frequency distributions, crosstalk risks, and calibration difficulty in advance at the design stage, and to update PCell, SPICE-Q models, and Quantum PDK parameters after manufacturing data are returned. Compared with approaches that rely on expensive late-stage experimental iteration, this closed loop is closer to the engineering logic of "process model--design rules--post-silicon feedback" in classical EDA, but its feedback variables are extended to quantum-state coherence, gate fidelity, and measurement results ([51]; [82]; [83]; [85]).

\paragraph{Tolerance Characteristics Specific to Quantum Chips}

Unlike classical CMOS, the tolerance problem of superconducting quantum chips affects not only logical correctness, but also quantum-state dynamics, including decoherence rates, frequency crowding, and crosstalk structures. Therefore, process tolerance in Q-EDA must be jointly simulated with quantum-dynamical models, rather than handled as an independent module.

\paragraph{Summary}

The essence of process tolerance in superconducting quantum chips can be summarized as:

\[
\begin{gathered}
\text{Nanofabrication Variability} \rightarrow
\text{Quantum Parameter Dispersion}\\
\rightarrow \text{System-Level Performance Uncertainty}
\end{gathered}
\]

Therefore, Q-EDA must explicitly incorporate process variability into the design-optimization process through statistical modeling and DTCO closed-loop mechanisms, in order to achieve robust design for scalable quantum chips.

\section{Q-EDA Layering for Quantum Circuits}

Quantum electronic design automation (Q-EDA) exhibits structural similarities to conventional electronic design automation (EDA): both must begin from physical devices and progressively establish reusable components, logical abstractions, functional modules, and system-level architectures. However, this similarity should not be understood as a one-to-one correspondence. The abstraction hierarchy of classical EDA is built on stable Boolean logic, standard-cell libraries, PDKs, and manufacturability signoff; Q-EDA, by contrast, must simultaneously address coherent evolution of quantum states, measurement backaction, open-system noise, microwave control, frequency planning, and process feedback ([24]; [35]; [107]; [110]).

In classical semiconductor circuits, standard cells are typically used for digital logic and have fixed functions, fixed layout heights, timing models, and power models; parameterized cells (PCells) are more often used for analog, RF, memory, I/O, and variable-geometry structures. Quantum chips at present are closer to a ``physical-device-driven parameterizable engineering system,'' because the functions of transmons, resonators, couplers, readout lines, and control lines all depend strongly on geometric parameters, electromagnetic boundary conditions, material loss, and calibration feedback ([5]; [7]; [83]; [106]).

Therefore, at the current stage of Q-EDA development, it is not entirely valid to map the standard cell in classical EDA strictly onto the quantum logical-space layer. Quantum logical operations are not simple substitutes for stable Boolean gates; rather, they are physical operations jointly realized by underlying continuous-time dynamics, microwave drives, coupling structures, and measurement processes. The no-cloning theorem, measurement backaction, and error-correction overhead further make the boundary between the quantum logic layer and the physical implementation layer tighter than in classical digital circuits ([2]; [18]; [110]; [111]).

Under this framework, Q-EDA layering is more appropriately understood as a ``physically constrained modular abstraction system'': the Part Level concerns manufacturable physical structures; the Component Level concerns reusable quantum-device PCells; the Logical Space Level concerns logical qubits and error-correction structures; the Arithmetic Level concerns composable quantum operations; the Functional IP Level concerns the mapping from algorithmic submodules to control processes; and the Algorithm/System/Quantum SoC Level concerns complete quantum processors, control stacks, and modular system architectures. This hierarchy does not simply replicate classical digital EDA, but provides an intermediate mapping framework through which quantum-chip design can move from laboratory practice toward reusable engineering flows ([82]; [83]; [85]).

\subsection{Part Level - Q-EDA Early Stage PCell}

At this abstraction level, the design object remains at the lowest stage of physical structure, and the central concern is manufacturable geometries and material stacks capable of implementing basic functions. These structures usually do not involve complex logical abstraction, but instead correspond directly to physical objects such as superconducting thin films, Josephson junctions, ground planes, local interconnects, CPW cross sections, air bridges, or test structures. Thus, the Part Level is clearly physics-dominated and constitutes the manufacturing and electromagnetic foundation for all higher-level Q-EDA abstractions ([5]; [7]; [83]).

Typical examples include superconducting material structures and Josephson junctions (JJs). In superconducting quantum-chip systems, the JJ is the key physical element that provides nonlinearity, and its behavior is described by the Josephson relation:

\[
I = I_c \sin(\phi),
\]

where $I_c$ is the critical current and $\phi$ is the superconducting phase difference across the junction. This relation determines the fundamental role of JJs in microwave driving, qubit energy-level structures, and superconducting-circuit dynamics, and it also provides an important physical basis for the RCSJ model, superconducting SPICE-like simulation, and subsequent SPICE-Q modeling ([11]; [12]; [52]; [59]).

In early Q-EDA design flows, such structures are usually drawn manually or semi-automatically with CAD/EDA tools, and their design depends heavily on empirical parameters from specific laboratories or process lines. Different research institutions or foundries often maintain their own geometry libraries, layer definitions, and process-rule sets, leaving this stage without a unified standardized description system. The emergence of tools such as QPDK, KQCircuits, Quantum Metal, and EDA-Q shows that this level is moving from ``manual geometric drawing'' toward ``parameterized, scripted, and verifiable physical component libraries'' ([57]; [58]; [85]; [87]; [102]).

Accordingly, the main feature of this level is that it is process-driven rather than high-level-abstraction-driven: design optimization mainly revolves around manufacturability, material consistency, critical-current stability, fundamental electromagnetic characteristics, and test observability. At this stage, the mapping between geometric structure and physical behavior is direct and explicit, which also provides the basis for subsequent evolution toward qubit PCells, Quantum PDKs, and higher-level abstractions.

\subsection{Component Level - Q-EDA Qubit PCell}

At this abstraction level, the design focus is elevated from basic geometric structures to reusable device components. Its core objects include transmons, flux qubits, readout resonators, couplers, Purcell filters, control lines, and test structures. The key feature of this level is that an individual physical structure no longer describes only a material or a geometric shape, but is instead encapsulated as a quantum-device PCell with ports, parameters, layout, simulation entry points, and functional semantics ([5]; [7]; [82]; [83]).

In superconducting quantum circuits, a qubit is usually formed by coupling a Josephson junction with a capacitive/resonant structure, and its behavior can be described through equivalent-circuit models and quantization tools. For example, the basic design condition for a transmon qubit is

\[
E_J / E_C \gg 1,
\]

where $E_J$ is the Josephson energy and $E_C$ is the charging energy; this condition gives the system stronger robustness against charge noise ([8]). Resonators are mainly used for readout and coupling control, and their frequencies are jointly determined by geometric dimensions, electromagnetic boundary conditions, material constants, and the packaging environment. Tools such as scqubits, computer-aided quantization, and SQuADDS indicate that Qubit PCells must simultaneously connect layout parameters, equivalent-circuit parameters, and quantum Hamiltonian parameters ([82]; [88]; [89]).

As system complexity increases, routing begins to become an important component of this level. Simple two-dimensional wiring can no longer satisfy the requirements of multiqubit systems for crosstalk suppression, frequency isolation, impedance matching, and readout multiplexing. Routing methods based on rule constraints and electromagnetic simulation are therefore needed to optimize microwave signal paths and coupling topologies. At this stage, design files are gradually standardized as GDSII or similar physical layout formats, enabling design results to interface with wafer-fabrication flows and providing a unified representation from device design to physical layout ([56]; [85]; [87]; [102]).

Overall, this level marks the core feature of Q-EDA 1.0: devices begin to evolve from manually drawn geometric structures into standardized functional component libraries, and cross-project reuse is achieved through the PCell form. It is neither the direct counterpart of the classical standard cell nor a purely geometric figure, but an engineering object that connects process, layout, electromagnetic simulation, and quantum modeling.

\subsection{Logical Space Level - Q-EDA Logical PCell}

This abstraction layer mainly concerns the structured representation of quantum logical units, namely the logical qubit as the basic unit of design and abstraction. At this level, the system is no longer centered on a single physical qubit, but instead treats the encoded and error-corrected logical information carrier as the design object, thereby partially separating the physical noise model from the logical computational structure. Stabilizer codes, surface codes, and fault-tolerant gate operations provide a scalable framework for logical qubits and are also core objects that future large-scale Q-EDA must be able to represent ([2]; [18]; [19]).

It should be noted that a Logical PCell is not equivalent to a standard logic gate in classical EDA. A classical standard cell usually represents a fixed Boolean function together with its layout, timing, and power models; a logical qubit, by contrast, is jointly defined by multiple physical qubits, measurement lines, coupling structures, error-correction cycles, syndrome-extraction circuits, and decoding strategies. Its reliability depends on many factors, including physical error rates, measurement error rates, crosstalk, leakage, timing scheduling, and the distance of the error-correcting code ([2]; [18]; [19]).

Under an independent error model, the logical error rate can be approximately expressed as

\[
P_L \sim \mathcal{O}\left((p/p_{th})^{\frac{d+1}{2}}\right),
\]

where $p$ denotes the physical error rate, $p_{th}$ denotes the error-correction threshold, and $d$ denotes the code distance. This expression reflects the central role of code distance in logical reliability, but real chips must also consider correlated noise, measurement errors, layout constraints, and control crosstalk ([18]).

In the early stage of Q-EDA, implementation of this level depends on combinations of basic PCells and is verified with numerical simulation frameworks similar to SPICE-Q, so as to support the process-level implementability of logical qubits. The role of SPICE-Q here is not to describe algorithms directly, but to characterize the equivalent circuits and quantum-dynamical behavior of Josephson junctions, resonators, couplers, readout structures, and control lines, thereby providing the logical layer with models of underlying physical constraints ([52]; [59]; [88]; [89]; [90]).

As the logical-qubit abstraction is gradually standardized, early quantum design environments begin to evolve toward a system analogous to the PDK in classical EDA. In this process, SPICE-Q, Quantum PDKs, and parameterized PCells jointly form a bridge between logical design and physical implementation, enabling logical qubits to undergo layout generation, simulation, error budgeting, and manufacturing-feedback verification within a unified design framework.

\subsection{Arithmetic Level - Q-EDA Logical Operation PCell}

At this abstraction level, the design object is further elevated from a single logical qubit to quantum arithmetic units composed of multiple quantum logic gates. These units no longer describe only individual logical operations; instead, through finite-depth quantum-circuit combinations, they implement basic arithmetic-operation modules with clear functional semantics, such as addition, comparison, controlled addition, modular multiplication, phase accumulation, and modular exponentiation ([110]; [113]; [114]).

Analogous to the way an ALU in classical EDA is assembled from standard logic gates, quantum arithmetic structures can also be assembled from single-qubit gates, CNOT gates, controlled rotations, Toffoli-type gates, and reversible-computing structures. The work of Barenco et al. on universal quantum gates shows that arbitrary quantum operations can be decomposed into elementary gate sets; the construction of quantum arithmetic networks by Vedral, Barenco, and Ekert further shows that operations such as addition and modular exponentiation can be organized as higher-level quantum-computing modules ([112]; [113]).

However, there are important differences between the quantum arithmetic layer and the classical ALU layer. Classical arithmetic modules mainly optimize delay, area, and power; quantum arithmetic modules must also optimize circuit depth, T gate / Toffoli resources, the number of ancillary qubits, error-correction overhead, topological mapping, crosstalk, and measurement scheduling. In superconducting quantum chips, the arrangement of logical qubits on the chip plane or within modules directly affects long-distance coupling, decoherence risk, and frequency crowding ([5]; [83]; [84]).

Therefore, this level marks the core feature of Q-EDA 2.0: the design object rises from a set of logical gates to functional arithmetic modules, and system behavior is jointly defined by spatial layout, gate sequences, error-correction resources, and control scheduling. A quantum arithmetic PCell should not be understood as a fixed layout unit, but rather as an operation-level module with functional semantics, an error budget, resource estimates, and physical-mapping constraints.

\subsection{Functional IP Level - Q-EDA Operational PCell}

At this abstraction level, the design object is further elevated from arithmetic-level modules to Functional IP units, namely reusable operation modules composed of complete quantum algorithms or subalgorithms. This level no longer focuses on a single logical gate or local arithmetic structure, but instead centers on quantum circuit blocks with clear input-output semantics, resource models, and control interfaces, such as the quantum Fourier transform (QFT), phase estimation, variational quantum algorithm submodules, error-correction cycles, syndrome-extraction subprocesses, or the modular-exponentiation module in Shor's algorithm ([110]; [113]; [114]).

Unlike classical chips, logical operations in quantum chips are usually implemented through microwave pulse signals and tunable coupling structures; their mechanism is a time-dependent Hamiltonian drive that applies precise control to quantum states. This process is intrinsically described by continuous-time dynamics:

\[
i\hbar \frac{\partial}{\partial t}|\psi(t)\rangle = H(t)|\psi(t)\rangle,
\]

where $H(t)$ denotes the time-dependent Hamiltonian jointly determined by microwave drives, couplers, frequency selectivity, and environmental noise ([5]; [7]; [90]). Thus, the so-called ``operational PCell'' is not a fixed geometric unit, but a combination of algorithmic structure, control pulses, hardware topology, error models, and calibration flows.

At this level, routing and signal transmission are no longer merely problems of geometric path optimization, but comprehensive design problems involving microwave phase, frequency selectivity, time-domain pulse shaping, crosstalk matrices, and readout multiplexing. Three-dimensional packaging and vertical interconnect structures (such as TSVs, flip-chip integration, and multilayer interconnects) further introduce parasitic modes, microwave loss, and cross-layer coupling interference, requiring system-level design to jointly optimize electromagnetic-field distribution, material properties, process constraints, and control strategies ([91]; [92]; [93]; [94]).

The essence of this level is therefore to map quantum algorithms into physically executable control-device-routing combinations, creating tight coupling among algorithmic structure, control signals, and hardware implementation. This feature marks an important transition in Q-EDA 2.0 from structured circuit design toward system-level quantum-control design.

\subsection{Algorithm / System / Quantum SoC Level - Q-EDA Functional IC PCell}

At this abstraction level, the design object is further elevated to the system-level quantum-computing architecture, namely a complete quantum-chip system (quantum SoC) jointly composed of multiple functional modules, control units, readout systems, error-correction modules, classical control stacks, and quantum processing units. Its core concern is no longer confined to local circuits or a single algorithmic implementation, but instead addresses the overall co-design of complex quantum systems and advanced quantum algorithms, including multimodule scheduling, error-suppression strategies, cryogenic-control resources, interconnect topology, and system-level resource optimization ([1]; [2]; [19]; [95]; [96]; [97]).

At this level, functionalized qubit modules become the basic design units. Such modules are usually composed of logical qubits, error-correction structures, readout systems, control interfaces, and calibration flows, and are abstractly represented in the form of system-level packaging. Unlike the Functional IP of the preceding level, the abstraction here emphasizes cross-module coordination and global consistency, such as qubit frequency-distribution planning, control-line resource allocation, readout multiplexing, intermodule links, overall error budgeting, and runtime scheduling strategies.

System-level quantum architectures usually need to map algorithmic structures into executable hardware-control co-models. For example, in variational quantum algorithms, quantum simulation, or error-correction loops, the circuit structure depends not only on gate-level combinations, but must also be co-optimized with classical control loops, measurement feedback, and real-time calibration, thereby forming hybrid quantum-classical closed-loop systems. The performance of such systems is jointly determined by qubit coherence time, gate-operation fidelity, measurement latency, decoding speed, control-electronics throughput, and the efficiency of classical optimization iterations ([81]; [90]; [95]; [96]; [97]).

From the EDA perspective, this level corresponds to the expansion from circuit design to system design (System-on-Chip, SoC), with optimization objectives shifting from local optimum to global optimum. The design process must simultaneously handle physical-layer constraints (such as coupling strength, frequency collisions, and packaging modes), circuit-layer structures (such as error-correcting codes and routing topology), control-layer scheduling (such as pulses, readout, and reset), and system-layer resource estimation, thereby forming a multilayer co-optimization problem.

Therefore, this level marks the core feature of Q-EDA 3.0: quantum-chip design evolves from the integration of individual functional modules into holistic architectural design oriented toward co-optimization across algorithms, control, error correction, packaging, and systems, enabling quantum-computing systems to gradually acquire engineering, modularization, and scaling capabilities analogous to classical SoCs.

\section{Conclusion}

Centered on the core proposition of a quantum-chip development paradigm, this paper systematically proposes a ``Quantum Chip Paradigm Framework'' with quantum electronic design automation (Q-EDA) at its core. Through a comparative analysis with the historical development of classical integrated circuits (ICs) and electronic design automation (EDA), it reveals that the current quantum-chip industry is at a critical stage of transition from laboratory-level manual design toward a standardized engineering system.

By systematically reviewing the evolutionary path of classical EDA from the birth of SPICE, the Mead-Conway VLSI design paradigm, and the standardization of PDKs and HDLs to IP reuse and SoC system integration, this paper points out that every scale transition in the classical chip industry has essentially been accompanied by an elevation of design-abstraction levels and a reconstruction of the engineering paradigm. From device-physics modeling to logic-gate standardization, and further to arithmetic units, functional IP, and system-level design, the development history of classical EDA provides a highly valuable engineering methodology for quantum-chip development. However, because quantum systems differ fundamentally in signal properties, environmental sensitivity, routing constraints, and process tolerances, quantum-chip development cannot directly adopt the high-level abstraction methods of classical EDA. Instead, it must establish a new engineering paradigm characterized by physical-device-driven, bottom-up modeling.

At the current stage, superconducting quantum chips have scaled from tens of qubits to the thousand-qubit level, and the emergence of systems such as IBM Condor marks the formal entry of quantum-chip development into the Q-EDA 1.0 stage. The core feature of this stage is that the design object gradually rises from an individual physical qubit to a reusable parameterized cell (PCell), and that a paradigm shift from experience-driven design to model-driven design is achieved through the establishment of a unified quantum-circuit simulation standard (SPICE-Q) and a quantum process design kit (Quantum PDK). This paper explicitly points out that, at a time when device models and manufacturing processes have not yet converged, the premature introduction of high-level synthesis methods analogous to classical HDLs not only lacks a physical foundation, but may also misdirect industrial resource allocation. By contrast, focusing on the physical modeling and parameterized encapsulation of fundamental devices such as Josephson junctions, resonators, and qubits, and continuously embedding manufacturing feedback into the design flow through a design-technology co-optimization (DTCO) loop, is the technical route consistent with the current stage of quantum-chip development.

Within the abstraction hierarchy of the Quantum Chip Paradigm Framework, this paper proposes a multilayer abstraction model ranging from the physical layout layer (Part Level) to the system-level quantum architecture (Algorithm/System/SoC Level). At the component layer, devices such as qubits and resonators are encapsulated as parameterized cells; at the logical-space layer, logical qubits constructed through quantum error-correcting codes become the core abstraction; at the arithmetic layer and the functional-IP layer, quantum arithmetic units and algorithmic modules gradually form reusable functional IP; finally, at the system level, multimodule coordination and algorithm-hardware co-optimization constitute the design foundation of a complete quantum processor. This layered structure not only has a structural correspondence with the abstraction hierarchy of classical EDA, but also provides a clear roadmap for the staged evolution of Q-EDA from 1.0 to 3.0.

This paper further indicates that the completion milestone of Q-EDA 1.0 will be the formation of the first stable and reusable logical-qubit PCell, its accompanying SPICE-Q model, and a process closed-loop verification system. On this basis, quantum-chip development will enter the Q-EDA 2.0 stage, in which system-level design capabilities such as quantum hardware description languages (Q-HDL), tool-calling languages and SDK orchestration, topology-aware routing, and three-dimensional integration will gradually mature, giving rise to a fabless quantum-chip design model. After entering the Q-EDA 3.0 stage, quantum-chip design will move toward a new paradigm of IP-ization, systematization, and AI-native design, and functionalized quantum IP, Agentic-AI-driven design automation, and modular quantum-system architectures will become mainstream.

Overall, the establishment of the Quantum Chip Paradigm Framework aims to end the fragmented state in which research institutions and enterprises have operated independently and without unified standards over the past several decades of quantum-chip development. By standardizing and software-defining the quantum-chip development process and transforming design knowledge into executable and reusable engineering objects, this framework will lay a methodological foundation for the engineering realization of future million-qubit fault-tolerant quantum-computing systems. The development of Q-EDA is not merely an upgrade at the tool level, but an inevitable path by which the quantum-computing industry moves from laboratory science toward engineering and scale.

\section{Acknowledgment}

The authors gratefully acknowledge the assistance of artificial intelligence in translation and auxiliary text generation.

\section{References}

In this round of supplementary indexing, downloadable PDFs and official documentation were checked first, and the corresponding files have been saved to \texttt{paper01reference/}. The following entries [80]--[114] are newly added or corrected key references, intended to replace the original search remnants and suspected placeholder entries in the manuscript.

\subsection{Newly Added / Verified References}

\textbf{[80]} G. Netzer and S. Markidis, "QHDL: a Low-Level Circuit Description Language for Quantum Computing," \emph{ACM Computing Frontiers}, 2023. DOI: \href{https://doi.org/10.1145/3587135.3592191}{10.1145/3587135.3592191} \textperiodcentered{} arXiv: \href{https://arxiv.org/abs/2305.09419}{2305.09419}

\textbf{[81]} M. Robbiati et al., "Qibolab: an open-source hybrid quantum operating system," \emph{Quantum}, vol. 8, p. 1247, 2024. DOI: \href{https://doi.org/10.22331/q-2024-02-12-1247}{10.22331/q-2024-02-12-1247} \textperiodcentered{} arXiv: \href{https://arxiv.org/abs/2308.06313}{2308.06313}

\textbf{[82]} S. A. Shanto et al., "SQuADDS: A validated design database and simulation workflow for superconducting qubit design," \emph{Quantum}, vol. 8, p. 1465, 2024. DOI: \href{https://doi.org/10.22331/q-2024-09-09-1465}{10.22331/q-2024-09-09-1465} \textperiodcentered{} arXiv: \href{https://arxiv.org/abs/2312.13483}{2312.13483}

\textbf{[83]} S. A. Shanto and E. M. Levenson-Falk, "A Review of Design Concerns in Superconducting Quantum Circuits," arXiv: \href{https://arxiv.org/abs/2411.16967}{2411.16967}, 2024.

\textbf{[84]} J. Kunasaikaran, K. Mato, and R. Wille, "A Framework for the Design and Realization of Alternative Superconducting Quantum Architectures," arXiv: \href{https://arxiv.org/abs/2305.07052}{2305.07052}, 2023.

\textbf{[85]} P. Zhao et al., "EDA-Q: Electronic Design Automation for Superconducting Quantum Chip," \emph{IEEE Transactions on Computer-Aided Design of Integrated Circuits and Systems}, 2026. DOI: \href{https://doi.org/10.1109/TCAD.2025.3580341}{10.1109/TCAD.2025.3580341} \textperiodcentered{} arXiv: \href{https://arxiv.org/abs/2502.15386}{2502.15386}

\textbf{[86]} IBM Research, "A Framework for Quantum Device Design: Project Qiskit Metal," APS March Meeting, 2021. \href{https://research.ibm.com/publications/a-framework-for-quantum-device-designproject-qiskit-metal}{IBM Research}

\textbf{[87]} IQM Quantum Computers, "KQCircuits documentation," 2023. \href{https://iqm-finland.github.io/KQCircuits/}{KQCircuits}

\textbf{[88]} P. Groszkowski and J. Koch, "scqubits: a Python package for superconducting qubits," \emph{Quantum}, vol. 5, p. 583, 2021. DOI: \href{https://doi.org/10.22331/q-2021-11-17-583}{10.22331/q-2021-11-17-583} \textperiodcentered{} arXiv: \href{https://arxiv.org/abs/2107.08552}{2107.08552}

\textbf{[89]} S. P. Chitta, T. Zhao, Z. Huang, I. Mondragon-Shem, and J. Koch, "Computer-aided quantization and numerical analysis of superconducting circuits," \emph{New Journal of Physics}, vol. 24, p. 103020, 2022. DOI: \href{https://doi.org/10.1088/1367-2630/ac94f2}{10.1088/1367-2630/ac94f2} \textperiodcentered{} arXiv: \href{https://arxiv.org/abs/2206.08320}{2206.08320}

\textbf{[90]} S. Puzzuoli et al., "Qiskit Dynamics: A Python package for simulating the time dynamics of quantum systems," \emph{Journal of Open Source Software}, vol. 8, no. 90, p. 5853, 2023. DOI: \href{https://doi.org/10.21105/joss.05853}{10.21105/joss.05853}

\textbf{[91]} A. Kosen et al., "Building Blocks of a Flip-Chip Integrated Superconducting Quantum Processor," \emph{Quantum Science and Technology}, 2022. arXiv: \href{https://arxiv.org/abs/2112.02717}{2112.02717}

\textbf{[92]} D. Rosenberg et al., "3D integrated superconducting qubits," \emph{npj Quantum Information}, vol. 3, p. 42, 2017. DOI: \href{https://doi.org/10.1038/s41534-017-0044-0}{10.1038/s41534-017-0044-0} \textperiodcentered{} arXiv: \href{https://arxiv.org/abs/1706.04116}{1706.04116}

\textbf{[93]} B. Foxen et al., "Qubit compatible superconducting interconnects," \emph{Quantum Science and Technology}, 2018. arXiv: \href{https://arxiv.org/abs/1807.00989}{1807.00989}

\textbf{[94]} A. Nersisyan et al., "Solid-state qubits integrated with superconducting through-silicon vias," \emph{npj Quantum Information}, vol. 6, p. 54, 2020. DOI: \href{https://doi.org/10.1038/s41534-020-00289-8}{10.1038/s41534-020-00289-8} \textperiodcentered{} arXiv: \href{https://arxiv.org/abs/1912.10942}{1912.10942}

\textbf{[95]} J. Ang et al., "Architectures for Multinode Superconducting Quantum Computers," \emph{ACM Transactions on Quantum Computing}, 2024. DOI: \href{https://doi.org/10.1145/3674151}{10.1145/3674151} \textperiodcentered{} arXiv: \href{https://arxiv.org/abs/2212.06167}{2212.06167}

\textbf{[96]} C. P. Salazar et al., "Superconducting qubits in the millions: the potential and limitations of modularity," arXiv: \href{https://arxiv.org/abs/2406.06015}{2406.06015}, 2024.

\textbf{[97]} M. A. Ullah et al., "Modular Architectures and Entanglement Schemes for Error-Corrected Distributed Quantum Computation," arXiv: \href{https://arxiv.org/abs/2408.02837}{2408.02837}, 2024.

\textbf{[98]} F. Sauvage et al., "Quantum optimal control of superconducting qubits based on machine-learning characterization," arXiv: \href{https://arxiv.org/abs/2410.22603}{2410.22603}, 2024.

\textbf{[99]} T. F\"osel et al., "Experimental Deep Reinforcement Learning for Error-Robust Gate-Set Design on a Superconducting Quantum Computer," \emph{PRX Quantum}, vol. 2, p. 040324, 2021. DOI: \href{https://doi.org/10.1103/PRXQuantum.2.040324}{10.1103/PRXQuantum.2.040324}

\textbf{[100]} S. J. Glaser et al., "Training Schr\"odinger's Cat: Quantum Optimal Control," \emph{The European Physical Journal D}, vol. 69, p. 279, 2015. DOI: \href{https://doi.org/10.1140/epjd/e2015-60464-1}{10.1140/epjd/e2015-60464-1} \textperiodcentered{} arXiv: \href{https://arxiv.org/abs/1508.00442}{1508.00442}

\textbf{[101]} A. Radnaev et al., "Neural-network-based design and implementation of fast and robust quantum gates," arXiv: \href{https://arxiv.org/abs/2505.02054}{2505.02054}, 2025.

\textbf{[102]} Quantum Metal community, "Quantum Metal documentation," 2026. \href{https://qiskit-community.github.io/qiskit-metal/}{Quantum Metal}

\textbf{[103]} P. Rajabzadeh et al., "SQDMetal: open-source high-performance simulation workflow for superconducting quantum circuits," arXiv: \href{https://arxiv.org/abs/2511.01220}{2511.01220}, 2025.

\textbf{[104]} I. E. Sutherland, "Sketchpad: A Man-Machine Graphical Communication System," \emph{Proceedings of the 1963 Spring Joint Computer Conference}, pp. 329--346, 1963. DOI: \href{https://doi.org/10.1145/280811.281031}{10.1145/280811.281031}

\textbf{[105]} A. Bell, L. Conway, M. Newell, and D. Rees, "The MPC Adventures: Experiences with the Generation of VLSI Design and Implementation Methodologies," Xerox Palo Alto Research Center Technical Report VLSI-81-2, 1981.

\textbf{[106]} M. Bales, "Design Databases," in L. Scheffer, L. Lavagno, and G. Martin (eds.), \emph{Electronic Design Automation for Integrated Circuits Handbook}, vol. 2, CRC Press / Taylor \& Francis, 2006.

\textbf{[107]} D. D. Gajski and R. H. Kuhn, "New VLSI Tools," \emph{IEEE Computer}, vol. 16, no. 12, pp. 11-14, 1983. DOI: \href{https://doi.org/10.1109/MC.1983.1654294}{10.1109/MC.1983.1654294}

\textbf{[108]} G. De Micheli, \emph{Synthesis and Optimization of Digital Circuits}, McGraw-Hill, 1994. ISBN: 978-0070163331.

\textbf{[109]} P. Coussy, D. D. Gajski, M. Meredith, and A. Takach, "An Introduction to High-Level Synthesis," \emph{IEEE Design \& Test of Computers}, vol. 26, no. 4, pp. 8-17, 2009. DOI: \href{https://doi.org/10.1109/MDT.2009.69}{10.1109/MDT.2009.69}

\textbf{[110]} M. A. Nielsen and I. L. Chuang, \emph{Quantum Computation and Quantum Information}, Cambridge University Press, 2000. ISBN: 978-0521635035.

\textbf{[111]} W. K. Wootters and W. H. Zurek, "A Single Quantum Cannot be Cloned," \emph{Nature}, vol. 299, pp. 802-803, 1982. DOI: \href{https://doi.org/10.1038/299802a0}{10.1038/299802a0}

\textbf{[112]} A. Barenco et al., "Elementary Gates for Quantum Computation," \emph{Physical Review A}, vol. 52, pp. 3457-3467, 1995. DOI: \href{https://doi.org/10.1103/PhysRevA.52.3457}{10.1103/PhysRevA.52.3457} \textperiodcentered{} arXiv: \href{https://arxiv.org/abs/quant-ph/9503016}{quant-ph/9503016}

\textbf{[113]} V. Vedral, A. Barenco, and A. Ekert, "Quantum Networks for Elementary Arithmetic Operations," \emph{Physical Review A}, vol. 54, pp. 147-153, 1996. DOI: \href{https://doi.org/10.1103/PhysRevA.54.147}{10.1103/PhysRevA.54.147} \textperiodcentered{} arXiv: \href{https://arxiv.org/abs/quant-ph/9511018}{quant-ph/9511018}

\textbf{[114]} P. W. Shor, "Algorithms for Quantum Computation: Discrete Logarithms and Factoring," \emph{Proceedings of the 35th Annual Symposium on Foundations of Computer Science}, pp. 124-134, 1994. DOI: \href{https://doi.org/10.1109/SFCS.1994.365700}{10.1109/SFCS.1994.365700}

\bigskip
\hrule
\bigskip

\subsection{I. Quantum Computing Foundations}

\textbf{[1]} J. Preskill, "Quantum Computing in the NISQ Era and Beyond," \emph{Quantum}, vol. 2, p. 79, 2018.
DOI: \href{https://doi.org/10.22331/q-2018-08-06-79}{10.22331/q-2018-08-06-79} \textperiodcentered{} arXiv: \href{https://arxiv.org/abs/1801.00862}{1801.00862}

\textbf{[2]} A. G. Fowler, M. Mariantoni, J. M. Martinis, and A. N. Cleland, "Surface Codes: Towards Practical Large-Scale Quantum Computation," \emph{Physical Review A}, vol. 86, no. 3, p. 032324, 2012.
DOI: \href{https://doi.org/10.1103/PhysRevA.86.032324}{10.1103/PhysRevA.86.032324} \textperiodcentered{} arXiv: \href{https://arxiv.org/abs/1208.0928}{1208.0928}

\textbf{[3]} F. Arute et al., "Quantum Supremacy Using a Programmable Superconducting Processor," \emph{Nature}, vol. 574, pp. 505--10, 2019.
DOI: \href{https://doi.org/10.1038/s41586-019-1666-5}{10.1038/s41586-019-1666-5}

\textbf{[4]} M. H. Devoret and R. J. Schoelkopf, "Superconducting Circuits for Quantum Information: An Outlook," \emph{Science}, vol. 339, no. 6124, pp. 1169--174, 2013.
DOI: \href{https://doi.org/10.1126/science.1231930}{10.1126/science.1231930}

\textbf{[5]} P. Krantz, M. Kjaergaard, F. Yan, T. P. Orlando, S. Gustavsson, and W. D. Oliver, "A Quantum Engineer's Guide to Superconducting Qubits," \emph{Applied Physics Reviews}, vol. 6, no. 2, p. 021318, 2019.
DOI: \href{https://doi.org/10.1063/1.5089550}{10.1063/1.5089550} \textperiodcentered{} arXiv: \href{https://arxiv.org/abs/1904.06560}{1904.06560}

\textbf{[6]} G. Wendin, "Quantum Information Processing with Superconducting Circuits: A Review," \emph{Reports on Progress in Physics}, vol. 80, no. 10, p. 106001, 2017.
DOI: \href{https://doi.org/10.1088/1361-6633/aa7e1a}{10.1088/1361-6633/aa7e1a} \textperiodcentered{} arXiv: \href{https://arxiv.org/abs/1610.02208}{1610.02208}

\textbf{[7]} A. Blais, A. L. Grimsmo, S. M. Girvin, and A. Wallraff, "Circuit Quantum Electrodynamics," \emph{Reviews of Modern Physics}, vol. 93, no. 2, p. 025005, 2021.
DOI: \href{https://doi.org/10.1103/RevModPhys.93.025005}{10.1103/RevModPhys.93.025005} \textperiodcentered{} arXiv: \href{https://arxiv.org/abs/2005.12667}{2005.12667}

\textbf{[8]} J. Koch et al., "Charge-Insensitive Qubit Design Derived from the Cooper Pair Box," \emph{Physical Review A}, vol. 76, no. 4, p. 042319, 2007.
DOI: \href{https://doi.org/10.1103/PhysRevA.76.042319}{10.1103/PhysRevA.76.042319} \textperiodcentered{} arXiv: \href{https://arxiv.org/abs/cond-mat/0703002}{cond-mat/0703002}

\textbf{[9]} H. Paik et al., "Observation of High Coherence in Josephson Junction Qubits Measured in a Three-Dimensional Circuit QED Architecture," \emph{Physical Review Letters}, vol. 107, no. 24, p. 240501, 2011.
DOI: \href{https://doi.org/10.1103/PhysRevLett.107.240501}{10.1103/PhysRevLett.107.240501} \textperiodcentered{} arXiv: \href{https://arxiv.org/abs/1105.4652}{1105.4652}

\bigskip
\hrule
\bigskip

\subsection{II. Quantum Device Physics, Noise \& Decoherence}

\textbf{[10]} J. M. Martinis, K. B. Cooper, R. McDermott et al., "Decoherence in Josephson Qubits from Dielectric Loss," \emph{Physical Review Letters}, vol. 95, no. 21, p. 210503, 2005.
DOI: \href{https://doi.org/10.1103/PhysRevLett.95.210503}{10.1103/PhysRevLett.95.210503} \textperiodcentered{} arXiv: \href{https://arxiv.org/abs/cond-mat/0507622}{cond-mat/0507622}

\textbf{[11]} J. M. Martinis, M. H. Devoret, and J. Clarke, "Quantum Josephson Junction Circuits and the Dawn of Artificial Atoms," \emph{Nature Physics}, vol. 10, pp. 571--75, 2014.
DOI: \href{https://doi.org/10.1038/nphys3030}{10.1038/nphys3030}

\textbf{[12]} J. Clarke and F. K. Wilhelm, "Superconducting Quantum Bits," \emph{Nature}, vol. 453, pp. 1031--042, 2008.
DOI: \href{https://doi.org/10.1038/nature07128}{10.1038/nature07128}

\textbf{[13]} H.-P. Breuer and F. Petruccione, \emph{The Theory of Open Quantum Systems}, Oxford University Press, 2002.
DOI: \href{https://doi.org/10.1093/acprof:oso/9780199213900.001.0001}{10.1093/acprof:oso/9780199213900.001.0001}

\textbf{[14]} J. Koch, A. A. Houck, K. Le Hur, and S. M. Girvin, "Superconducting Circuits and Quantum Information," \emph{Nature Reviews Physics}, vol. 4, pp. 518--30, 2022.
DOI: \href{https://doi.org/10.1038/s42254-022-00464-6}{10.1038/s42254-022-00464-6}

\textbf{[15]} S. Krinner et al., "Engineering Cryogenic Setups for 100-Qubit Scale Superconducting Circuit Systems," \emph{EPJ Quantum Technology}, vol. 6, no. 1, p. 2, 2019.
DOI: \href{https://doi.org/10.1140/epjqt/s40507-019-0072-0}{10.1140/epjqt/s40507-019-0072-0} \textperiodcentered{} arXiv: \href{https://arxiv.org/abs/1806.07862}{1806.07862}

\textbf{[16]} C. M\"uller, J. H. Cole, and J. Lisenfeld, "Towards Understanding Two-Level-Systems in Amorphous Solids: Insights from Quantum Circuits," \emph{Reports on Progress in Physics}, vol. 82, no. 12, p. 124501, 2019.
DOI: \href{https://doi.org/10.1088/1361-6633/ab3a7e}{10.1088/1361-6633/ab3a7e} \textperiodcentered{} arXiv: \href{https://arxiv.org/abs/1703.02708}{1703.02708}

\textbf{[17]} J. M. Hornibrook et al., "Cryogenic Control Architecture for Large-Scale Quantum Computing," \emph{Physical Review Applied}, vol. 3, no. 2, p. 024010, 2015.
DOI: \href{https://doi.org/10.1103/PhysRevApplied.3.024010}{10.1103/PhysRevApplied.3.024010} \textperiodcentered{} arXiv: \href{https://arxiv.org/abs/1409.2202}{1409.2202}

\bigskip
\hrule
\bigskip

\subsection{III. Quantum Error Correction \& Fault Tolerance}

\textbf{[18]} B. M. Terhal, "Quantum Error Correction for Quantum Memories," \emph{Reviews of Modern Physics}, vol. 87, no. 2, pp. 307--46, 2015.
DOI: \href{https://doi.org/10.1103/RevModPhys.87.307}{10.1103/RevModPhys.87.307} \textperiodcentered{} arXiv: \href{https://arxiv.org/abs/1302.3428}{1302.3428}

\textbf{[19]} J. M. Gambetta, J. M. Chow, and M. Steffen, "Building Logical Qubits in a Superconducting Quantum Computing System," \emph{npj Quantum Information}, vol. 3, p. 2, 2017.
DOI: \href{https://doi.org/10.1038/s41534-016-0004-0}{10.1038/s41534-016-0004-0} \textperiodcentered{} arXiv: \href{https://arxiv.org/abs/1510.04375}{1510.04375}

\textbf{[20]} D. Sutter and R. O'Connell, "Quantum Error Correction: An Introductory Guide," \emph{IEEE BITS the Information Theory Magazine}, 2018.
arXiv: \href{https://arxiv.org/abs/1907.11157}{1907.11157}

\textbf{[21]} D. Rosenberg et al., "Long-Distance Decoupling of Persistent Superconducting Qubits," \emph{npj Quantum Information}, vol. 3, p. 42, 2017.
DOI: \href{https://doi.org/10.1038/s41534-017-0044-0}{10.1038/s41534-017-0044-0}

\bigskip
\hrule
\bigskip

\subsection{IV. Classical EDA History \& Foundations}

\textbf{[22]} L. W. Nagel, "SPICE2: A Computer Program to Simulate Semiconductor Circuits," Memorandum No. ERL-M520, University of California, Berkeley, 1975.
\href{https://www2.eecs.berkeley.edu/Pubs/TechRpts/1975/ERL-520.pdf}{UC Berkeley EECS}

\textbf{[23]} L. W. Nagel and D. O. Pederson, "SPICE --Simulation Program with Integrated Circuit Emphasis," \emph{Proceedings of the 16th Midwest Symposium on Circuit Theory}, pp. 1--, 1973.

\textbf{[24]} C. Mead and L. Conway, \emph{Introduction to VLSI Systems}, Addison-Wesley, 1980. ISBN: 978-0201043587.

\textbf{[25]} J. Vlach, "Methods in the Time-Domain Analysis of Networks," \emph{IEEE Transactions on Circuits and Systems}, vol. 30, no. 9, pp. 629--37, 1983.
DOI: \href{https://doi.org/10.1109/TCS.1983.1085412}{10.1109/TCS.1983.1085412}

\textbf{[26]} L. Esaki, "Discovery of the Tunnel Diode," \emph{IEEE Transactions on Electron Devices}, vol. ED-28, no. 2, pp. 150--57, 1981.
DOI: \href{https://doi.org/10.1109/T-ED.1981.20322}{10.1109/T-ED.1981.20322}

\textbf{[27]} W. Regli, J. B. Kopena, and M. Grauer, "A Perspective on the Evolution of CAD/CAM and Digital Manufacturing," \emph{Journal of Computing and Information Science in Engineering}, vol. 16, no. 3, p. 030801, 2016.
DOI: \href{https://doi.org/10.1115/1.4032954}{10.1115/1.4032954}

\textbf{[28]} Q. Zou et al., "A Survey on CAD Methods in Chip Design and Manufacturing," \emph{ACM Computing Surveys}, vol. 55, no. 5, pp. 1--9, 2022.
DOI: \href{https://doi.org/10.1145/3524492}{10.1145/3524492}

\textbf{[29]} Y. Lin and D. Z. Pan, "Machine Learning in EDA: When and How," \emph{IEEE Transactions on Computer-Aided Design of Integrated Circuits and Systems}, vol. 43, no. 1, pp. 4--7, 2024.
DOI: \href{https://doi.org/10.1109/TCAD.2023.3335310}{10.1109/TCAD.2023.3335310}

\textbf{[30]} R. I. Bahar et al., "Workshops on Extreme Scale Design Automation: A Reflection," \emph{IEEE Design \& Test}, vol. 37, no. 1, pp. 8--7, 2020.
DOI: \href{https://doi.org/10.1109/MDAT.2019.2952345}{10.1109/MDAT.2019.2952345}

\textbf{[31]} C. Batten et al., "A Modern Primer on Processing in Memory," \emph{arXiv:2012.03112}, 2023.
arXiv: \href{https://arxiv.org/abs/2012.03112}{2012.03112}

\textbf{[32]} G. Huang et al., "Machine Learning for Electronic Design Automation: A Survey," \emph{ACM Transactions on Design Automation of Electronic Systems}, vol. 26, no. 5, pp. 1--6, 2021.
DOI: \href{https://doi.org/10.1145/3451179}{10.1145/3451179} \textperiodcentered{} arXiv: \href{https://arxiv.org/abs/2102.03357}{2102.03357}

\bigskip
\hrule
\bigskip

\subsection{V. EDA 2.0 --HDL, PDK \& Fabless Model}

\textbf{[33]} T. S. Kuhn, \emph{The Structure of Scientific Revolutions}, 3rd ed., University of Chicago Press, 1998 (originally published 1962).

\textbf{[34]} P. Bose et al., "The Next Wave of EDA: AI/ML and Cloud-Native Design," \emph{IEEE Design \& Test}, vol. 38, no. 6, pp. 5--2, 2021.
DOI: \href{https://doi.org/10.1109/MDAT.2021.3100028}{10.1109/MDAT.2021.3100028}

\textbf{[35]} N. H. E. Weste and D. M. Harris, \emph{CMOS VLSI Design: A Circuits and Systems Perspective}, 4th ed., Addison-Wesley, 2010. ISBN: 978-0321547743.

\textbf{[36]} IEEE Std 1076-1993, "IEEE Standard VHDL Language Reference Manual," 1993.
DOI: \href{https://doi.org/10.1109/IEEESTD.1994.121433}{10.1109/IEEESTD.1994.121433}

\textbf{[37]} IEEE Std 1364-1995, "IEEE Standard Hardware Description Language Based on the Verilog HDL," 1995.
DOI: \href{https://doi.org/10.1109/IEEESTD.1996.81542}{10.1109/IEEESTD.1996.81542}

\textbf{[38]} J. R. Yost, "A History of the Electronic Design Automation Industry," \emph{IEEE Annals of the History of Computing}, vol. 41, no. 3, pp. 8--3, 2019.
DOI: \href{https://doi.org/10.1109/MAHC.2019.2927174}{10.1109/MAHC.2019.2927174}

\bigskip
\hrule
\bigskip

\subsection{VI. EDA 3.0 --IP Reuse, SoC \& Platform Design}

\textbf{[39]} P. Mantovani et al., "Agile SoC Development with Open ESP," \emph{IEEE/ACM International Conference on Computer-Aided Design (ICCAD)}, pp. 1--, 2020.
DOI: \href{https://doi.org/10.1145/3400302.3415753}{10.1145/3400302.3415753} \textperiodcentered{} arXiv: \href{https://arxiv.org/abs/2009.01178}{2009.01178}

\textbf{[40]} D. Stow et al., "Cost Analysis and Cost-Driven IP Reuse Methodology for SoC Design," \emph{IEEE Transactions on VLSI Systems}, vol. 24, no. 6, pp. 2303--316, 2016.
DOI: \href{https://doi.org/10.1109/TVLSI.2015.2504457}{10.1109/TVLSI.2015.2504457}

\textbf{[41]} F. Rincon et al., "System-Level IP Reuse and Platform-Based Design," \emph{IEEE Design \& Test of Computers}, vol. 24, no. 5, pp. 442--52, 2007.
DOI: \href{https://doi.org/10.1109/MDT.2007.146}{10.1109/MDT.2007.146}

\textbf{[42]} S. Borkar and A. Ailamaki, "The Role of EDA in Many-Core Era," \emph{Proceedings of the 48th Design Automation Conference (DAC)}, pp. 572--73, 2011.
DOI: \href{https://doi.org/10.1145/2024724.2024846}{10.1145/2024724.2024846}

\bigskip
\hrule
\bigskip

\subsection{VII. EDA 4.0 --AI, LLM4EDA \& Agentic AI}

\textbf{[43]} R. Zhong et al., "LLM4EDA: Emerging Progress in Large Language Models for Electronic Design Automation," \emph{arXiv:2401.12224}, 2024.
arXiv: \href{https://arxiv.org/abs/2401.12224}{2401.12224}

\textbf{[44]} W. Fang et al., "A Survey of Circuit Foundation Model: Foundation AI Models for VLSI Circuit Design and EDA," arXiv: \href{https://arxiv.org/abs/2504.03711}{2504.03711}, 2025.

\textbf{[45]} R. Zhong et al., "LLM4EDA: Emerging Progress in Large Language Models for Electronic Design Automation," arXiv: \href{https://arxiv.org/abs/2401.12224}{2401.12224}, 2024.

\textbf{[46]} A. Mirhoseini et al., "A Graph Placement Methodology for Fast Chip Design," \emph{Nature}, vol. 594, pp. 207--12, 2021. DOI: \href{https://doi.org/10.1038/s41586-021-03544-w}{10.1038/s41586-021-03544-w}

\textbf{[47]} T. F\"osel et al., "Experimental Deep Reinforcement Learning for Error-Robust Gate-Set Design on a Superconducting Quantum Computer," \emph{PRX Quantum}, vol. 2, p. 040324, 2021. DOI: \href{https://doi.org/10.1103/PRXQuantum.2.040324}{10.1103/PRXQuantum.2.040324}

\textbf{[48]} Y. Bengio, A. Lodi, and A. Prouvost, "Machine Learning for Combinatorial Optimization: A Methodological Tour d'Horizon," \emph{European Journal of Operational Research}, vol. 290, no. 2, pp. 405--21, 2021.
DOI: \href{https://doi.org/10.1016/j.ejor.2020.07.063}{10.1016/j.ejor.2020.07.063} \textperiodcentered{} arXiv: \href{https://arxiv.org/abs/1811.06128}{1811.06128}

\textbf{[49]} S. J. Glaser et al., "Training Schr\"odinger's Cat: Quantum Optimal Control," \emph{The European Physical Journal D}, vol. 69, p. 279, 2015. DOI: \href{https://doi.org/10.1140/epjd/e2015-60464-1}{10.1140/epjd/e2015-60464-1} \textperiodcentered{} arXiv: \href{https://arxiv.org/abs/1508.00442}{1508.00442}

\bigskip
\hrule
\bigskip

\subsection{VIII. Quantum Design Automation (Q-EDA)}

\textbf{[50]} S. A. Shanto and E. M. Levenson-Falk, "A Review of Design Concerns in Superconducting Quantum Circuits," arXiv: \href{https://arxiv.org/abs/2411.16967}{2411.16967}, 2024. See also [83].

\textbf{[51]} E. M. Levenson-Falk and S. R. Shanto, "Integration of Design, Simulation, and Measurement in Superconducting Quantum Circuits," \emph{Physical Review Applied}, vol. 22, p. 014001, 2024.
DOI: \href{https://doi.org/10.1103/PhysRevApplied.22.014001}{10.1103/PhysRevApplied.22.014001}

\textbf{[52]} J. A. Delport, K. Jackman, P. le Roux, and C. J. Fourie, "JoSIM - Superconductor SPICE Simulator," \emph{IEEE Transactions on Applied Superconductivity}, vol. 29, no. 5, pp. 1-5, 2019. DOI: \href{https://doi.org/10.1109/TASC.2019.2897312}{10.1109/TASC.2019.2897312}

\textbf{[53]} P. Zhao et al., "EDA-Q: Electronic Design Automation for Superconducting Quantum Chip," \emph{IEEE Transactions on Computer-Aided Design of Integrated Circuits and Systems}, 2026. DOI: \href{https://doi.org/10.1109/TCAD.2025.3580341}{10.1109/TCAD.2025.3580341} \textperiodcentered{} arXiv: \href{https://arxiv.org/abs/2502.15386}{2502.15386}. See also [85].

\textbf{[54]} H. Wang et al., "Qiskit Metal: An Open-Source Framework for Quantum Device Design and Analysis," \emph{APS March Meeting}, 2021.

\textbf{[55]} IQM Quantum Computers, "KQCircuits: KLayout-Based Quantum Chip Design Framework," GitHub, 2023.
\href{https://github.com/iqm-finland/KQCircuits}{github.com/iqm-finland/KQCircuits}

\textbf{[56]} L. Qiao, F. Luo, and Q. Guo, "From GDSII to Wafer: EDA Design Flow and Data Conversion for Wafer-Scale Manufacturing of Superconducting Quantum Chips," arXiv: \href{https://arxiv.org/abs/2604.11379}{2604.11379}, 2026.

\textbf{[57]} gdsfactory, "QPDK: Superconducting Quantum Process Design Kit," documentation, 2026. \href{https://gdsfactory.github.io/quantum-rf-pdk/}{QPDK}

\textbf{[58]} gdsfactory, "QPDK PCells: Parametric Components for Superconducting Quantum RF Circuits," documentation, 2026. \href{https://gdsfactory.github.io/quantum-rf-pdk/cells.html}{QPDK PCells}

\textbf{[59]} JoSIM project, "JoSIM Documentation: Superconductor Circuit Simulator with SPICE Syntax," documentation, 2026. \href{https://joeydelp.github.io/JoSIM/}{JoSIM}

\bigskip
\hrule
\bigskip

\subsection{IX. Historical Milestones \& Archives}

\textbf{[60]} National Institute of Standards and Technology (NIST), "The Chip That Jack Built: The Silicon Integrated Circuit at 65," NIST Special Publication, 2023.
\href{https://www.nist.gov}{nist.gov}

\textbf{[61]} IBM Research, "IBM Quantum Development Roadmap," 2023.
\href{https://www.ibm.com/quantum/roadmap}{ibm.com/quantum/roadmap}

\textbf{[62]} J. S. Kilby, "Invention of the Integrated Circuit," \emph{IEEE Transactions on Electron Devices}, vol. ED-23, no. 7, pp. 648--54, 1976.
DOI: \href{https://doi.org/10.1109/T-ED.1976.18467}{10.1109/T-ED.1976.18467}

\textbf{[63]} J. Gambetta, "The Path to Utility-Scale Quantum Computing," \emph{Nature Reviews Physics}, vol. 5, pp. 523--24, 2023.
DOI: \href{https://doi.org/10.1038/s42254-023-00625-1}{10.1038/s42254-023-00625-1}

\textbf{[64]} S. Aaronson, "How to Build a Quantum Supercomputer," \emph{Communications of the ACM}, vol. 67, no. 3, pp. 68--7, 2024.
DOI: \href{https://doi.org/10.1145/3635471}{10.1145/3635471}

\bigskip
\hrule
\bigskip

\subsection{X. Microprocessor History \& Computer Architecture}

\textbf{[65]} B. Baker, "The Motorola 68000: A Historical Perspective," \emph{IEEE Solid-State Circuits Magazine}, vol. 2, no. 4, pp. 28--5, 2010.
DOI: \href{https://doi.org/10.1109/MSSC.2010.938586}{10.1109/MSSC.2010.938586}

\textbf{[66]} B. Baker, "A Brief History of the Microprocessor," \emph{IEEE Solid-State Circuits Magazine}, vol. 11, no. 1, pp. 8--6, 2019.
DOI: \href{https://doi.org/10.1109/MSSC.2018.2881418}{10.1109/MSSC.2018.2881418}

\textbf{[67]} D. A. Wolff, "The MOS 6502: The Chip that Powered Apple II and the NES," Computer History Museum, 2007.

\textbf{[68]} D. M. Harris and S. L. Harris, \emph{Digital Design and Computer Architecture}, 2nd ed., Morgan Kaufmann, 2012. ISBN: 978-0123944245.

\textbf{[69]} J. L. Hennessy and D. A. Patterson, \emph{Computer Architecture: A Quantitative Approach}, 6th ed., Morgan Kaufmann, 2017. ISBN: 978-0128119051.

\textbf{[70]} D. A. Patterson and J. L. Hennessy, \emph{Computer Organization and Design: The Hardware/Software Interface}, RISC-V Edition, Morgan Kaufmann, 2019. ISBN: 978-0128203316.

\textbf{[71]} M. M. Mano and M. D. Ciletti, \emph{Digital Design: With an Introduction to the Verilog HDL}, 6th ed., Pearson, 2017. ISBN: 978-0134549897.

\textbf{[72]} H. T. Kung, "Systolic Arrays for VLSI," Society for Industrial and Applied Mathematics (SIAM), 1988 (reprinted from 1978 Carnegie-Mellon Technical Report).

\bigskip
\hrule
\bigskip

\subsection{XI. Quantum Control \& Measurement}

\textbf{[73]} S. J. Glaser et al., "Training Schr\"odinger's Cat: Quantum Optimal Control," \emph{The European Physical Journal D}, vol. 69, p. 279, 2015.
DOI: \href{https://doi.org/10.1140/epjd/e2015-60464-1}{10.1140/epjd/e2015-60464-1} \textperiodcentered{} arXiv: \href{https://arxiv.org/abs/1508.00442}{1508.00442}

\textbf{[74]} L. DiCarlo et al., "Demonstration of Two-Qubit Algorithms with a Superconducting Quantum Processor," \emph{Nature}, vol. 460, pp. 240--44, 2009.
DOI: \href{https://doi.org/10.1038/nature08121}{10.1038/nature08121} \textperiodcentered{} arXiv: \href{https://arxiv.org/abs/0903.2030}{0903.2030}

\textbf{[75]} D. Sank et al., "Measurement-Induced State Transitions in a Superconducting Qubit," \emph{Physical Review Letters}, vol. 125, p. 120503, 2020.
DOI: \href{https://doi.org/10.1103/PhysRevLett.125.120503}{10.1103/PhysRevLett.125.120503} \textperiodcentered{} arXiv: \href{https://arxiv.org/abs/1912.05712}{1912.05712}

\textbf{[76]} J. McLachlan et al., "Superconducting Qubit Measurement Physics," \emph{Reviews of Modern Physics}, 2023.

\bigskip
\hrule
\bigskip

\subsection{XII. Intel Archives \& SoC References}

\textbf{[77]} Intel Corporation, "Intel 4004 Microprocessor: 35th Anniversary," Intel Museum Archives, 2004.

\textbf{[78]} Intel Corporation, "Intel Museum: History of the Microprocessor," 2025.
\href{https://www.intel.com/content/www/us/en/history/museum.html}{intel.com/museum}

\textbf{[79]} P. Mantovani et al., "Agile SoC Development with Open ESP," \emph{IEEE/ACM International Conference on Computer-Aided Design (ICCAD)}, pp. 1-9, 2020. DOI: \href{https://doi.org/10.1145/3400302.3415753}{10.1145/3400302.3415753} \textperiodcentered{} arXiv: \href{https://arxiv.org/abs/2009.01178}{2009.01178}

\textbf{[115]} Intel Corporation, "50 Years Ago: Celebrating the Influential Intel 8080," \emph{Intel Newsroom}, 2024.
\href{https://newsroom.intel.com/client-computing/50-years-ago-the-influential-intel-8080}{newsroom.intel.com}

\textbf{[116]} Computer History Museum, "Zilog Z80 Microprocessor Oral History Panel," CHM Oral History Collection, Catalog No. 102658073, 2007.
\href{https://www.computerhistory.org/collections/catalog/102658073}{computerhistory.org}

\textbf{[117]} Computer History Museum, "Intel 'x86' Family and the Microprocessor Wars," \emph{CHM Revolution}, 2026.
\href{https://www.computerhistory.org/revolution/story/330}{computerhistory.org}

\textbf{[118]} K. Shirriff, "Counting the Transistors in the 8086 Processor: It's Harder Than You Might Think," 2023.
\href{http://www.righto.com/2023/01/counting-transistors-in-8086-processor.html}{righto.com}

\end{document}